\documentclass{article}
\usepackage{jheppub_1}
\usepackage[english]{babel}

\usepackage{amsmath,amsfonts,amssymb}
\usepackage{graphicx}
\usepackage{siunitx}
\usepackage{xspace}
\usepackage{braket}
\usepackage{mathtools}
\usepackage[colorlinks=true, allcolors=magenta]{hyperref}
\usepackage[capitalize]{cleveref}
\binoppenalty=\maxdimen
\relpenalty=\maxdimen 

\graphicspath{{./graphics/}}

\newcommand*{\GW}{\ensuremath{\mathrm{GW}}\xspace}
\newcommand*{\Neff}{\ensuremath{N_\mathrm{eff}}\xspace}
\newcommand*{\Trh}{\ensuremath{T_\mathrm{rh}}\xspace}
\newcommand*{\Reff}{\ensuremath{R_\mathrm{eff}}\xspace}
\newcommand*{\Req}{\ensuremath{R_\mathrm{eq}}\xspace}
\newcommand*{\rhoR}{\ensuremath{\rho_{R}}\xspace}
\newcommand*{\geff}{\ensuremath{g_{*}}\xspace}
\newcommand*{\gs}{\ensuremath{g_{s}}\xspace}

\DeclareSIUnit\parsec{pc}
\crefname{section}{section}{sections}
\Crefname{section}{Section}{Sections}
\crefname{appendix}{appendix}{appendices}
\Crefname{appendix}{Appendix}{Appendices}
\crefname{table}{Tab.}{Tabs.}
\Crefname{table}{Table}{Tables}

\makeatletter
\gdef\@fpheader{\strut}
\makeatother

\DeclareMathOperator{\erfc}{erfc}

\title{Primordial gravitational waves in the nano-Hertz regime and PTA data --- towards solving the GW inverse problem}

\preprint{MITP-23-029}

\author[a]{Eric~Madge,}
\author[b,c,d]{Enrico~Morgante,}
\author[b]{Cristina~Puchades-Ib\'a\~nez,}
\author[b]{Nicklas~Ramberg,}
\author[a]{Wolfram~Ratzinger,}
\author[b]{Sebastian~Schenk,}
\author[b]{and Pedro~Schwaller}

\affiliation[a]{Department of Particle Physics and Astrophysics, Weizmann Institute of Science,\\ Herzl Street 234, Rehovot, 7610001, Israel}
\affiliation[b]{PRISMA$^+$  Cluster of Excellence and Mainz Institute for Theoretical Physics,\\ Johannes Gutenberg-Universit\"at  Mainz,\\ Staudingerweg 7, 55099 Mainz, Germany}
\affiliation[c]{Dipartimento di Fisica, Università di Trieste,\\ Strada Costiera 11, I-34151 Trieste, Italy}
\affiliation[d]{INFN, Sezione di Trieste,\\ Via Valerio 2, 34127 Trieste, Italy}

\emailAdd{eric.madge-pimentel@weizmann.ac.il}
\emailAdd{enrico.morgante@units.it}
\emailAdd{crpuchad@uni-mainz.de}
\emailAdd{nramberg@uni-mainz.de}
\emailAdd{wolfram.ratzinger@weizmann.ac.il}
\emailAdd{sebastian.schenk@uni-mainz.de}
\emailAdd{pedro.schwaller@uni-mainz.de}

\abstract{In recent years, several pulsar timing array collaborations have reported first hints for a stochastic gravitational wave background at nano-Hertz frequencies. Here we elaborate on the possibility that this signal comes from new physics that leads to the generation of a primordial stochastic gravitational wave background. We propose a set of simple but concrete models that can serve as benchmarks for gravitational waves sourced by cosmological phase transitions, domain wall networks, cosmic strings, axion dynamics, or large scalar fluctuations. These models are then confronted with pulsar timing data and with cosmological constraints. With only a limited number of free parameters per model, we are able to 
identify viable regions of parameter space and also make predictions for future astrophysical and laboratory tests that can help with model identification and discrimination. 
}

\begin{document}
% %%%%%%%%%%%%%%%%%%%%%%%%%%%%%%%%%%%%%%%%%%%%%%%%%%%%%%%%%%%%%%%%%%%%%%

\maketitle\flushbottom
\newpage

% ======================================================================
\section{Introduction}
% ======================================================================

The discovery of the first gravitational wave~(GW) signal by the LIGO/VIRGO~\cite{LIGOScientific:2016aoc} collaboration has started a new era in astrophysics and cosmology. GWs offer a new avenue to explore the physics of the very early universe, since they travel almost entirely undisturbed through space-time. While LIGO and other ground based GW detectors are most sensitive to GWs with kilo-Hertz frequencies, here we focus on much longer wavelengths down to nano-Hertz~(nHz). GWs in that frequency range are searched for by pulsar timing array~(PTA) experiments such as the European Pulsar Timing Array~(EPTA)~\cite{Kramer:2013kea,Desvignes:2016yex}, the North American Nanohertz Observatory for Gravitational Waves~(NANOGrav)~\cite{McLaughlin:2013ira,NANOGrav:2020gpb} and the Parkes Pulsar Timing Array~(PPTA)~\cite{Hobbs:2013aka,Kerr:2020qdo}, as well as their joint collaboration, the International Pulsar Timing Array~(IPTA)~\cite{Manchester:2013ndt,Perera:2019sca}.

In fact, there is now mounting evidence for a stochastic gravitational wave background~(SGWB) in the nHz range from the observation and timing of pulsars by several PTA collaborations~(NANOGrav~\cite{NANOGrav:2020bcs,NANOGrav:2021flc}, PPTA~\cite{Goncharov:2021oub}, EPTA~\cite{Chen:2021rqp} and IPTA~\cite{Antoniadis:2022pcn}). Such a signal is expected from GW emission from supermassive black hole binaries~(SMBHBs)~\cite{Becsy:2022pnr,Ellis:2023owy,DeRocco:2023qae}, but it can also be due to so far unknown new physics sources such as cosmological first order phase transitions~(FOPTs)~\cite{Nakai:2020oit, Ratzinger:2020koh, NANOGrav:2021flc, Ashoorioon:2022raz, Freese:2022qrl, Morgante:2022zvc, Bringmann:2023opz}, annihilating domain wall~(DW) networks~\cite{Ferreira:2022zzo}, cosmic strings~(CSs)~\cite{Blasi:2020mfx,Ellis:2020ena,Buchmuller:2020lbh, Samanta:2020cdk, Lazarides:2021uxv, Lazarides:2022jgr}, axion dynamics~\cite{Ratzinger:2020koh,Wang:2022rjz}, or induced by large scalar fluctuations~\cite{Vaskonen:2020lbd,Kohri:2020qqd,DeLuca:2020agl,Zhao:2022kvz,Dandoy:2023jot}. 

An interesting aspect of the PTA frequency range is that, due to causality, it requires sources which are active down to cosmological temperatures of a few Mega-electronvolt~(MeV). This is exactly the time when we start to have, at least indirect, constraints on the content and dynamics of the universe, from observations of the cosmic microwave background~(CMB), big bang nucleosynthesis~(BBN) and neutrino decoupling. Thus, while there are many mechanisms known to produce sufficiently large GW signals to explain the PTA data, it is non-trivial to write down concrete models which satisfy all existing constraints. 

The main goal of our paper is therefore to identify well defined benchmark models that can be confronted with both the GW data and cosmological constraints. Our focus here is on simplicity, i.e.\ to introduce as few new fields, parameters and couplings as possible, that nevertheless allow a consistent cosmological history. This also reduces the penalty for having too many parameters in Bayesian model comparison, which can in particular plague models of FOPTs, for which the signal parameterisation can be quite complicated even for simple underlying models. 
Other open questions that motivate our work are whether the frequency spectrum of the PTA observations can already discriminate between different models, whether the data from different PTAs can be consistently explained within these models, and which other cosmological observables can be used to identify or rule out different explanations of the signal. 

In \cref{sec:sources}, we introduce our benchmark models and their GW spectra. These are a simple abelian Higgs model with classical scale invariance as a benchmark for GWs from supercooled FOPTs, several axion and axion-like particle (ALP) models that serve as benchmarks for GWs from global CSs, DWs and from bosonic instabilities, and a single field inflationary model for secondary GWs from large density fluctuations. The GW spectra of these models are confronted with the PTA data using the \texttt{enterprise} software suite~\cite{enterprise,enterprise_extensions} and the rapid fitting tool \texttt{ceffyl}~\cite{Lamb:2023jls}, as discussed in more detail in \cref{sec:fit}. Cosmological constraints in general and also separately for each model are discussed in \cref{sec:results}. There we also present the results of our fits to the PTA data and identify the regions of parameter space which satisfy all cosmological constraints. We conclude with some thoughts about the necessary steps for addressing the GW inverse problem~\cite{Friedrich:2022cak}, in anticipation of upcoming data releases by the PTA collaborations. 

% ======================================================================
\section{Primordial sources of GWs in the nano-Hertz regime}
\label{sec:sources}
% ======================================================================

A variety of models and scenarios can give rise to a primordial SGWB, see e.g.\ Ref.~\cite{Caprini:2018mtu} for a review. Our goal here is not a comprehensive overview, but to study a selection of simple, but concrete, models that can produce a GW signal compatible with the current PTA data. This allows us to also confront the models with cosmological and laboratory constraints which provide complementary insight into their viable parameter space. 

An important consideration is the magnitude and frequency range of the signal. To fully explain the observed excess noise, the GWs should today have a fractional energy density $\Omega_\GW h^2 \sim 10^{-9}$ in the \SIrange{e-9}{e-8}{\Hz} range. GWs of this frequency are produced before matter-radiation equality, and thus subject to dilution by a redshift factor ${\cal F} \sim 10^{-5}$ between production and today. The largest possible amplitude is further constrained by the fraction of the total energy density carried by the source at the time of production $\Omega_S$, and by the ratio of the typical length scale of the source divided by the Hubble radius at production, $(LH_*)$, to some power. Thus, the expressions for the peak amplitude can all be cast in a form 
\begin{equation}
    \Omega_{\GW}h^2 \sim {\cal F}\, \Omega_S^2 (L H_*)^n\,,
\end{equation}
where the power $n$ depends on the type of source and the expression should be evaluated at the time of GW production. 
Consistency and causality imply that these factors cannot exceed unity, but they can easily suppress the signal by several orders of magnitude in realistic models. Said differently, most scenarios that give rise to primordial GWs will fail to produce a sufficiently large signal. 

We therefore focus on models and scenarios that are known to produce large GW signals at least in parts of their parameter space, which are also the models that will first be observed, or constrained, by future GW searches.\footnote{In the context of LHC searches for new physics, the term "supermodels" was used for such scenarios~\cite{Bauer:2009cc}.} Since it will appear frequently below, let us also define the reduced Planck scale here as~$M_{\rm P} = \SI{2.4e18}{\GeV}$. 

% ----------------------------------------------------------------------
\subsection{Strong first order phase transitions}
\label{sec:FOPT}
% ----------------------------------------------------------------------

Spontaneously broken symmetries are typically restored at high temperatures~\cite{Kirzhnits:1972ut}.
As a consequence, the early universe probably went through at least two cosmological phase transitions~(PTs) related to the breaking of the electroweak gauge symmetry and chiral symmetry breaking.
While both transitions in the Standard Model~(SM) are crossovers~\cite{DOnofrio:2015gop,HotQCD:2018pds} and hence do not generate GWs, new physics effects can modify the SM transitions to first order, or induce additional symmetry breaking transitions, which can then generate a SGWB~\cite{Witten:1984rs,Hogan:1986qda}.

Cosmological FOPTs proceed via the nucleation of spherical bubbles of the true vacuum within the false vacuum~\cite{Coleman:1977py,Callan:1977pt,Linde:1980tt,Linde:1981zj}.
The latent heat released in the transition drives the expansion of the bubbles, which then collide and merge and eventually fill the entire universe with the true vacuum. 
Interactions of the vacuum bubbles with the ambient primordial plasma of the universe further induce bulk motion of the fluid, forming fluid shells around the bubble walls and producing sound waves as well as turbulent motion in the plasma.
Collisions of the vacuum bubbles and fluid shells as well as sound waves and turbulence can then act as sources of a SGWB~\cite{Witten:1984rs,Hogan:1986qda,Kamionkowski:1993fg}.

A FOPT can be characterized by four parameters: the temperature~$T_*$ at which it occurs, the transition strength~$\alpha$, its characteristic timescale~$\beta^{-1}$, and the wall velocity~$v_w$ at which the bubbles expand.
The transition temperature is approximately given by the percolation temperature~\cite{Athron:2022mmm,Athron:2023xlk}, $T_* \sim T_p$, at which the probability to be in the false vacuum falls below $e^{-0.34}$~\cite{Guth:1982pn}.
The strength parameter~$\alpha$ is often defined as the latent heat release normalized to the energy density~\rhoR in radiation at the time of the transition.\footnote{%
    Note, however, that different definitions are used throughout the literature, see e.g.\ Ref.~\cite{Giese:2020rtr} for a discussion.
}
We here use the approximation $\alpha \approx \Delta V(T_*)/\rhoR(T_*)$, where $\Delta V(T_*)$ is the potential difference between the minima, which holds for strong transitions.
If the energy released in the transition is non-negligible compared to the radiation energy density, and the transitioning field couples to the plasma, the universe is subsequently reheated to the temperature $\Trh \sim T_p (1+\alpha)^{1/4}$.
The timescale~$\beta = \dot{\Gamma}/\Gamma$ is related to the time derivative of the nucleation rate~$\Gamma$. 
The wall velocity~$v_w$, last but not least, depends on the pressure on the wall exerted by the plasma, and hence on the particle physics model under consideration.
In the following, we will focus on PTs that release sufficient energy to accelerate the bubbles close to the speed of light, i.e.\ $v_w=1$. 

The corresponding GW spectrum can be extracted from numerical simulations and is approximated as a single or double broken power-law for the bubble and fluid shell collision~\cite{Lewicki:2020jiv,Ellis:2020nnr,Lewicki:2020azd,Lewicki:2022pdb} or sound wave~\cite{Hindmarsh:2019phv,Jinno:2020eqg,Jinno:2022mie} and turbulence~\cite{Auclair:2022jod,RoperPol:2022iel} contribution, respectively.
The parametric dependence of the peak frequency and peak amplitude on the PT parameters is given by
\begin{align}
    f_p \propto T_* \frac{\beta}{H_*} \,, \qquad \Omega_{\GW}h^2 \propto \mathcal{F} \left(\frac{\kappa\, \alpha}{1+\alpha}\right)^p \left(\frac{H_*}{\beta}\right)^q \,,
    \label{eq:pt_gw_scaling}
\end{align}
where $H_*$ is the Hubble rate at the temperature $T_*$.
Furthermore, we assumed $v_w=1$, and $\mathcal{F}$ is a redshift factor that depends on the transition temperature only indirectly through the effective number of degrees of freedom.
The powers $p$ and $q$ as well as the efficiency factors~$\kappa$ for converting vacuum energy into the respective source depend on the contribution under consideration (cf.\ e.g.\ Ref.~\cite{Breitbach:2018ddu}).
The collision contributions for instance scale as $p=q=2$.

As can be inferred from the scaling of \cref{eq:pt_gw_scaling}, observable GW signals typically require strong transitions with $\alpha \gtrsim 1$.
We hence here focus on transitions that exhibit a large amount of supercooling, where the energy released in the tunneling can be comparable or even exceed the energy of the radiation bath, and the plasma friction can be insufficient to prevent the bubble walls from accelerating until collision. 
In this case, the dominant contributions to the GW spectrum come from the collisions of the vacuum bubbles as well as the relativistic fluid shells~\cite{Lewicki:2022pdb}.

\bigskip 

A simple class of models known to feature strong FOPTs are scenarios of near-conformal dynamics undergoing a supercooled PT~\cite{Jinno:2016knw,Iso:2017uuu,Levi:2022bzt,Marzola:2017jzl,Azatov:2019png,Borah:2021ftr,Kierkla:2022odc,Freese:2022qrl,Sagunski:2023ynd}. Arguably the simplest model is that of a classically scale invariant scalar field with radiative breaking of conformal symmetry, the famous Coleman-Weinberg~(CW) model~\cite{Coleman:1973jx}.
At the classical level the CW theory is defined by the Lagrangian density
\begin{equation}
    {\cal L} = -\frac{1}{4}F^2_{\mu\nu}+D_{\mu}\Phi^\dagger D^{\mu}\Phi - V(\Phi,T)
\end{equation}
where $D^{\mu}\Phi=\partial^\mu\Phi-igA^\mu \Phi$ is the covariant derivative, and the tree-level potential of the conformal theory is $V_0 = \lambda |\Phi|^4$.
Loop corrections to the potential then induce radiative symmetry breaking.
The corresponding PT can be of first order, provided that the potential exhibits a sufficiently flat direction, i.e.\ $\lambda \ll g^4$.
To simplify the analysis, we hence neglect the tree-level contribution and focus on the potential $V_1(\phi,T)$ generated at one-loop level for the order parameter $\phi$ of the theory, such that $\Phi = (\phi + i G )/\sqrt{2}$, and where $G$ denotes the Goldstone boson that becomes the longitudinal mode of the gauge boson after symmetry breaking. 
The theory is then described by only two parameters: the gauge coupling $g$, and the renormalization scale~$\mu_R$ at which the coupling is given.

At zero-temperature, the potential is given by the CW one-loop potential~\cite{Coleman:1973jx}. 
The field develops a global minimum at $\langle\phi\rangle= e^{1/6} \mu_R/g$, breaking the gauge symmetry spontaneously.
The gauge bosons then acquire a mass $m_A = g \langle\phi\rangle$, while the mass of the scalar mode is loop-suppressed, $m_\phi = \sqrt{3/2} g^2 / (2\pi) \langle\phi\rangle$.

The finite-temperature effective potential at one-loop can be expressed in the high-temperature expansion as
\begin{equation}
    V_1(\phi,T)=\frac{m^2(T)}{2}\phi^2-\frac{\delta(T)}{3}\phi^3+\frac{\lambda(T)}{4}\phi^4 \,.
    \label{eq:V1loop_approx}
\end{equation}
In this equation $m^2(T)$, $\delta(T)$ and $\lambda(T)$ are functions of the temperature $T$ and of other possible couplings or mass scales of the theory. 
For the CW model, we find the following form for these parameters~\cite{Levi:2022bzt}
\begin{align}
    m^2(T)&=N_b \frac{g^2 T^2}{12}\,, &
    \delta(T)&=N_b \frac{g^3 T}{4 \pi}\,, &
    \lambda(T)&=N_b \frac{g^4}{8\pi}\log\left(\frac{T}{M}\right) \,,
\end{align}
where $N_b=3$ is the number of the degrees of freedom of the (massive) gauge boson. Following~\cite{Levi:2022bzt} we introduce the mass scale $M$ of the theory as
\begin{equation}
    \label{eq:CW_mass_scale}
    M\equiv e^{-\frac{1}{3}+\gamma_E}\mu_R/4\pi\,,
\end{equation}
such that the zero-temperature vacuum expectation value becomes $v_\phi = \langle \phi \rangle \approx 11.6\, M/g$, and $m_\phi \approx 2.3\, g M$. 
Note that \cref{eq:V1loop_approx} includes both, the finite-temperature and zero-temperature contributions, see \cref{app:CW_details} for further details.

Using the expressions above, we can computationally determine the parameters associated with the PT (see also \cref{app:CW_details} for further details on the determination of the PT parameters). 
The parameters are then used to determine the corresponding SGWB, cf.\ \cref{app:FOPT_SGWB}. 
This then allows a comparison between our predicted spectra and empirical data from PTAs. 

\begin{figure}
    \centering
    \includegraphics[width=\textwidth]{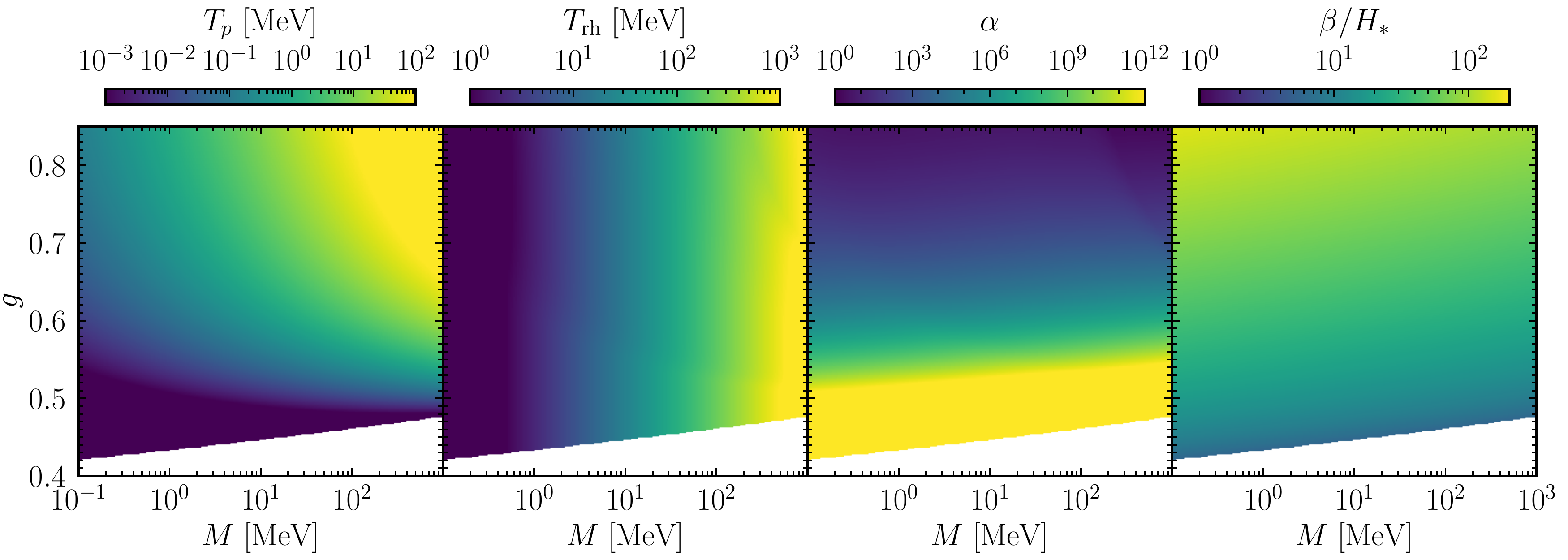}
    \caption{%
        Percolation~($T_p$) and reheating temperature~($T_\mathrm{rh}$) as well as transition strength~$\alpha$ and timescale parameter~$\beta/H_*$ as a function of the mass scale~$M$ and the gauge coupling~$g$ in the CW model.
        In the white region, the PT does not complete.
    }
    \label{fig:cw_params}
\end{figure}

\Cref{fig:cw_params} depicts the percolation and reheating temperatures $T_p$ and \Trh as well as the transition strength~$\alpha$ and inverse timescale~$\beta$ in the CW model as a function of the mass scale~$M$ and gauge coupling $g$. 
A significant amount of supercooling with percolation temperatures orders of magnitude below the mass scale, $T_p \ll M$, can be achieved in a large part of the parameter space, in particular at low gauge couplings.
If the gauge coupling becomes too low, the field does not tunnel and the universe remains stuck in the false minimum, leading to eternal inflation, indicated by the white region.
The supercooling goes along with strong transitions as well as large amount of reheating, with the reheating temperature on the order of the mass scale. 
As the reheating temperature eventually determines the frequency scales of the corresponding SGWB, $M\sim\si{\MeV}$ is required for a signal in the NANOGrav frequency band.
We further find $\alpha > 1$ throughout most of the parameter space.
The largest values of $\alpha$ are obtained close to the region where the field gets stuck in the false vacuum, however, which is also the region where our calculation is unreliable as higher-order corrections%
\footnote{See e.g.\ Refs.~\cite{Croon:2020cgk,Schicho:2022wty,Ekstedt:2022bff} for recent works regarding improvements in the calculation of the effective potential.}
such as Daisy resummation become important.
The inverse timescale parameter is typically of the order of $\beta/H_* \sim$~10 to~100.
Note that for $\beta/H_*<3$ bubbles cannot percolate efficiently due to the expansion for the universe, and for $\beta/H_*<10$, the GW spectrum is likely overestimated as the expansion during the transition has to be taken into account~\cite{Freese:2022qrl,Jinno:2022mie,Bringmann:2023opz}.

% ----------------------------------------------------------------------
\subsection{Meta-stable topological defects, remnants of symmetry breaking}
% ----------------------------------------------------------------------

Spontaneous symmetry breaking in the early universe is accompanied, in many extensions of the SM, by the production of topological defects such as CSs and DWs~\cite{Vilenkin:1984ib,Saikawa:2017hiv}.
If symmetry breaking happens after inflation, these topological defects can carry a significant fraction of the total energy density in the early universe, and they give rise to GW spectra with distinct features.

Of particular relevance for GW production are networks of CSs from the breaking of a global U(1) symmetry, or networks of annihilating DWs from the breaking of a discrete symmetry. 
A simple and well studied model that gives rise to CSs is that of 
ALPs with post-inflationary Peccei-Quinn~(PQ) breaking. For DWs, we consider two scenarios. In aligned axion models~\cite{Higaki:2016jjh,Chiang:2020aui} the QCD axion is accompanied by several heavier partners, which necessarily couple to photons and gluons and thus reheat the visible sector after DW annihilation. Instead in heavy ALP models~\cite{Blasi:2022ayo} the ALPs could also decay to dark radiation after DW annihilation, thus leading to a ``dark DW" scenario~\cite{Babichev:2021uvl}. In the following we introduce these scenarios in more detail. 

\subsubsection{Global (ALP) Strings} 
\label{sec: Global strings}

CSs are one-dimensional objects originating from the spontaneous breaking of a U(1) symmetry. Here we will solely discuss global strings, i.e.\ strings stemming from breaking a global U(1) symmetry, for instance, an ALP string network. The core of the CS has a typical size that is of the order of the inverse of the symmetry-breaking scale, usually much smaller than the horizon scale. This allows us to employ the Nambu-Goto approximation, where CSs are one-dimensional infinitely thin objects whose tension~$\mu$ is in units of energy per unit length. The string tension of ALP strings is given by 
\begin{equation}
 \mu = 2 \pi n\, f_{a}^{2} \, \log (f_{a} t) \, ,
\end{equation}  
where $f_{a}$ is the symmetry-breaking scale for an ALP string network with winding number~$n=1$. 
The network forms when the temperature of the universe is of the order of the PQ symmetry-breaking scale through the Kibble mechanism~\cite{Kibble:1976sj},
\begin{equation}
    T_{f} \simeq f_{a}\,.
\end{equation}
This implies that there is a UV cut-off for the network associated with the time of formation, but this is in frequency ranges far above our region of interest. 
Once the network is formed it evolves towards an attractor solution that is independent of the initial conditions, the so called scaling regime. This is the result of equilibration between the competing effects of string recombination and the Hubble expansion. 

GWs from strings are primarily radiated off loops. For global strings, this happens shortly after the loops are produced.  
GWs are only a subcomponent of the total emission from CS networks, with the main radiation emitted from long strings and loops going into axions. We base our quantitative analysis on the GW spectrum that was obtained in Ref.~\cite{Gorghetto:2020qws} from numerical simulations. In the scaling regime, it takes the form
\begin{equation}\label{eq:string_GW_spec}
     \Omega_{\GW}^\mathrm{scaling}(f)h^2 \simeq \num{0.8e-15} \left(\frac{f_{a}}{\SI{e14}{\GeV}}\right)^{4}\left(1 + 0.12\, \log\left[\left(\frac{f_{a}}{\SI{e14}{\GeV}}\right) \left(\frac{\SI{e-8}{\Hz}}{f} \right)^{2} \right] \right)^{4} \,.
\end{equation}  
If the U(1) symmetry was exact, the string network would exist until today and would be subject to strong bounds from the non-observation of gravitational imprints on the CMB. We therefore consider a scenario where the U(1) is only approximately realized. The amount of symmetry breaking can be parameterized by the axion mass $m_a$ (the Goldstone boson associated with the symmetry). Once the Hubble rate falls below the mass, DWs form and lead to the rapid annihilation of the string network given that there is only one minimum in the axion potential (see below for multiple minima). The time of the decay can be related to a pivot frequency today at which the spectrum deviates from the form given above and transitions to a $\propto f^3$ slope in the IR.
The pivot frequency can be expressed in terms of the mass as
\begin{equation}
    f_\mathrm{pivot} \sim \SI{3e-9}{\Hz} \, \left(\frac{m_a}{\SI{e-14}{\eV}}\right)^{\frac{1}{2}}\,,
\end{equation}
which is the annihilation frequency of the string network. We use a smooth step function for interpolating the GW spectra between the regime $f>f_{\mathrm{pivot}}$ described by \cref{eq:string_GW_spec} and a $\propto f^{3}$ dependence in the IR regime.

Before moving on, let us note that in the literature there is an ongoing debate about whether the GW spectrum features the log enhancement that is apparent in \cref{eq:string_GW_spec}. When these simulations are carried out, technical constraints force one to only consider times close to the creation of the network when $\log(f_a/H)=\mathcal{O}(\numrange{1}{10})$. While there are many simulations in this regime~\cite{Gorghetto:2018myk,Gorghetto:2021fsn,Buschmann:2021sdq,Figueroa:2020lvo} that agree up to order one factors, there is disagreement on how one should extrapolate these results to $\log(f_a/H)=\mathcal{O}(\numrange{10}{100})$ which is the relevant regime for predictions on GWs in PTAs. Suggestions reach from an exactly scale invariant spectrum~\cite{Figueroa:2020lvo,Hindmarsh:2021vih} to the $\log^4$ enhanced one~\cite{Gorghetto:2018myk,Gorghetto:2021fsn} that we are showing. The analytic observation that the string tension is expected to have a $\log$ enhancement together with other considerations on the scaling of the number of strings per Hubble patch leads semi-analytic approaches  to suggest an in-between behavior~\cite{Gouttenoire:2021jhk, Gouttenoire:2019kij, Chang:2021afa}. 

The main consequence of this choice is that it changes the value of $f_a$ towards larger values by a factor of up to $\mathcal{O}(10)$ to explain the amplitude.
When comparing the PTA fit to the strength of other constraints these effects partially cancel, however, since also other observables such as the axion abundance are predicted to be $\log$ enhanced if the GW amplitude is (see \cref{sec: Global strings results} for more details).  A minor effect is that a $\log$ enhancement leads to a small slope in the scaling regime.

\subsubsection{Annihilating ALP/Axion DWs}
\label{sec:ALP_DW}

DWs are two-dimensional topological defects that form due to the spontaneous breaking of a discrete symmetry. We refer the reader to Ref.~\cite{Saikawa:2017hiv} for a review of their cosmology, which we briefly summarize here. After the breaking of the symmetry, different patches in the universe populate the different degenerate minima. At the intersections of the patches the field has to interpolate between these ground states, which leads to a concentration of energy in thin two-dimensional sheets, the DWs. They carry a characteristic energy per area, or surface tension, $\sigma$. Subsequently, these networks evolve to minimize the area of the DWs, while radiating the majority of the released energy as particles and a small fraction as GWs.

Numerical simulations show that the evolution rapidly reaches a scaling regime, with roughly one DW per Hubble volume.
The  typical curvature radius is on the order of $\mathcal{O}(H^{-1})$ and the energy density is given by
\begin{equation}
    \rho_\mathrm{DW} \simeq \mathcal{A} \, \sigma H\,,
\end{equation}
where $H$ is the Hubble rate and the area parameter $\mathcal{A} \simeq \mathcal{O}(1)$ is a numerical prefactor extracted from simulations. In a model with a single real scalar field with broken $Z_2$ symmetry, its value is $\mathcal{A}= 0.8\pm 0.1$~\cite{Hiramatsu:2013qaa}.  For larger numbers of degenerate minima, $N_\mathrm{DW}>2$, it is slightly bigger~\cite{Kawasaki:2014sqa}, but we will restrict our discussions to the case of $N_\mathrm{DW}=2$ below. 

As the universe cools down and the Hubble rate decreases, the total energy $\propto H^2$ is depleted faster than the one in the DW network. Thus the energy fraction in the DW network is growing as
\begin{equation}
    \Omega_\mathrm{DW} \simeq \frac{\mathcal{A} \, \sigma}{3 M_{\rm P}^{2}H} \,,
\end{equation}
 and eventually comes to dominate the energy density at a temperature
\begin{equation}
    T_\mathrm{dom}^{2} = \frac{\mathcal{A}\,\sigma}{\pi M_{\rm P}}\sqrt{\frac{10}{\geff(T_\mathrm{dom})}}\,.
\end{equation}
In order not to overclose the universe, DWs must be unstable and decay at temperatures above $T_\mathrm{dom}$. This requires an explicitly symmetry-breaking term in the Lagrangian, lifting the degeneracy between the vacua. We denote by $V_\mathrm{bias}$ the energy difference between minima. This difference generates a volume pressure which leads to the annihilation of the network once the pressure becomes comparable to the surface tension in the walls. In a radiation-dominated universe, the network then annihilates at a temperature~\cite{Kawasaki:2014sqa}
\begin{equation}
    T_\mathrm{ann} \simeq \SI{20}{\MeV}\, \left(\frac{\sigma}{\si{\TeV}^3}\right)^{\!-\frac{1}{2}} \left(\frac{V_\mathrm{bias}}{\si{\MeV}^4} \right)^{\frac{1}{2}}\,.
    \label{eqn:DW_tann}
\end{equation}
 In order to avoid overclosure, $V_\mathrm{bias}$ has to satisfy
$
V_\mathrm{bias}^{1/4} \gtrsim \SI{0.03}{\MeV}\,\left(\sigma/{\si{\TeV}}^{3} \right)^{1/2}$. 
The network's energy is located in the DWs that are moving at relativistic velocities. Additionally, the scalar particles that get radiated from the DWs also possess large inhomogeneities. Both contribute to an anisotropic stress that leads to the emission of GWs. The majority of GWs get emitted right before the network annihilates, when its energy $\Omega_\mathrm{DW}$ is largest. The GW spectrum can be found from numerical simulations~\cite{Saikawa:2017hiv} and its peak amplitude and frequency are found to be
\begin{align}\begin{aligned}\label{eq:DWGW}
    \Omega_\mathrm{GW}^\mathrm{peak}h^2
    & \simeq \num{2.8e-18} \, \left(\frac{\sigma}{\si{\TeV}^3} \right)^2 \left(\frac{T_\mathrm{ann}}{\SI{10}{\MeV}} \right)^{-4} \,,\\
    f_{\mathrm{peak}} & \simeq \SI{1.1e-9}{\Hz} \left(\frac{T_\mathrm{ann}}{\SI{10}{\MeV}} \right)\,.
\end{aligned}\end{align}
Here we included redshift factors assuming that annihilation takes place during radiation domination at temperatures close to \SI{10}{\MeV} that lead to the correct frequencies for PTAs.
We follow Ref.~\cite{Ferreira:2022zzo} and assume that away from the peak the spectral shape is given as
\begin{equation}
    \Omega_\mathrm{GW}(f)h^2 = 
    \Omega_\mathrm{GW}^\mathrm{peak}h^2\times S(f/f_{\mathrm{peak}})\,,
    \qquad 
    S(x) = \frac{4}{x^{-3} + 3 \, x}\,.
\end{equation}
The behavior for frequencies below the peak as $f^3$ is dictated by causality, while the $f^{-1}$ above the peak is found in numerical simulations. Additionally, there will be a cutoff at much larger frequencies that is correlated with the time when the network is first formed. These frequencies are, however, way too large to be relevant for the PTA data we consider.

\bigskip 

So far, the discussion has been fairly model-independent. We will now explore two explicit models, where the DWs arise in a generic ALP model or in a clockwork realization of the QCD axion. While the two scenarios lead to almost identical GW spectra, their cosmological signatures are quite distinct. 

We first consider the case of an ALP as the pseudo Nambu-Goldstone boson generated from the spontaneous breaking of an anomalous U(1) symmetry, which we will refer to as a Peccei-Quinn symmetry, at a scale $f_a$. 
The Lagrangian takes the form
\begin{equation}\label{eq:ALPLagrangian}
    \mathcal{L}= \partial_{\mu}\Phi^{\dagger}\partial^{\mu}\Phi - \lambda\left(\Phi^{\dagger}\Phi - \frac{v_{a}^2}{2} \right)^{2} -V(a)\,, 
\end{equation}
where $\Phi = \rho/\sqrt{2} \exp{(i a/v_{a})}$ is a complex scalar field and the axion $a$ is its angular part. The potential of \cref{eq:ALPLagrangian} is such that
the U(1) symmetry is spontaneously broken, with a vev $\langle\Phi\rangle = v_a/\sqrt{2}$ and $a\in [0,2\pi v_{a})$. The term $V(a)$ in \cref{eq:ALPLagrangian} is the anomaly-induced U(1) breaking under the influence of a strongly coupled gauge theory with dynamical scale $\Lambda \simeq \sqrt{m_{a} f_{a}}$.
This explicitly breaks the U(1) symmetry into its $\mathcal{Z}_{N_{\rm DW}}$ subgroup.
The conventional form of such explicit breaking at zero temperature is
\begin{equation}
    V(a) = \Lambda^{4}\left(1-\cos{\frac{a}{f_{a}}} \right)\,,
\end{equation}
where $f_{a} = v_{a}/N_\mathrm{DW}$. 
The tension of ALP DWs in the absence of finite temperature effects can be estimated as~\cite{Saikawa:2017hiv}
\begin{equation}
    \label{eq:DW_ALP_wall_tension}
    \sigma= 8 m_{a} f_{a}^2\,.
\end{equation}

We will work under the assumptions that $v_{a}<T_\mathrm{rh}$, such that the U(1) symmetry is restored after inflation and the network forms as the universe cools down, and that there is a large separation of scales between $v_{a}$ and $\Lambda$. This initially leads to the formation of a CS network which persists until the time of DW formation when $H\approx m_a$.  Since $N_\mathrm{DW}\geq2$, there are multiple DWs attached to every string and the network is stable.
Shortly after this time, the combined network is dominated by the dynamics of the DWs and one can neglect any effect the remnant strings have on the evolution. GWs produced by strings, as well as cosmological constraints such as those coming from $N_\mathrm{eff}$, are negligible with respect to the contributions from DWs.
This is because, as will be clear from \cref{sec:ALP_DW_results}, the decay constant $f_a$ is much lower than in the CS scenario discussed above. 

In addition, the global U(1) symmetry is expected to be broken quantum gravity effects.
Therefore, additional breaking terms, if not accidentally aligned with the anomalous breaking, lift the degeneracy between the minima. They provide the necessary $V_\mathrm{bias}$ for the network to annihilate, with the temperature of annihilation given by \cref{eqn:DW_tann}.

\bigskip 

In addition to the generic ALP model, we consider DWs in models of the QCD axion, i.e. models that solve the strong CP problem. One such scenario is that of axion alignment~\cite{Higaki:2016jjh} realized by a clockwork mechanism.
Here a collection of $N$ axions that individually respect a shift symmetry
\begin{equation}
\phi_{i} \rightarrow \phi_{i} + C_{i}\,,    
\end{equation} 
is considered,
where $C_{i}$ is a real-valued transformation parameter. One then assumes that  $N-1$ of these shift symmetries get explicitly broken into their discrete subgroups, giving rise to the potential for $N-1$ linear combinations of the axions. The remaining flat direction is then identified as the QCD axion with its associated gluon coupling in, for instance, the Kim-Shifman-Vainshtein-Zakharov~(KSVZ) model. 

The main advantage of this scenario is that it gives a light QCD axion with an exponentially enhanced effective decay constant $F_a \sim f_a e^{N}$, while the actual symmetry breaking scales $f_a$ can be much lower, e.g.~around the TeV-PeV scale, thus making the model testable at particle physics experiments. This also ensures that the symmetry breaking can take place after reheating, and thus a DW network, made from the $N-1$ heavy axions predicted by the model, can form. 
Ref.~\cite{Higaki:2016jjh} found that the DWs are long-lived and survive until the QCD axion potential becomes relevant. For simplicity we take equal masses $m_a$ and equal decay constants $f_a$ for all the heavy axions. In terms of these, the DW tension is again given by \cref{eq:DW_ALP_wall_tension}. 

Different from the generic ALP model, here the network is destabilized by QCD instantons at the time of the QCD phase transition. This lifts the degeneracy between the different minima by $\Delta V \simeq \Lambda_\mathrm{QCD}^{4}$, and the annihilation temperature can be predicted as~\cite{Higaki:2016jjh}
\begin{equation}
    T_\mathrm{ann} \sim \SI{1}{\GeV}\, \left(\frac{\geff(T_\mathrm{ann})}{80}\right)^{-\frac{1}{4}}\left(\frac{\Lambda_\mathrm{QCD}}{\SI{400}{\MeV}}\right)^{2}\left(\frac{\SI{e7}{\GeV}}{f_a}\right) \sqrt{\frac{\SI{10}{\GeV}}{m_a}}\,.
\end{equation}
Up to order-one factors, the GW spectrum obtained in Ref.~\cite{Higaki:2016jjh} agrees with~\cref{eq:DWGW}, and we will therefore use the latter for both models. The main difference instead is that in the aligned axion model, also the heavy axions couple to the SM via the usual axion couplings~\cite{Higaki:2015jag}. Shortly after DW annihilation, the energy is therefore transferred back to the visible sector via decays of the heavy axions into SM particles. As discussed in more detail in \cref{sec:ALP_DW_results}, these two DW scenarios are therefore subject to very different cosmological constraints. 

% ----------------------------------------------------------------------
\subsection{Bosonic instabilities and late preheating}
\label{sec:audible_axion}
% ----------------------------------------------------------------------

Instabilities in bosonic fields can lead to the exponential amplification of fluctuations, which then act as sources for GWs. Such mechanism are best known in the context of preheating~\cite{Greene:1997fu,Kofman:1994rk,Shtanov:1994ce,Kofman:1997yn,Dufaux:2007pt,Figueroa:2016wxr} but can also take place in a purely gravitationally coupled sector contributing a sub-dominant fraction of energy~\cite{Machado:2018nqk,Chatrchyan:2020pzh,Kitajima:2020rpm,Banerjee:2021oeu,Madge:2021abk,Eroncel:2022vjg}. In these models the energy is initially stored in one homogeneous field, typically a scalar. The self-couplings of the field or couplings to other bosons then induce a time-varying effective mass that leads to the exponential amplification of fluctuations in the secondary field. Once the fluctuations start to dominate the energy density, the distribution of the energy also becomes highly inhomogeneous. With this transition, a large anisotropic stress get generated that leads to the emission of GWs.

The peak wavelength of the produced GWs is given by the fastest growing mode. Due to causality this wavelength is always smaller than the horizon at the time of production. On the other hand, the wavelength cannot be much larger in order for the mechanism to efficiently source GWs, establishing a firm relation between the peak frequency of the GW spectrum and the time of production. Similar to the PTs discussed above, any such process must therefore occur at temperatures of $T\sim\SIrange{10}{100}{\MeV}$ in order to account for the PTA signal.

The amplitude of the GW signal is mainly set by the fraction of energy in the bosonic sector. Explaining the PTA signal requires energies that cannot easily be hidden from other cosmological probes. We can distinguish two possible scenarios: If the GW production is highly efficient, a purely gravitationally coupled sector might still be able to evade the bounds on dark radiation and matter. If this is not possible, all or part of the energy in the dark sector must be depleted into the visible sector via direct couplings. Such energy transfers would need to occur before BBN. A detailed discussion of the bounds can be found in \cref{sec: Constraints AA}.

\bigskip 

As a concrete example, we consider a model where an ALP is coupled to a dark photon~\cite{Anber:2009ua,Barnaby:2012xt,Machado:2018nqk}. The dark sector consists of a CP odd scalar $\phi$ with mass $m_a$ and decay constant $f_a$ and a dark photon $X^{\mu}$. The two are coupled through
\begin{align}
    \mathcal{L}\supset \frac{\alpha}{4f_a} \phi X_{\mu\nu}\tilde X^{\mu\nu}~,
\end{align}
where $\alpha$ denotes a dimensionless coupling constant while $X_{\mu\nu}$ and $\tilde X_{\mu\nu}$ denote the dark photon field strength and its dual. We assume the axion is initially displaced by $\phi_0=\theta f_a$ from the minimum of its potential that we take as $V(\phi)=m_a^2 f_a^2 (1-\cos(\phi/f_a))$ to be explicit. Once the Hubble rate drops below the axion mass, $H(t_\mathrm{osc})=m_a$, the axion starts oscillating around the minimum of the potential. The non-vanishing velocity of the axion $\dot\phi\neq 0$ leads to a tachyonic instability in one of the chiral polarizations of the dark photon induced by the coupling.  As a result, fluctuations with momenta $k\sim \alpha\theta m_{a}$ grow exponentially. In order for the tachyonic production to be efficient the product $\alpha\theta$ needs to be larger than 20, although larger values lead to more efficient emission of GWs. For simplicity we will fix $\theta=1$ and $\alpha=100$.\footnote{Such large couplings can be obtained in a technically natural way e.g.\ via the clockwork mechanism \cite{Choi:2014rja,Kaplan:2015fuy}.} The tachyonic growth is very rapid as compared to the process of parametric resonance occurring in other models (cf.\ e.g.\ Refs.~\cite{Figueroa:2016wxr,Chatrchyan:2020pzh}) and therefore leads to more efficient GW emission. 

The GW peak frequency in this case is largely controlled by the mass~$m_a$, since it determines the time at which the axion starts to oscillate. The peak amplitude at emission is controlled by the energy in the dark sector which is given as $\Omega_{d,\mathrm{osc}}\approx(\phi_0/M_{\rm P})^2$. In principle one needs to distinguish here between $\phi_0\gtrsim M_{\rm P}$ and $\phi_0\lesssim M_{\rm P}$. In the first case the axion dominates the energy density and possibly even drives a period of inflation. It must then necessarily reheat the SM. Also the expansion is primarily driven by the bosonic sector and should be self-consistently taken into account in simulations~\cite{Cuissa:2018oiw,Adshead:2018doq}. In the other case the bosonic sector only comprises a small amount of the total energy and can be purely gravitationally coupled~\cite{Machado:2018nqk,Kitajima:2017peg}. 

Using the data from Ref.~\cite{Ratzinger:2020oct} we update the template for the GW spectrum given in Ref.~\cite{Machado:2019xuc}. We find that the peak frequency and amplitude of the signal today are given by
\begin{align}
    f_\mathrm{peak}&\approx \SI{3.9e-9}{\Hz}\ \left(\frac{\alpha \theta}{100}\right)^{2/3} \left(\frac{m_a}{\SI{e-15}{\eV}}\right)^{1/2}\,,\\
    \Omega_\mathrm{GW}^\mathrm{peak}h^2&\approx \num{1.2e-7} \left(\frac{ \theta^2}{\alpha}100\right)^{2/3}\left(\frac{ f_a}{M_{\rm P}}\right)^{4}~.
\end{align}
The shape of the spectrum can be characterized by a broken power-law of the form
\begin{align}
    \Omega_\mathrm{GW}(f) \,h^2 =\Omega_\mathrm{GW}^\mathrm{peak} \,h^2 \, \left(\frac{f}{f_\mathrm{peak}}\right)^{n_1}\left[\frac{1}{2}\left(1+\left(\frac{f}{f_\mathrm{peak}}\right)^{1/\Delta}\right)\right]^{(n_2-n_1)\Delta}~,
    \label{eqn:AAfit}
\end{align}
where we determined the values of $n_1=0.73,\ n_2=-4.96$ and $\Delta=0.24$ by fitting. The small prefactor here includes the usual redshift, while the main additional suppression of the signal is due to the energy density of the source scaling as $(f_a/M_{\rm P})^2$. Thus, an observable signal requires the scale $f_a$ to be close to the Planck scale. 

% ----------------------------------------------------------------------
\subsection{Scalar-induced GWs}
\label{sec:scalar_induced}
% ----------------------------------------------------------------------

When scalar perturbations generated during inflation reenter the horizon, they act as a second-order source term for GWs, schematically through the relation $\Omega_\mathrm{GW}(k) = \iint \mathcal{T} P_\zeta P_\zeta $, where $\mathcal{T}$ is an appropriate transfer function~\cite{Acquaviva:2002ud, Mollerach:2003nq, Ananda:2006af, Baumann:2007zm} (see Ref.~\cite{Domenech:2021ztg} for a review). The predicted GW abundance for a power-law spectrum compatible with CMB observations is $\Omega_\mathrm{GW}\simeq 10^{-21}$ at NANOGrav frequencies~\cite{Ananda:2006af}. If scalar perturbations are enhanced at scales smaller than the ones probed by CMB, they could instead generate a visible signal that may explain the NANOGrav data. This may happen typically because of some feature in the inflaton potential which leads to slow-roll violation and the onset of an ultra slow-roll phase~\cite{Garcia-Bellido:2017mdw, Ballesteros:2017fsr}, or because of the dynamics of a second field in multi-field or spectator field scenarios~\cite{Cheong:2022gfc, Espinosa:2018eve}. Other, more exotic options are also possible, including e.g.\ the violation of the null energy condition (see Ref.~\cite{Cai:2023uhc} and references therein). At the same time, such large scalar fluctuations are expected to induce a population of primordial black holes~(PBHs). Imposing that these PBHs do not overclose the universe, and that they comply with astrophysical bounds, strongly constrains the amplitude of the primordial scalar fluctuations~\cite{Dandoy:2023jot}.

A conventional parametrization for the curvature power spectrum is a log-normal function centered around some momentum $k_*$ and of width $\Delta$.
Assuming standard expansion history and that the modes reenter the horizon during radiation dominance, analytic expressions can be obtained for the induced GW spectrum~\cite{Pi:2020otn, Dandoy:2023jot} which, around the peak, is well approximated by a log-normal function with width $\Delta/\sqrt{2}$.
The problem can be studied in a more general background~\cite{Kohri:2018awv, Domenech:2019quo}, but we will stick to radiation dominance for simplicity, as it is expected in the standard cosmological history.

In inflationary models featuring an inflection point, the log-normal parametrization discussed above does not reproduce the qualitative behaviour of the curvature power spectrum, which, in this case, is rather asymmetric around the peak, with a steep tail in the IR and a relatively flat (although red-tilted) plateau above the peak. We found that a reasonably good parametrization around the peak region can be written as a log-normal function below the peak and a red-tilted power spectrum above it,
\begin{align}\label{eq:broken ps scalar}
    P_\zeta (k) & = \frac{A_\zeta}{\sqrt{2\pi} \Delta} \exp \left[ - \frac{\log^2(k/k_*)}{2\Delta^2}\right] \theta(e^{n\Delta^2} k_*-k) + \frac{A_\zeta}{\sqrt{2\pi} \Delta} e^{n^2\Delta^2/2} \left( \frac{k}{k_*}\right)^{-n} \theta(k- e^{n\Delta^2} k_*)
\end{align}
where we have stitched the two pieces with a $\theta$ function centred at $k/k_* = e^{n\Delta^2}$, ensuring a continuous first derivative. Following conventions, the peak amplitude in this parametrization is $A_\zeta/(\sqrt{2\pi}\Delta)$.

From \cref{eq:broken ps scalar}, one can derive the GW abundance by integrating numerically $P_\zeta P_\zeta$ with the correct transfer function~\cite{Espinosa:2018eve, Kohri:2018awv}, or try to obtain an analytic approximation along the lines of Ref.~\cite{Pi:2020otn}.
As it turns out, due to the smallness of the cross terms, a good approximation is to simply glue the analytic expressions for the GW power spectrum obtained for the log-normal and for the power-law scalar spectra. These are given by the following expressions,
\begin{align}
\Omega_\mathrm{GW}^\mathrm{LN} h^2 & \simeq 10^{-5}\left(\frac{\geff(T_\star)}{17.25}\right)\left(\frac{\gs(T_\star)}{17.25}\right)^{-\frac{4}{3}}\left(\frac{\Omega_{r,0}h^2}{4\times 10^{-5}}\right) \nonumber\\
& \frac{4}{5\sqrt{\pi}} \alpha^3 \frac{e^{\frac{9\Delta^2}{4}}A_{\zeta}^2}{\Delta} \left\{\left[\left(\log^2 K +\frac{\Delta^2}{2}\right)\erfc\left(\frac{\log K+\frac{1}{2}\log\frac{3}{2}}{\Delta}\right)-\frac{\Delta}{\sqrt{\pi}} \exp\left( -\frac{\left(\log K+\frac{1}{2}\log\frac{3}{2}\right)^2}{\Delta^2}\right) \right. \right. \nonumber\\
& \qquad \left. \times \left(\log K-\frac{1}{2}\log\frac{3}{2}\right) \right] + \frac{0.0659}{\Delta^2}\alpha^2e^{\Delta^2}\exp\left(-\frac{\left(\log\alpha+\Delta^2-\frac{1}{2}\log\frac{4}{2}\right)^2}{\Delta^2}\right) \nonumber\\
& \qquad \left. + \frac{1}{3}\sqrt{\frac{2}{\pi}}\alpha^{-4}\frac{e^{8\Delta^2}}{\Delta}\exp\left(-\frac{\log^2\alpha}{2\Delta^2}\right)\erfc\left( \frac{4\Delta^2-\log\alpha/4}{\sqrt{2}\Delta}\right) \right\} \,, 
\end{align}
for the log-normal part, where $\alpha =f/f_\star$, $K=\alpha \,\exp\left(3\Delta^2/2\right)$, and we assumed $\Delta\gtrsim 0.5$~\cite{Pi:2020otn, Dandoy:2023jot}, and
\begin{align}
    \Omega_\mathrm{GW}^\mathrm{PL} h^2 \simeq 
    10^{-5}\left(\frac{\geff(T_\star)}{17.25}\right)\left(\frac{\gs(T_\star)}{17.25}\right)^{-\frac{4}{3}}\left(\frac{\Omega_{r,0}h^2}{\num{4e-5}}\right) \times 
    0.8 \left( \frac{A_\zeta}{\sqrt{2\pi} \Delta} e^{n^2\Delta^2/2} \right)^2 \left( \frac{f}{f_*}\right)^{-2n}
\end{align}
for the power-law part~\cite{Kohri:2018awv}. Putting things together, our ansatz for the GW power spectrum is
\begin{equation}\label{eq:broken ps scalar GW}
    \Omega_\mathrm{GW} h^2 \simeq \left(\frac{1}{2}+\frac{1}{2}\tanh \frac{\sqrt{2}\log(f/f_*)}{\Delta} \right) \Omega_\mathrm{GW}^\mathrm{LN} h^2
    +
    \left(\frac{1}{2}-\frac{1}{2}\tanh \frac{\sqrt{2}\log(f/f_*)}{\Delta} \right) \Omega_\mathrm{GW}^\mathrm{PL} h^2 \,,
\end{equation}
where a $\tanh$ function has been used to glue the two contributions together.
The GW frequencies~$f$ correspond to momenta~$k$ of the scalar power spectrum through the relation
\begin{equation}
    f = \frac{k}{2\pi} \approx 1.55 \times 10^{-9}\,\mathrm{Hz}\, \left( \frac{k}{10^6\,\mathrm{Mpc}^{-1}} \right)
\end{equation}
In \cref{sec:results} we will use \cref{eq:broken ps scalar GW} to fit PTA data, while \cref{eq:broken ps scalar} will be needed to discuss constraints from the overproduction of PBH and CMB spectral distortions.

\bigskip 

In order to motivate an interpretation of PTA data in terms of scalar-induced GWs, it is important to discuss models of inflation that could generate such a signal. Refs.~\cite{Dimopoulos:2019wew} and~\cite{Clesse:2015wea} discuss examples of this in the context of thermal inflation and two-fields inflation with a waterfall phase, respectively. The focus of these works is on PBH production, but the same scalar perturbations would induce a GW signal in the right ballpark. In both cases, a symmetric log-normal distribution seems to be a reasonable proxy for the power spectrum.
Another model is the one shown in Ref.~\cite{Ferrante:2023bgz}, in which the fluctuations of an axion-curvaton field are enhanced by the rolling phase of its radial counterpart. Interestingly, in this model the power spectrum can be enhanced for many efolds, generating a GW signal spanning from PTA up to the LISA range. An important consideration made in this paper concerns the role of non-gaussianities. In their model, non-gaussianities suppress the production of PBHs, while only mildly affecting the spectrum of GWs, thus avoiding the bound from the overproduction of PBHs. This highlights the fact that the correlation between the GWs and the PBH signals can only be discussed once non-gaussianities are taken into account.
Finally, in Refs.~\cite{Franciolini:2022pav, Franciolini:2022tfm}, a reverse-engineering approach is taken, in which a suitable power spectrum is generated by assuming a specific behaviour of the second slow-roll parameter as a function of the number of efolds during inflation $\eta(N)$, and then a slow-roll potential is obtained numerically.

Here we try a more direct approach, and consider whether a simple single-field inflationary model with an inflection point could reproduce our signal, while still respecting CMB bounds. We show below that this appears to be impossible.
The simplest example one can think of is a $\phi^4$ theory with a running coupling and non-minimal coupling to gravity~\cite{Ballesteros:2017fsr},
    \begin{equation}
        S = \int d^4 x \sqrt{-g} \left(-\frac{1}{2}(M_{\rm P}^2+\xi\phi^2)R + \frac{1}{2}g^{\mu\nu}\partial_\mu\phi \partial_\nu\phi - V(\phi) \right)\,,
    \end{equation}
    where
    \begin{equation}
        V(\phi) = \frac{\lambda_0}{4!} \left(1+b_1 \log\frac{\phi^2}{\phi_0^2}+b_2\left(\log\frac{\phi^2}{\phi_0^2}\right)^2+\ldots\right)\phi^4
        \quad
        \text{and}
        \quad
        \xi(\phi) = \xi_0 \left(1+b_3\log\frac{\phi^2}{\phi_0^2}\right) \,.
    \end{equation}
The parameters $b_1$, $b_2$, $b_3$ mimic the logarithmic running of the potential, and can be tuned in such a way to reproduce an inflection point around $\phi=\phi_0$. A simplified version of this model is obtained by setting $b_1 =0$, and reproduces a possible scenario for Higgs inflation with a reasonable choice of parameters~\cite{Garcia-Bellido:2017mdw, Drees:2019xpp}.
The potential for the field $\phi$ in the Einstein frame can be rewritten as
\begin{equation}
    V(x) = \frac{V_0 (1+a \log^2 x)x^4}{[1+c (1+b \log x)^2 x^2]}\,,
\end{equation}
with $ x= \phi/\phi_0$, $a = b_2/\lambda_0$, $b = b_3/\xi_0$, $c = \xi_0 \phi_0^2/M_P^2$ and $V_0 = \lambda_0 \phi_0^4 / 4$.
If now the parameters are tuned to
\begin{equation}
    a = \frac{4}{1 + c x_c^2 + 2 \log x_c - 4 \log^2 x_c } \,, \qquad
    b = (1-\beta ) \frac{2}{c x_c^2} \frac{1 + c x_c^2 + 2 c x_c^2 \log x_c + 4 \log x_c}{1 + c x_c^2 + 2 \log x_c - 4 \log^2 x_c }\,,
\end{equation}
with $\beta = 0$, the potential has an exact inflection point at $x = x_c$. The parameter $\beta \neq 0$ allows for a small deviation from an exact inflection point. We assume $\beta\ll 1$, and take it as a measure of the amount of fine-tuning needed.

The curvature power spectrum is obtained in the standard way by solving the equations of motion for the zero-mode field and the Mukhanov-Sasaki equation for the curvature power spectrum, for each choice of parameters. One should remember here that the field $\phi$ is not canonically normalized after going to the Einstein frame, and thus the equations of motion are not the usual ones.
In order to speed up the calculation, it is useful to adopt the approximate expression for the curvature power spectrum
\begin{equation}\label{eq:scalar PS SR}
P_\zeta \approx \frac{H^2}{8\pi^2 \epsilon} \,.    
\end{equation}
Strictly speaking, this expression is valid only at slow-roll, while a peak in the power spectrum is obtained precisely due to its violation. Nevertheless, it gives a reliable enough indication of the position of the peak in momentum space and of its amplitude. Once some relevant benchmarks are identified, one can then solve the Mukhanov-Sasaki equation for a more reliable calculation.

The parameter space of the model is scanned in the range reported in \cref{tab:scalar induced params} in~\cref{app:scalar_induced}.
We fix the amplitude of the scalar power spectrum at the pivot scale $k=\SI{0.05}{\per\mega\parsec}$ to $P_\zeta = \num{2.1e-9}$~\cite{Planck:2018vyg}, and compute the peak amplitude $P_{\zeta,\,\mathrm{max}}$, the position of the peak $k_*$, and the momentum $k_{0.1} < k_*$ at which the amplitude decreases by a factor $0.1$. The last quantity is important when the spectrum has a very flat plateau, in which case $k_{0.1}$ represents better than $k_*$ the point above which the power spectrum is enhanced. In order to compare with CMB observations, we also compute the scalar spectral index and the number of efolds from when the pivot scale exits the horizon to the end of inflation.

\begin{figure}
    \centering
    \includegraphics[width=.3\textwidth]{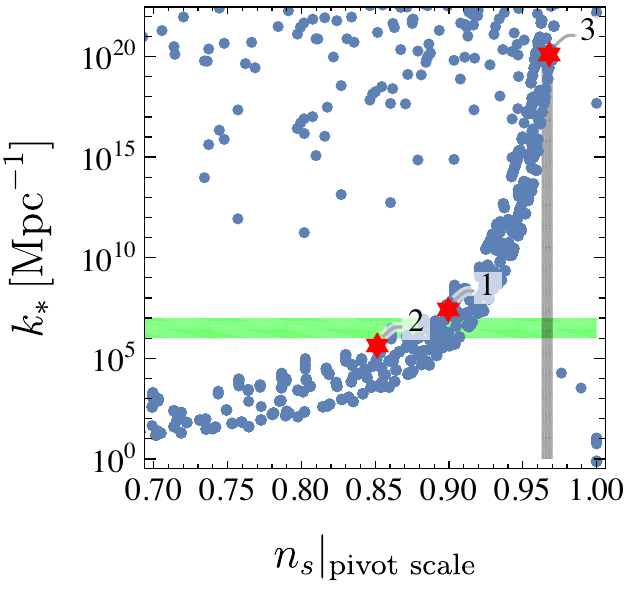}
    \includegraphics[width=.3\textwidth]{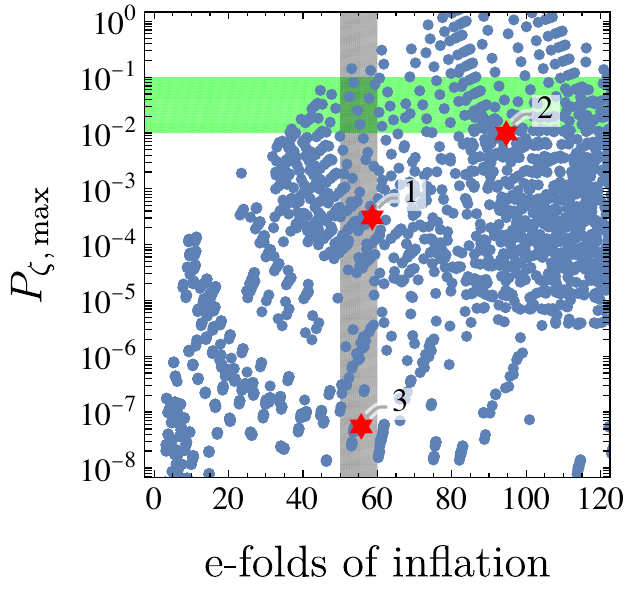}
    \includegraphics[width=.3\textwidth]{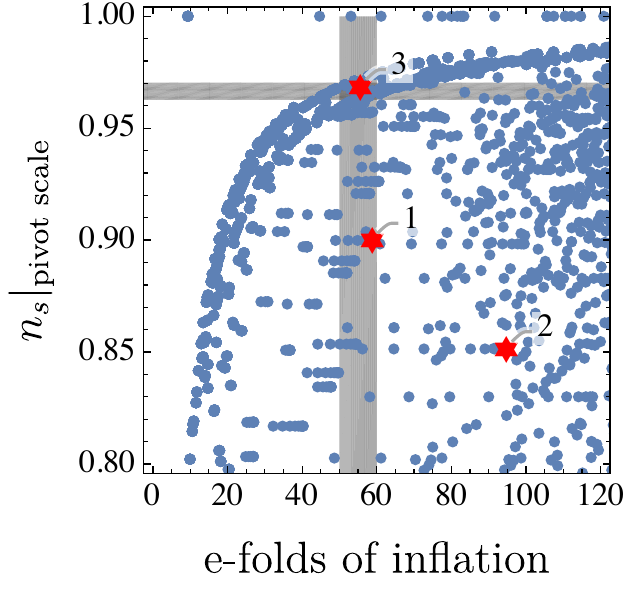}
    \caption{Scan of the parameter space of the inflationary model. The green horizontal bands identify the target region that could fit PTA data~\cite{Dandoy:2023jot}, while the gray bands correspond to the CMB measurements. The blue dots corresponds to the full scan. The three stars are the benchmark points discussed in \cref{app:scalar_induced}. Shown in the plots are the peak position~$k_*$, the peak amplitude~$P_{\zeta,\,\mathrm{max}}$, the spectral index at the pivot scale and the number of efolds from when the pivot scale exits the horizon until the end of inflation.}
    \label{fig:scalar induced scan}
\end{figure}

The results of the scan are shown in \cref{fig:scalar induced scan}.
The situation is particularly clear from the first plot. A peak in the power spectrum in the PTA region inevitably leads to a low spectral index at the pivot scale. This is ultimately due to how the enhancement is generated: in field space, the CMB pivot scale \SI{0.05}{\per\mega\parsec} and the PTA scale are too close to each other to have an important feature at the latter without affecting the former, unless the feature is generated via some kind of step function, which is difficult to justify in a single field model. We checked that this result does not depend on the finiteness of our scan region by extending it in all directions.

A similar tension is present even if one is interested in a signal in the LISA frequency band, that corresponds to $k_*\sim \SIrange{e12}{e14}{\per\mega\parsec}$. This can be generated in this scenario, at the price of a small $n_s$ and a large number of efolds between the horizon crossing of the CMB scale and the end of inflation~\cite{Drees:2019xpp}.

\bigskip 

The same exercise we discussed here can be done with a polynomial potential inspired by EFT considerations, of the form
\begin{equation}
    V(\phi) = V_0 \left(1 + \sum_n \left(\frac{\phi}{\Lambda}\right)^n\right) \,.
\end{equation}
Ref.~\cite{Bhaumik:2019tvl}, for example, considers such a potential including only even terms up to $n=6$, and mimics the inclusion of a non-minimal coupling to gravity by dividing $V$ by $(1+\xi\phi^2)^2$. Notice that they do not canonically normalize the field $\phi$, so this is not equivalent to what is done here and in Ref.~\cite{Ballesteros:2017fsr}. While they are able to find a peak corresponding to NANOGrav frequencies, they obtain $n_s\sim 0.9$ (see their Fig.~3), in accordance with our results and in $10\,\sigma$ tension with the $n_s$ measurement by Planck.

% ======================================================================
\section{PTA data and fitting method}
\label{sec:fit}
% ======================================================================

Since 2020, three of the currently operating PTA observatories, NANOGrav~\cite{NANOGrav:2020bcs}, EPTA~\cite{Chen:2021rqp} and PPTA~\cite{Goncharov:2021oub} have reported strong evidence for a common-spectrum red process across pulsars in their data.
Similar results were found by the joint IPTA~\cite{Antoniadis:2022pcn} collaboration combining previous data releases of the former three.
Although the evidence for the interpulsar Hellings-Downs correlations~\cite{Hellings:1983fr} required for establishing the observation of a GW signal is still not conclusive, it is intriguing to attribute this signal to a SGWB.

We here operate upon the assumption that the putative SGWB signal is of cosmological origin and assume that the potential SGWB from SMBHBs is subdominant in the frequency range we consider.
We fit the GW signals produced by the sources discussed in \cref{sec:sources} to pulsar timing data.
For PTA data this fit is typically done in terms of the timing-residual cross-power spectral density,
\begin{equation}
    S_{ab} = \Gamma_{ab} \frac{H_{100}^2}{8 \pi^4} \frac{\Omega_{\GW}h^2 (f)}{f^5} \, .
\end{equation}
Here, $\Gamma_{ab}$ denotes the overlap reduction function between two pulsars $a$ and $b$ and $H_{100} = \SI{100}{\kilo\meter\per\second\per\mega\parsec}$.
However, a thorough Bayesian analysis of this quantity for PTA data is a computationally expensive task.
Hence, fast refitting techniques have been developed to reinterpret free-spectrum fits (to the entire PTA or for each pulsar individually) in terms of arbitrary SGWBs, resulting in the development of the \texttt{ceffyl} analysis suite~\cite{Lamb:2023jls}.
In this vein, we perform a PTA free-spectrum refit in \texttt{ceffyl} to the NANOGrav~12.5~year dataset~\cite{NANOGrav:2020gpb} and the second data release~(DR2) of IPTA~\cite{Perera:2019sca,Antoniadis:2022pcn}. 
The kernel density estimator~(KDE) representation required by \texttt{ceffyl} for the NANOGrav dataset is provided by the collaboration~\cite{NANOGrav:KDErepresentation}.
For the IPTA data, the KDE is created reproducing the 30-frequency-bin free-spectrum fit with \texttt{enterprise}~\cite{enterprise} and \texttt{enterprise\_extensions}~\cite{enterprise_extensions}. We verified that \texttt{ceffyl} reproduces the best-fit regions for a simple power-law SGWB that were obtained by NANOGrav and IPTA. Furthermore we also checked that this remains true for broken power-law signals by comparing the best-fit regions obtained with \texttt{ceffyl} to those from a full \texttt{enterprise} run for the scenario~\cref{sec:audible_axion}.
We use \texttt{PTMCMCSampler}~\cite{ptmcmc} to sample the respective parameter spaces of our signal models, except for the CW model for which we evaluate the likelihood on a regular grid.
An overview of the priors we use in our Bayesian analysis can be found in \cref{app:triangle_plots}.

Following the NANOGrav searches for a general common-spectrum signal~\cite{NANOGrav:2020bcs} and for a SGWB from FOPTs~\cite{NANOGrav:2021flc}, we use only the five lowest-frequency bins of the NANOGrav data, corresponding to frequencies below $f \lesssim \SI{12.5}{\nano\Hz}$.
These bins give the strongest contribution to the signal-to-noise ratio~(SNR) of the NANOGrav signal~\cite{NANOGrav:2020bcs}.
For consistency, we cut the IPTA data at the same frequency.
Due to the longer observation period of 30~years, this corresponds to twelve bins in the IPTA 30-component free-spectrum fit.
Other choices regarding the IPTA frequency range are discussed in \cref{app:IPTA_bins}.

Intuitively, the most robust insight into the low-frequency regime can be obtained by combining data from different PTA collaborations.
However, this is a very challenging task.
The different datasets comprise pulsars observed by multiple collaborations,\footnote{%
    Currently, IPTA includes timing measurements of 65 millisecond pulsars, of which 24 are observed by more than one PTA.
} 
timed over various observation periods at different times and with different telescopes.
A considerable amount of care has to be taken to properly account for all correlations between the individual measurements due to overlaps in the timing data as well as instrument-specific noise contributions. 
In order to address this, EPTA, NANOGrav and PPTA, as well as the Indian Pulsar Timing Array~(InPTA) project, have formed the joint IPTA consortium.
While the current evidence in the individual PTAs could only be established with the respective latest data releases, IPTA was able to detect the common spectrum combining data available in 2016 and before.
Hence, the question arises what can be expected when all currently available measurements are taken into account.

Clearly, a forecast with a proper account of pulsar overlaps is beyond the scope of this work.
To approximate the signal reconstruction obtained in a future IPTA release, we here employ a naive combination by multiplying the likelihoods of the NANOGrav 12.5~year data and IPTA~DR2, where we again evaluate the likelihood of each experiment using \texttt{ceffyl}, i.e.\ treating the two datasets as independent.
In addition to the overlap discussed above, this procedure evidently double-counts pulsars, as IPTA includes the nine-year NANOGrav data.
However, not all pulsars in the nine-year dataset have been observed for a sufficient amount of time to contribute to the frequency bins we consider, somewhat reducing the number of double-counted pulsars.
Nonetheless, a significant overlap between the two datasets remains. 
Consequently, while we still present the naive combination as a rough indicator for how the results might change in the near future, the corresponding fit should be taken with caution.

% ======================================================================
\section{Cosmological constraints and final results}
\label{sec:results}
% ======================================================================

As discussed in \cref{sec:sources}, in order to explain the large observed signal, the GW sources should constitute a significant fraction of the total energy density in the universe at the time of GW production, and are therefore likely to be subject to other cosmological constraints. Furthermore, the frequency range puts the time of GW production close to that of nucleosynthesis and neutrino decoupling at $T\lesssim \SI{1}{\MeV}$ and $T\approx \SI{2}{\MeV}$, respectively. More precisely, the frequency today can be written as~\cite{Schwaller:2015tja}
\begin{equation}
    f_0 = \frac{a_*}{a_0} H_* \frac{f_*}{H_*} \approx \SI{e-9}{\Hz} \times \frac{T_*}{\SI{10}{\MeV}} \frac{f_*}{H_*}\,,
\end{equation}
where $T_*$ is the temperature of the universe at the time of production. 
Causality requires the last factor to be larger than unity, therefore GWs in the frequency range probed by NANOGrav cannot be produced much before the time of nucleosynthesis. 

Overall, the following constraints are considered for each model:
\begin{itemize}  
    \item The abundance of \textbf{relativistic degrees of freedom}, commonly expressed as $\Delta N_{\rm eff}$, should satisfy the current constraints $\Delta N_{\rm eff}^{\rm BBN} \leq 0.39$ at $2\sigma$ and $\Delta N_{\rm eff}^{\rm CMB} \leq 0.29$ at $2\sigma$ \cite{Planck:2018vyg} confidence level~(CL), at the relevant temperatures. 
    \item \textbf{Big Bang Nucleosynthesis} should not be affected significantly, which roughly implies that the universe should look standard model like at temperatures below \SI{1}{\MeV}. More precise constraints on the allowed amounts of energy influx into the SM thermal bath during and after BBN are available e.g.\ in Refs.~\cite{Kawasaki:2000en,Bai:2021ibt} and are taken into account when relevant. 
    \item If the energy in the GW source leads to significant \textbf{late time reheating}, then the universe should reheat to $T_{RH} \gtrsim \SI{2}{\MeV}$ such that also the neutrino sector is re-thermalised~\cite{Bai:2021ibt,deSalas:2015glj}. Furthermore the \textbf{dilution of previously produced baryon and dark matter~(DM) abundances} has to be taken into account. 
    \item \textbf{Spectral distortions of the CMB} can be induced by late decays of particles~\cite{Chluba:2011hw,Chluba:2013pya} or by large fluctuations in the plasma or even decoupled sectors~\cite{Chluba:2019kpb,Ramberg:2022irf}, and thus can both directly and indirectly constrain models that produce large GW signals at low frequencies. The current limit on $\mu$-type spectral distortions from COBE/FIRAS is $\mu < 4.7\times 10^{-5}$~\cite{Bianchini:2022dqh}, which could be improved to $\mu < 3\times 10^{-8}$ by future missions such as PIXIE~\cite{Kite:2020uix}.
    \item If the energy stored in the GW source is partially converted to non-relativistic \textbf{dark matter}, its abundance should not exceed the observed value today $\Omega_{\rm DM}h^2 \leq 0.12$. If this becomes a significant fraction of the total dark matter, also isocurvature constraints apply. 
    \item The large energy anisotropies typically required to produce GWs may lead to the formation of \textbf{primordial black holes or dark matter mini-clusters}. Their abundance should obviously agree with observational constraints. 
\end{itemize}
While above these cosmological observations are discussed as constraints, more precise measurements in the future will allow us to discriminate between different sources of GWs. Below we show this in particular for the case of constraints on $\Delta N_{\rm eff}$, which is expected to improve by an order of magnitude in the coming decade~\cite{CMB-S4:2016ple,SimonsObservatory:2018koc}, and for the case of CMB spectral distortions, which could be improved by several orders of magnitude in the not too distant future~\cite{Kogut:2011xw,Chluba:2019nxa}.

% ----------------------------------------------------------------------
\subsection{Strong first order phase transitions}
\label{sec:FOPT_result}
% ----------------------------------------------------------------------

Only very strong PTs can explain the observed GW signal, which implies that at the time of GW production the energy in the source is a sizable fraction of the total energy density of the universe. Fits of generic PT spectra have obtained $\alpha \gtrsim 0.1$~\cite{Ratzinger:2020koh,Bai:2021ibt,NANOGrav:2021flc}, which however requires pathologically small values of $\beta/H_*$ that might be difficult to realise in concrete models. Instead in our model we find $\alpha \gg 1$ for all viable parameter points. This implies that converting the energy into dark radiation and thus hiding it, as suggested in Ref.~\cite{Breitbach:2018ddu}, is not a viable option. 

\begin{figure}
    \includegraphics[width=.49\textwidth]{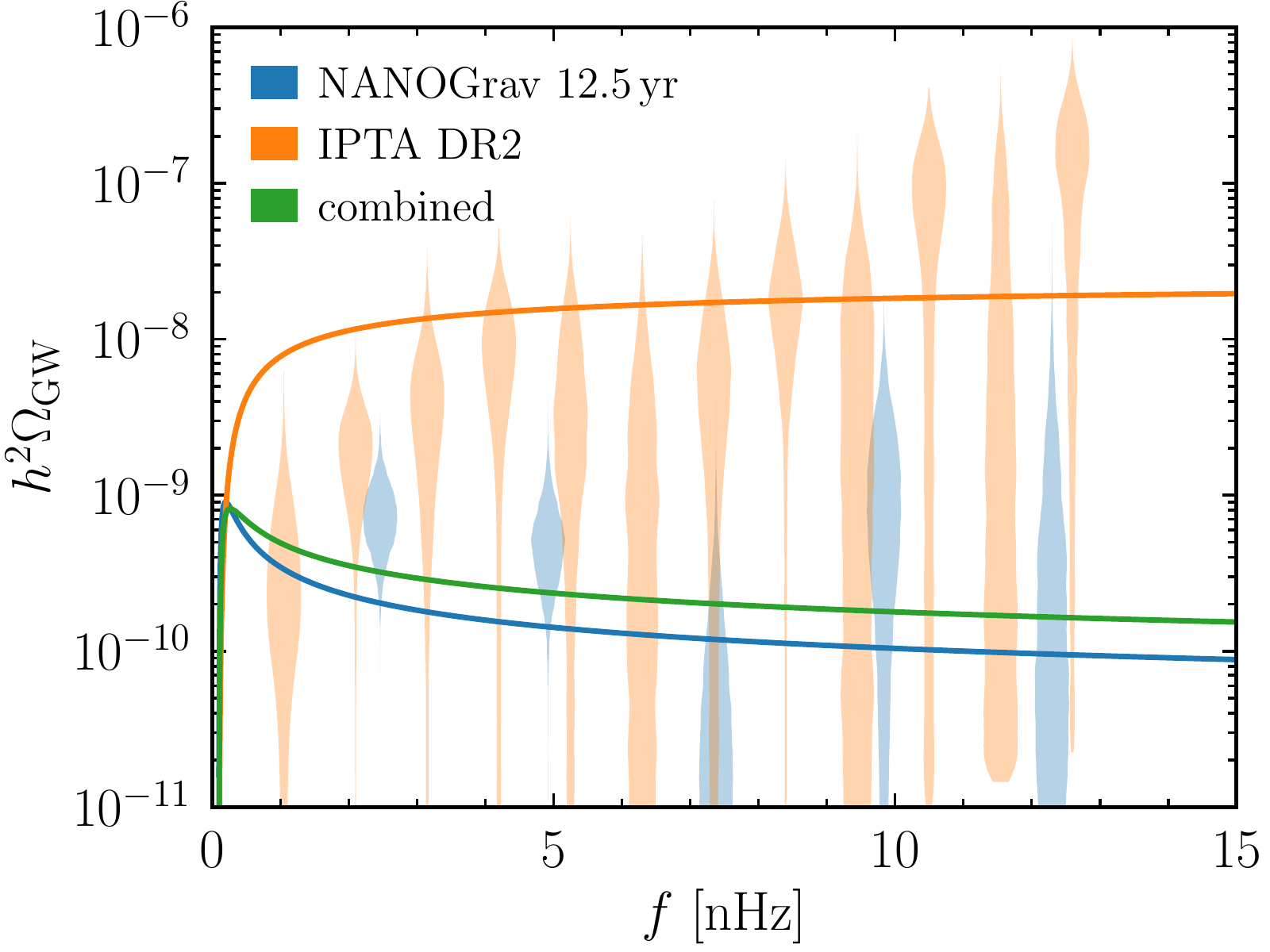}
    \hfill
    \includegraphics[width=.49\textwidth]{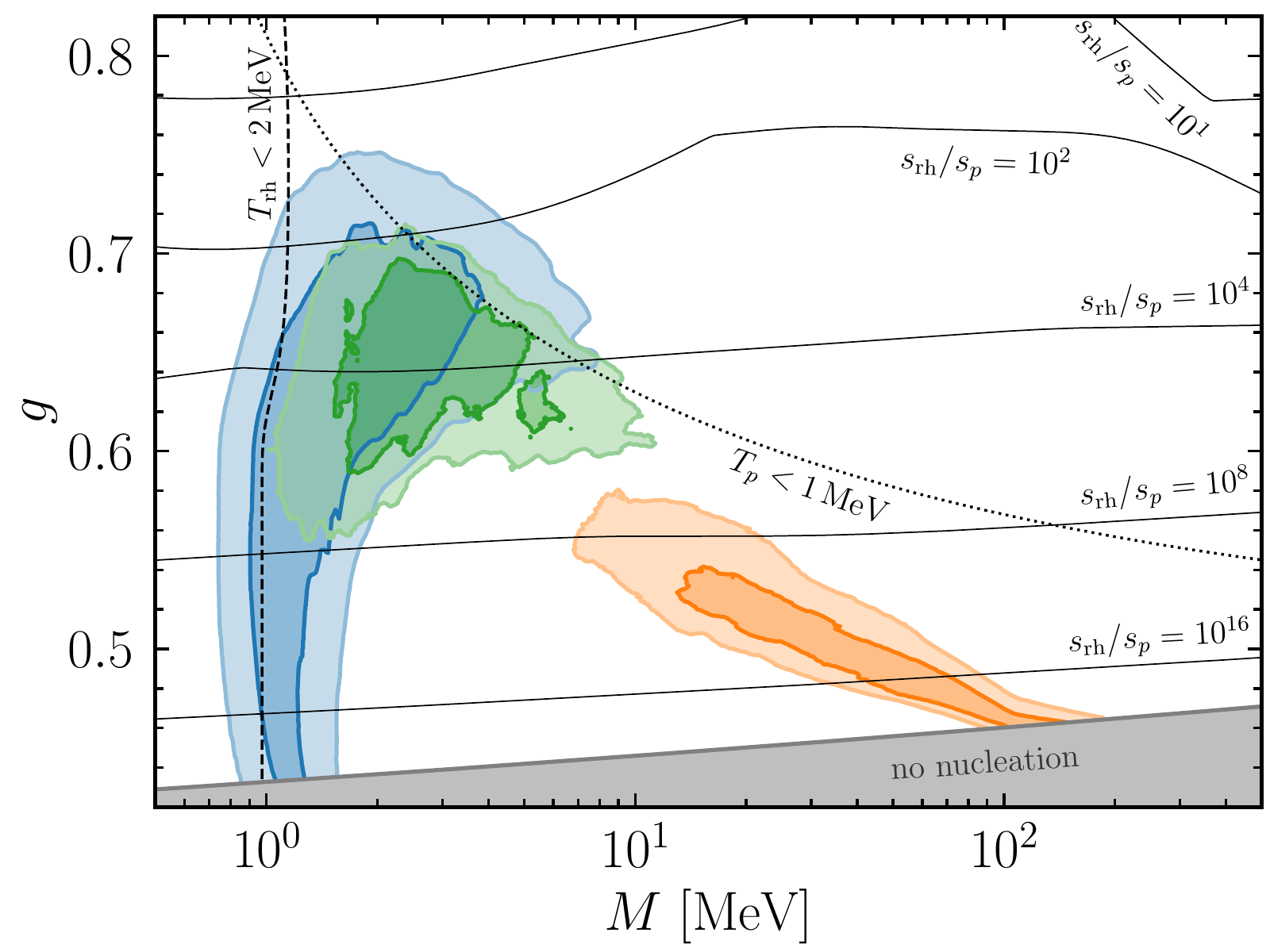}
    \caption{%
        Fit of the supercool dark sector model to the PTA data as discussed in \cref{sec:fit}. 
        Shown are the best-fit regions for the NANOGrav 12.5~year dataset~\cite{NANOGrav:2020bcs,NANOGrav:2020gpb}~(blue) and IPTA DR2 dataset~\cite{Antoniadis:2022pcn,Perera:2019sca}~(orange), as well as the best-fit of a naive combination of the two datasets~(green).
        \emph{Left:} Best-fit~(solid lines) of the GW spectrum. The violins depict the corresponding 30-bin free-spectrum fits used for the refitting.
        \emph{Right:} 2D posteriors for the coupling $g$ and mass scale $M$ of the model. Solid black contours show the dilution factor due to reheating after the PT. Below the dotted line the percolation temperature is below \SI{1}{\MeV}, while to the left of the dashed line the reheating temperature does not reach \SI{2}{\MeV}.
        The full triangle plot including 1D posteriors is shown in \cref{fig:cw_triangle}.
        \label{fig:fitPT}
    }
\end{figure}

Instead, the energy should be transferred back to the visible sector, which leads to a substantial amount of reheating. We quantify this using the ratio of entropies in the visible sector at percolation~$T_p$ and after reheating~$T_{\rm rh}$, $s_{\rm rh}/s_{\rm p}$. This is more than one order of magnitude for most parameter points, and can exceed $10^{10}$ in some regions, as shown in \cref{fig:fitPT}. This ratio is important because it corresponds to the amount by which previously generated abundances such as the baryon asymmetry or the dark matter density are diluted. A dilution factor of order $10^{10}$ would require an order one baryon asymmetry before the phase transition, so significantly smaller values are clearly preferred. 

Let us consider the results of the fit, shown in \cref{fig:fitPT}, in more detail. First we see that the NANOGrav data prefers a lower mass scale of $M \sim \SIrange{1}{10}{\MeV}$, while the IPTA data prefers a higher mass but also significantly more supercooling/reheating. This can easily be understood from the left figure, where one sees that IPTA data prefers to sit on the IR tail of a very large signal, which pushes the fit to larger masses and larger amplitudes, while NANOGrav prefers to have the GW peak at lower frequencies. This discrepancy depends quite sensitively on the number of frequency bins included for each dataset, as discussed in more detail in \cref{app:IPTA_bins}. Ultimately, additional data will hopefully resolve this situation. 

In fact most of the parameter range preferred by IPTA would be challenging to accommodate in standard baryogenesis scenarios.\footnote{Similarly any previously produced dark matter abundance would be almost completely erased, which can however be a feature too, see e.g.\ Ref.~\cite{Berlin:2016vnh}.} While one could imagine producing the baryon asymmetry directly in this supercooled PT, it would require a departure from the minimal model that we consider here. The reheating constraint is satisfied in most of the preferred parameter space. Another interesting aspect is the percolation temperature, which is the minimal temperature that the thermal bath reaches before the PT completes. In a large fraction of parameter space, this is below \SI{1}{\MeV}, which implies that nucleosynthesis begins, with the undiluted baryon asymmetry. A second phase of BBN then follows after reheating, with the correct baryon abundance. While reheating should dissolve most bound states, this is not immediately obvious for Helium-4 with a binding energy of \SI{28.3}{\MeV}. We plan to revisit potential constraints from this two stage of nucleosynthesis in the future. 

Finally, let us discuss how the reheating process can work in practice. From the Friedmann equations, one finds the reheating temperature 
\begin{equation}
    \Trh = 0.55\, \geff^{-1/4} \sqrt{M_{\rm P}\Gamma_\phi}\,,
\end{equation}
where $\geff \approx 10.75$, and $\Gamma_\phi$ is the decay width of $\phi$. Reheating above \SI{2}{\MeV} requires $\Gamma_\phi \gtrsim \SI{4e-20}{\MeV}$. The preferred range of $m_\phi$ is \SIrange{0.92}{6.9}{\MeV} for NANOGrav and \SIrange{11.5}{124}{\MeV} for IPTA. We can consider different decay operators that can satisfy these constraints. 

The simplest scenario is the Higgs portal, via the operator
\begin{equation}
    {\cal L} \supset \lambda_{h \Phi} |\Phi|^2 |H|^2\,, \label{eqn:HiggsPortal}
\end{equation}
which after symmetry breaking leads to mixing of $\phi$ with the Higgs boson with mixing angle $\theta \approx \lambda_{h \Phi} v_\phi v_{h}/m_h$, where $v_h = \SI{246}{\GeV}$ and $m_h=\SI{125}{\GeV}$ (see e.g.\ Ref.~\cite{Ferber:2023iso} for a recent study). This allows $\phi$ decays to electrons and photons, however both channels are suppressed by the small Higgs couplings to those states, requiring a Higgs mixing of order $10^{-4}$~\cite{Ibe:2021fed,Agrawal:2021dbo},\footnote{See also Ref.~\cite{Bringmann:2023opz} for a very recent discussion of this point.} and thus a large portal coupling $\lambda_{h \Phi}$. Unfortunately the operator in \cref{eqn:HiggsPortal} also contributes a large mass for the scalar after electroweak symmetry breaking, and is thus in conflict with our initial assumption of classical scale invariance. 

Alternatively we can consider a direct decay channel to electrons or photons, via couplings
\begin{equation}
    {\cal L} \supset c_{ee}\frac{|\Phi|^2}{\Lambda^2} L H \bar{e} + c_{\gamma \gamma} \frac{|\Phi|^2}{\Lambda^2} F_{\mu\nu}F^{\mu\nu}\,,
\end{equation}
where $\Lambda$ is some UV scale. 
These operators do not violate scale invariance at the tree level, and are otherwise not strongly constrained~\cite{NA64:2021xzo}. The main laboratory probes of our PT scenario therefore are searches for a light scalar in the \SIrange{1}{100}{\MeV} range which decay to electron or photon pairs. 
In fact there is an intriguing hint for a new boson with a mass around \SI{17}{\MeV}~\cite{Krasznahorkay:2015iga,Feng:2016ysn}, which could be either $\phi$ or the gauge boson of our model, and
which could be searched for in high intensity electron beam experiments such as MESA~\cite{Backens:2021qkv}. 

As far as astrophysical probes of this scenario are concerned, for very small $m_\phi$, close to the lowest preferred masses, late decays will continue to inject energy into the plasma during BBN and before CMB formation~\cite{Coffey:2020oir,Depta:2020zbh}, and could produce CMB spectral distortions that can be probed in the future~\cite{Chluba:2019nxa}, while acoustic waves \`a la Ref.~\cite{Ramberg:2022irf} are not expected to contribute here. 

% ----------------------------------------------------------------------
\subsection{Meta-stable topological defects, remnants of symmetry breaking}
% ----------------------------------------------------------------------

%
\subsubsection{Global (ALP) Strings}\label{sec: Global strings results}

\begin{figure}
    \centering
    \includegraphics[width=.49\textwidth]{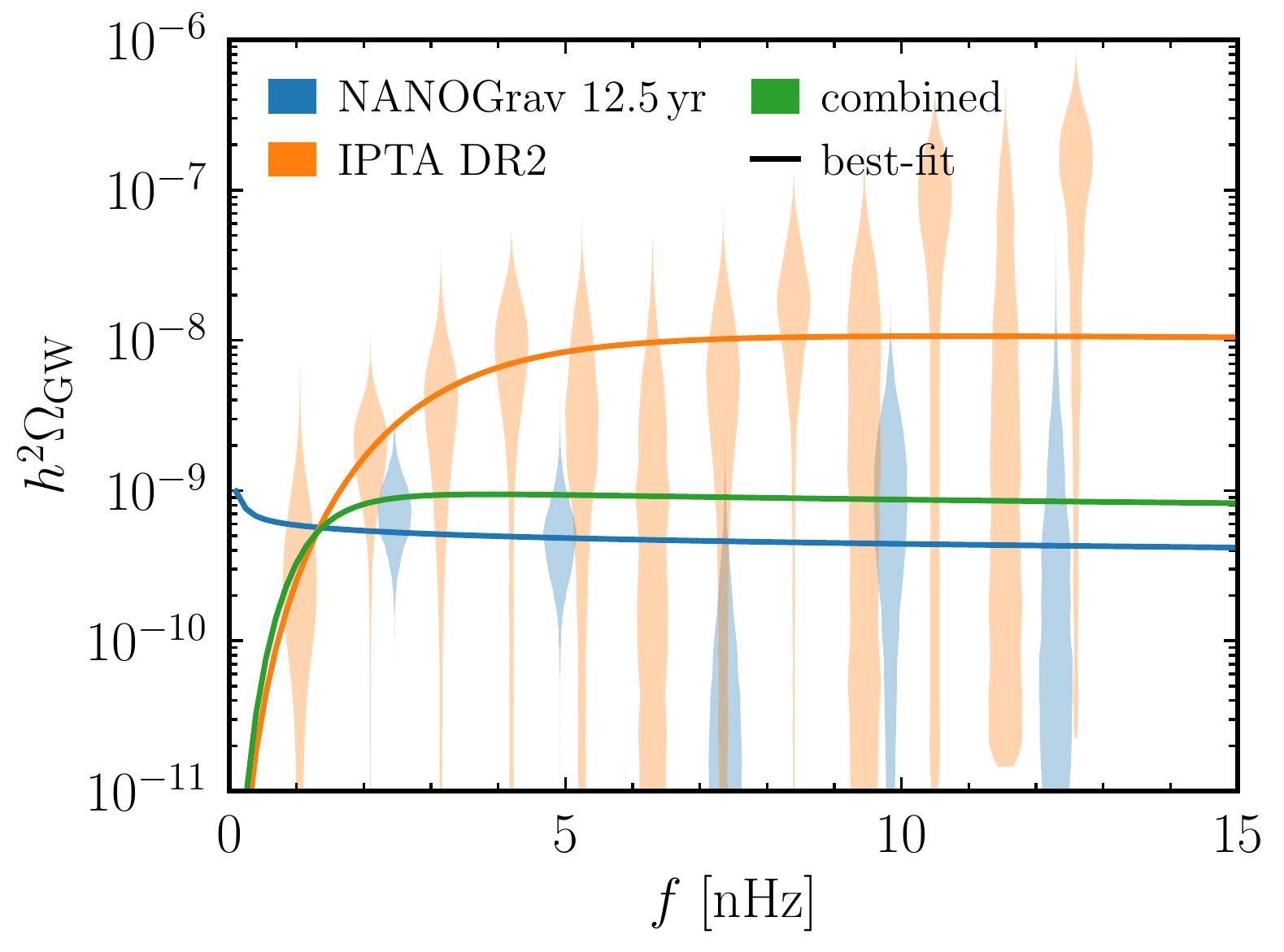}
    \hfill
    \includegraphics[width=.49\textwidth]{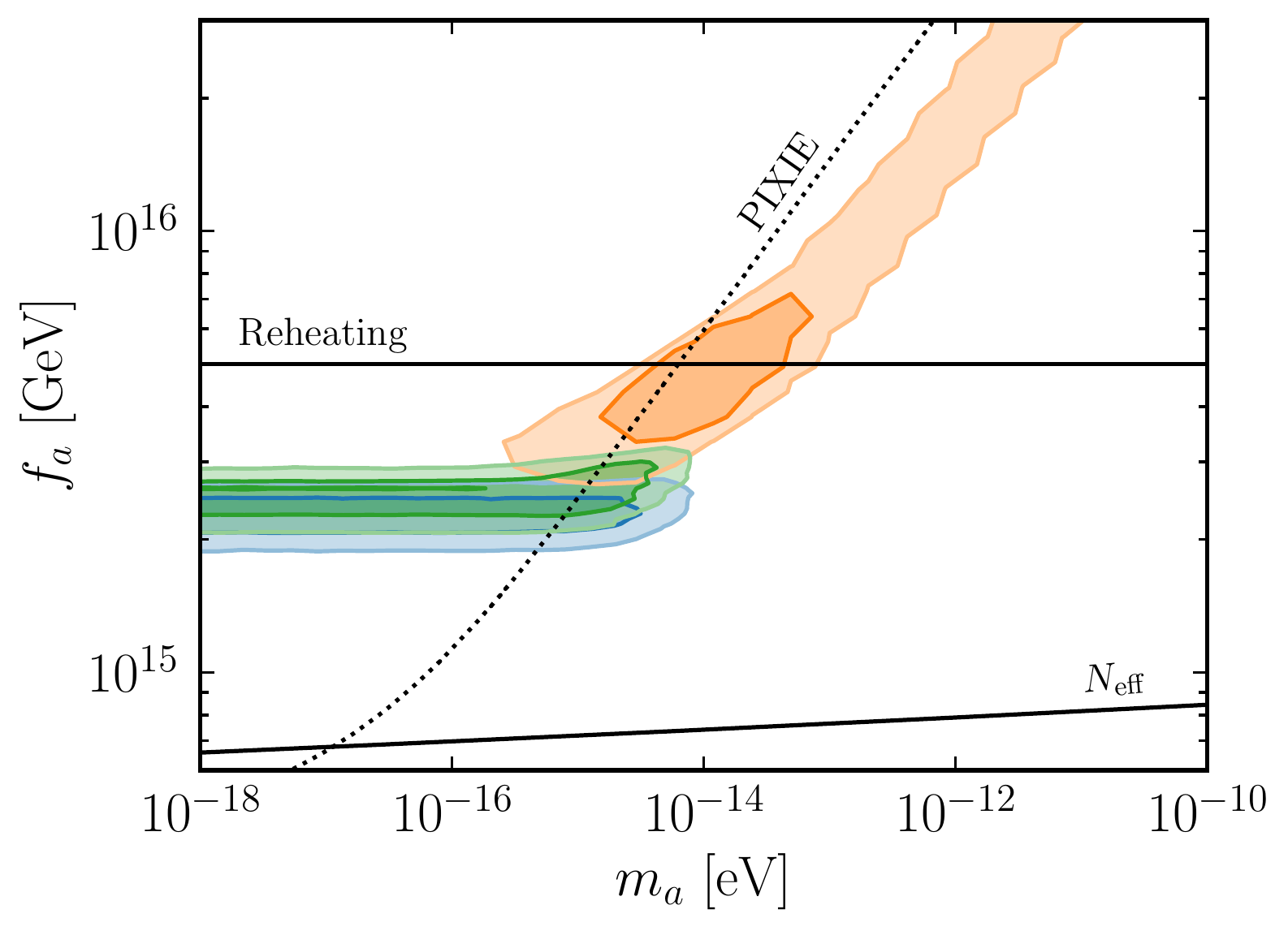}
    \caption{%
        Best-fit regions from fitting the GW signal from an ALP string network (cf.\ \cref{sec: Global strings}) to NANOGrav~(blue) and IPTA~(orange) data, as well the naive combination of both~(green).
        \emph{Left:} Fitted GW spectrum and input data for the refitting~(violins).
        \emph{Right:} Contours of the ALP mass~$m_a$ and decay constant~$f_a$. The regions above the solid lines are excluded by constraints from reheating and $N_{\rm eff}$, as discussed in the text. The projected sensitivity of PIXIE is indicated by the dotted line. The full triangle plot including 1D posteriors is shown in \cref{fig:cs_triangle}.
        \label{fig:fitCS}
    }
\end{figure}
In \cref{fig:fitCS} we show the preferred parameter space in which cosmic strings from the spontaneous breaking of a U(1) symmetry can explain the PTA signal. This region changes slope around $m_a\approx \SI{e-14}{\eV}$. For lower masses, the network decays at temperatures below $\SI{1}{\MeV}$ and the PTA signal is due to the almost scale invariant part of the spectrum, for larger masses the $\propto f^3$ tail of the spectrum that gets created in the decay of the network gives the signal. The NANOGrav data seems to prefer the former scenario, while IPTA favors the latter.  

In a minimal scenario ALPs in this parameter space are expected to be cosmologically stable with negligible SM interactions. 
As the ALP strings continuously radiate ALPs during the evolution of the network, they create an abundance of both relativistic and non-relativistic ALPs contributing to $\Neff$ and DM respectively.
For the contribution towards $\Neff$ we obtain using Eq.~(21) of Ref.~\cite{Gorghetto:2021fsn},
\begin{equation}
    \Delta \Neff = 0.4 \,\left(\frac{f_{a}}{\SI{e15}{\GeV}} \right)^{2} \left(\frac{\log(f_a/m_a)}{80} \right)^{3}\,.
\end{equation}

Using this relation, the present bound on $\Delta \Neff$ corresponds to an ALP decay constant of $f_{a}\lesssim \SI{e15}{\GeV}$. It is therefore at tension with all of the parameter space favored by the fit. At this point it is however worth mentioning that other predictions concerning the log scaling as discussed in \cref{sec: Global strings} might lead to a milder tension. 
Assuming that this is a viable explanation, the non-relativistic ALPs also contribute to DM.\footnote{This Ultralight DM is too fuzzy however to allow for the observed structure formation and can therefore only constitute a small fraction of the total dark matter~\cite{Gorghetto:2021fsn}.} In order to not overproduce DM, very small masses of $m_a\sim \SI{e-22}{\eV}$ are required. At least the NANOGrav data would still be consistent with such a scenario, our figure only ends at $m_a\sim \SI{e-18}{\eV}$ due to our choice of priors. Alternatively, also a period of late matter domination can improve the viability of the model, while still yielding observable GWs~\cite{Ramberg:2019dgi, Chang:2021afa, Gouttenoire:2021jhk,Ghoshal:2023sfa}. For completeness, let us also mention one further constraint, even though it is always weaker than $N_{\rm eff}$. In order for the string network to form, the global symmetry should be broken after inflation and reheating. The current Planck bound on the Hubble scale~\cite{Planck:2018vyg} at the end of inflation, $H_\mathrm{inf} \leq \SI{6e13}{\GeV}$, one finds that the highest temperature of the post-inflationary universe is $T_\mathrm{max} \sim \SI{5e15}{\GeV}$. The constraint $f_{a}\leq T_\mathrm{max}$ is therefore also indicated in \cref{fig:fitCS}. 

Overall, it appears that the global CS scenario is not a viable explanation for the current PTA data. Nevertheless, future GW probes in the nHz range will be able to test viable regions of parameter space (see e.g. Fig.~11 of~\cite{Ramberg:2022irf}), and it could easily be that most of the current signal is of astrophysical origin. A complementary probe of such a scenario would also be provided by future measurements of $\mu$-distortions, as shown by the dotted line in \cref{fig:fitCS}. We therefore decided to keep it in our list of simple benchmark models. 

\subsubsection{Annihilating ALP/Axion DWs}\label{sec:ALP_DW_results}
To explain the available PTA data with DWs, the network needs to annihilate close to BBN.
In order to produce a sufficient amount of GWs, it needs to comprise at least a fraction of the total energy, $\Omega_\mathrm{DW}\sim 0.1$. 
Annihilating DWs mainly decay into non-relativistic particles, which in our models are the heavy axions and ALPs. They necessarily have to decay further, in order to not overclose the universe. At this point we have to distinguish between the ALP and the aligned axion model. 

\bigskip 

In the ALP model, for simplicity, we assume that the ALPs subsequently decay to some form of dark radiation, which will contribute to $\Delta \Neff$. Assuming efficient annihilation, we can set $\rho_\text{DW}\simeq \rho_{d}$ where $\rho_{d}$ is the energy density in dark radiation, and obtain
  \begin{equation}
    \Delta \Neff \simeq 1.6\left(\frac{\gs(T)}{10}\right)^{5/6} \frac{m_{a}f_{a}^{2}}{M_{\rm P} T^2}\,,
\end{equation}
where the temperature should be taken at the time of annihilation, $T = T_{\rm ann}$.
Besides $\Delta\Neff$ constraints, DW networks that are decoupled from the SM were shown in Ref.~\cite{Ramberg:2022irf} to exhibit $\mu$-distortions. We follow their approach to compute the induced $\mu$-distortions and impose the constraints from FIRAS as well as the expected sensitivity of the PIXIE proposal. 
There are two more technical constraints that we impose. First we require that the DW network annihilates sufficiently before it would start to dominate the energy density of the universe, $T_{\rm dom}\lesssim 4\, T_{\rm ann}$. On the other hand we also require that $T_{\rm dom} \gtrsim T_{\rm ann}/4$, to ensure that plasma effects on the DWs are negligible~\cite{Blasi:2022ayo}. This should ensure that our results are in the region where the simulations of the GW spectra can be trusted. 

\begin{figure}
    \centering
    \includegraphics[width=.5\textwidth]{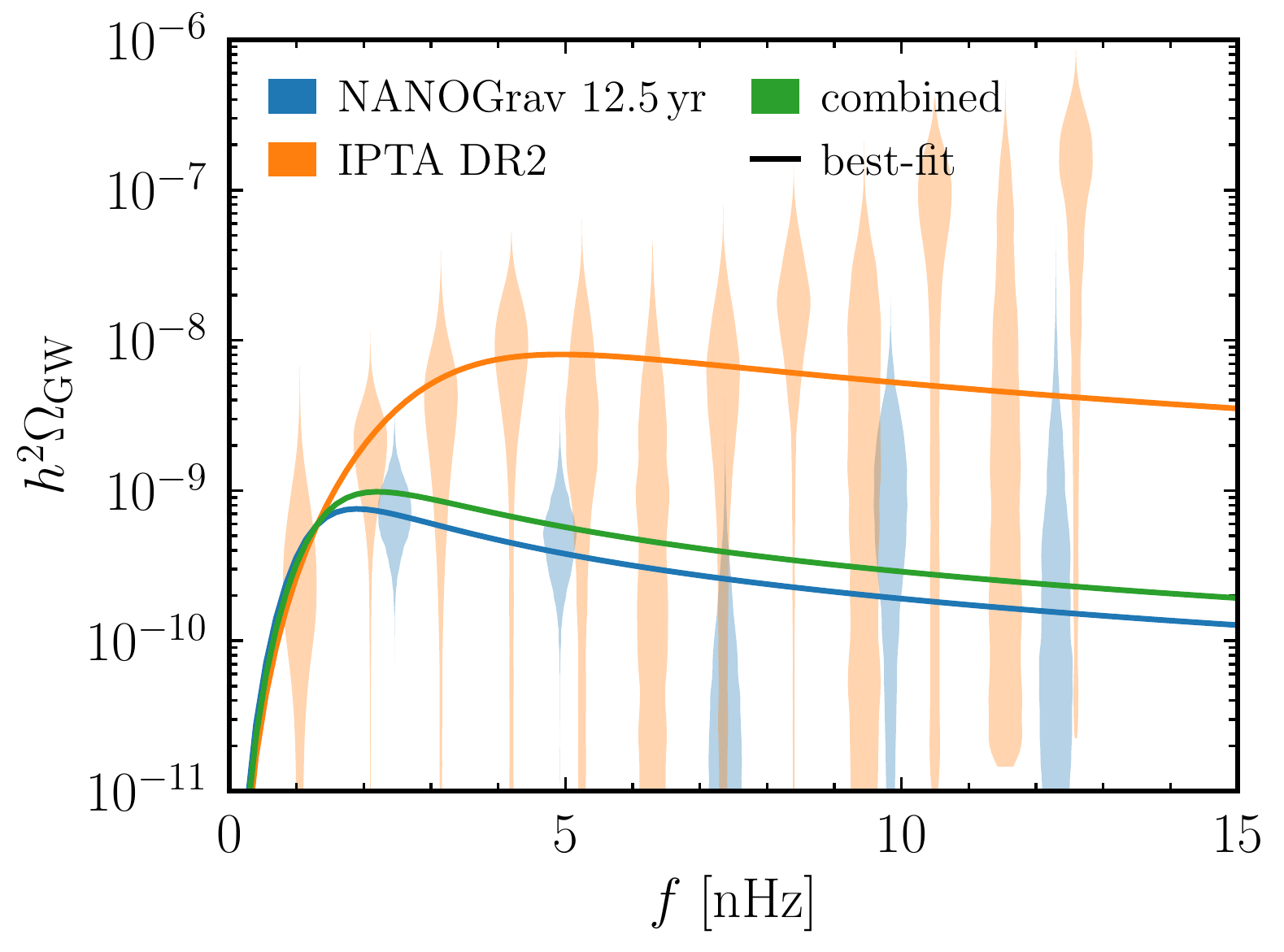}
    \hfill
    \includegraphics[width=.49\textwidth]{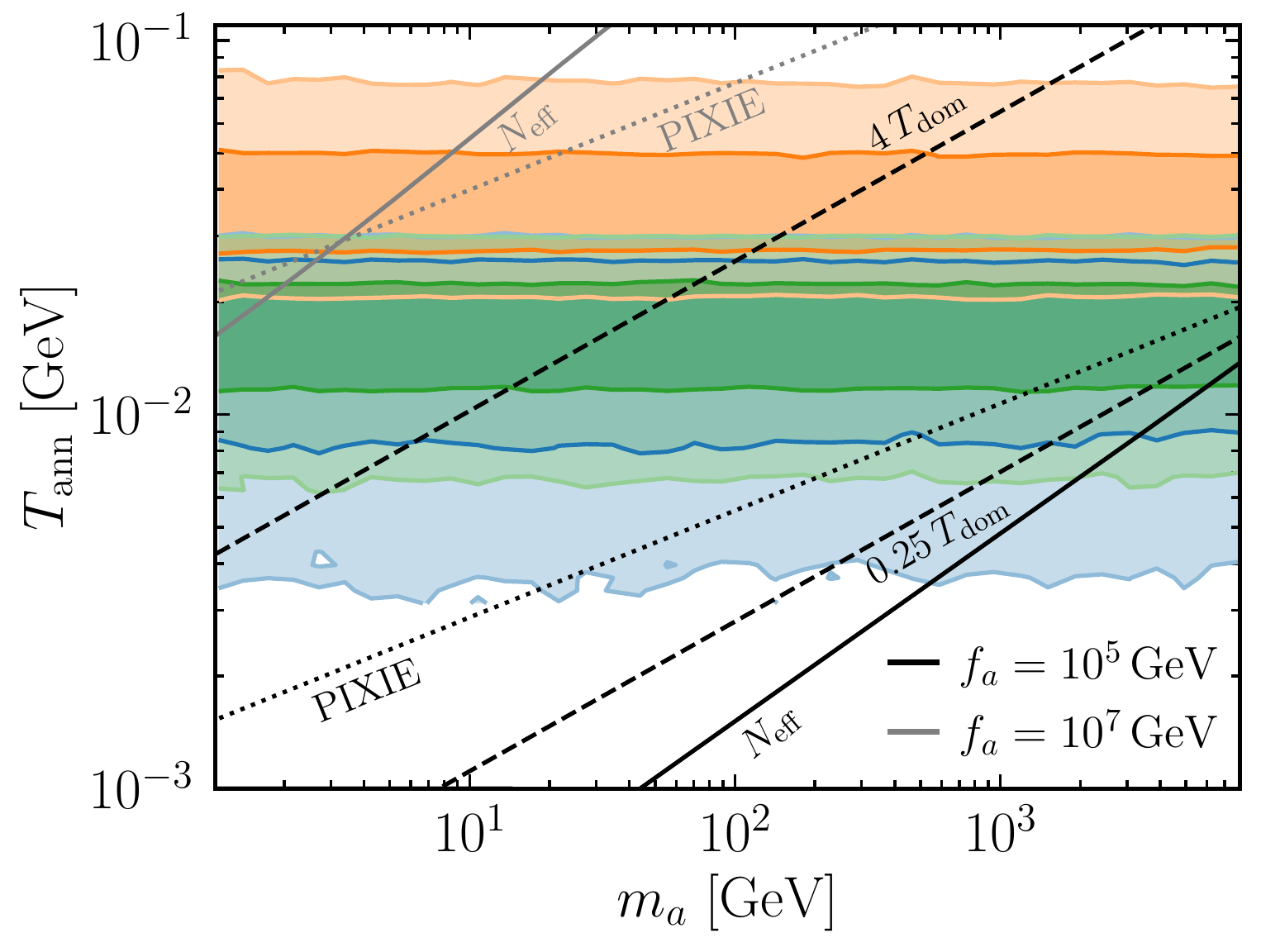}
    \caption{%
        Fit results of the ALP DW model from \cref{sec:ALP_DW} to NANOGrav~(blue), IPTA~(orange) and their combination~(green).
        \emph{Left:} Best-fit GW spectrum alongside the free-spectrum fit~(violins).
        \emph{Right:} 68\,\% and 95\,\%~CL fit region in terms of the axion mass~$m_a$ and annihilation temperature~$T_{\rm ann}$. In the region between the dashed lines, our description of the GW spectrum in terms of the scaling regime is valid. The region below the solid lines is excluded by $N_{\rm eff}$ for $f_a=10^{5}$~GeV (black) and for $f_a=10^{7}$~GeV (grey). The dotted line shows the projected sensitivity of PIXIE. 
        The full triangle plot including 1D posteriors is shown in \cref{fig:alpdw_triangle}.
        \label{fig:fitALPDW}
    }
\end{figure}

\Cref{fig:fitALPDW} shows the range of axion masses and annihilation temperatures $T_{\rm ann}$ preferred by the fit of the ALP DW GW spectrum to NANOGrav~(blue) and IPTA~(orange) data, as well as the combined fit~(green). Here $f_a$ takes values between \SI{5e4}{GeV} and \SI{e7}{GeV}, with smaller values preferred for larger $m_a$ and for NANOGrav, while IPTA prefers values closer to the upper end of the range, as can be seen in the full triangle plot in \cref{fig:alpdw_triangle}. Only a small part of the parameter space is disfavored by $N_{\rm eff}$ limits.
We indicate them for two characteristic values of $f_a$ in the figure. As discussed above, our estimate of the GW signal is only reliable in a certain window, which here is the region between the dashed lines. Finally PIXIE would be able to probe the region below the dotted line. It is apparent that especially at small annihilation temperatures $\mu$-distortions provide a strong independent probe of the model going much beyond the \Neff limit. Within the range of decay temperatures favored by the fit their reach is however limited.

\begin{figure}
    \centering
    \includegraphics[width=.5\textwidth]{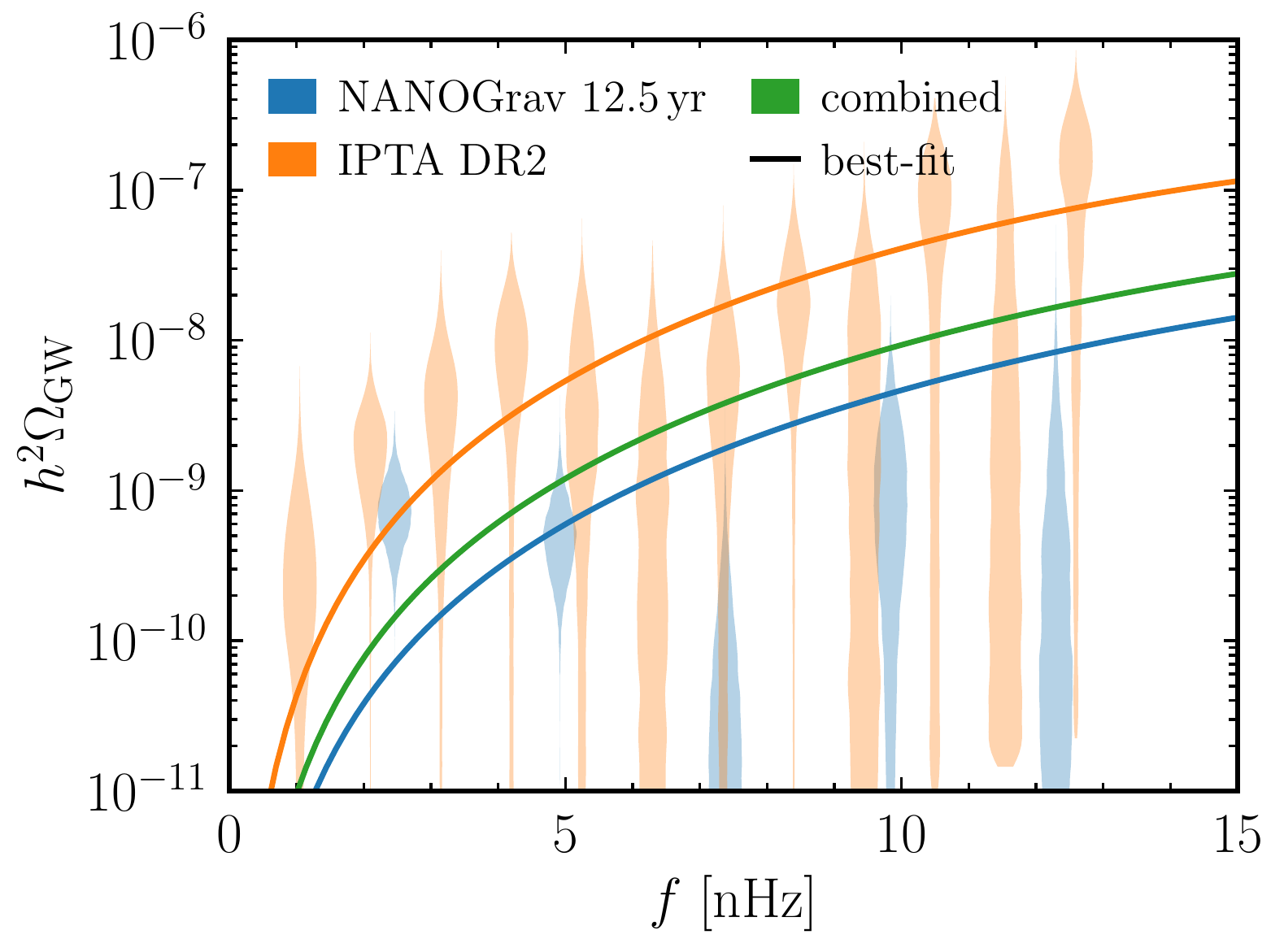}
    \hfill
    \includegraphics[width=.48\textwidth]{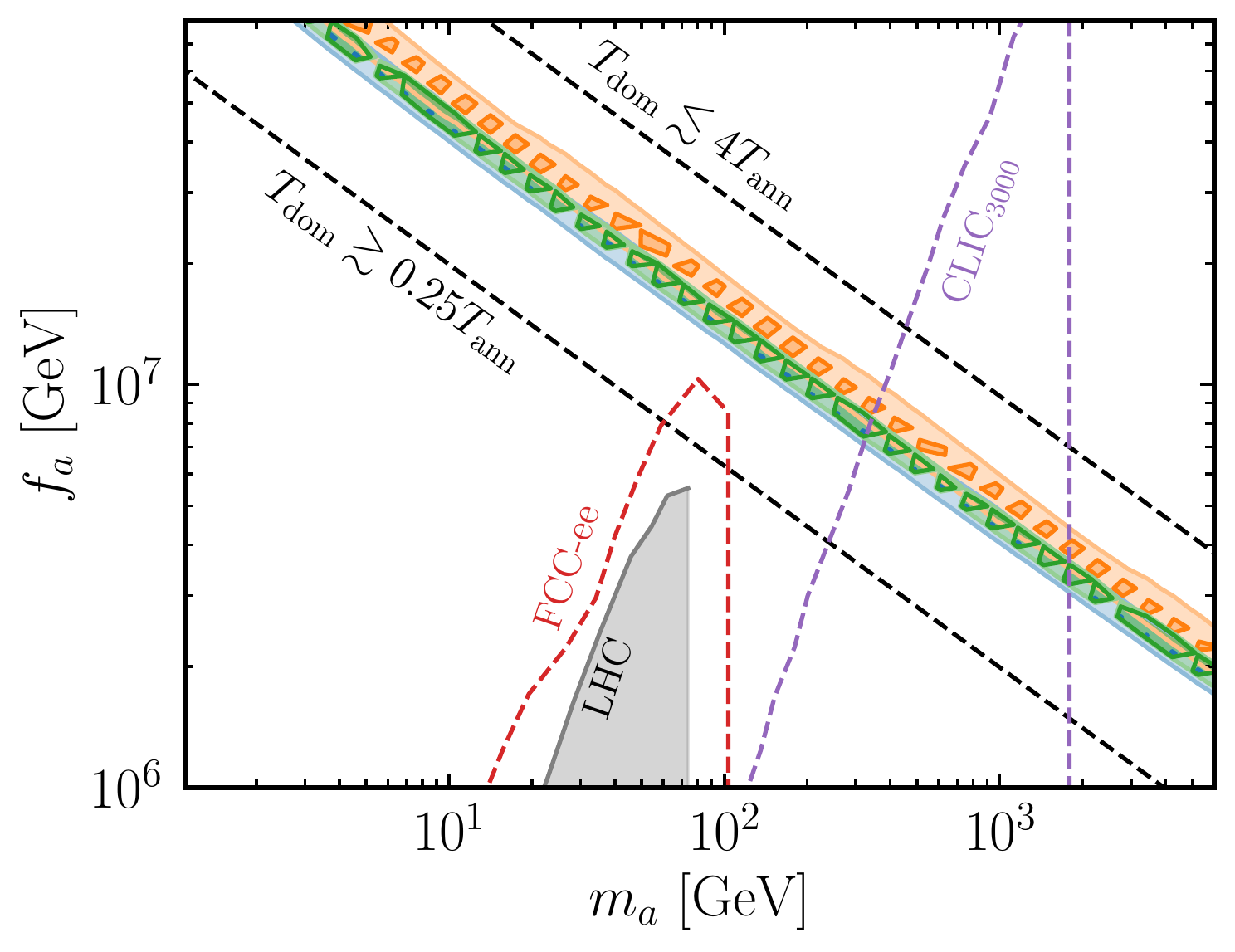}
    \caption{%
        Fit results of the aligned QCD axion DW model from \cref{sec:ALP_DW} to NANOGrav~(blue), IPTA~(orange) and their combination~(green).
        \emph{Left:} Best-fit GW spectrum alongside the free-spectrum fit~(violins).
        \emph{Right:} 68\,\% and 95\,\%~CL fit region in terms of the axion mass~$m_a$ and decay constant~$f_a$. In between the dashed lines our description of the GW spectrum in terms of the scaling regime is valid. The full triangle plot including 1D posteriors is shown in \cref{fig:haa_triangle}. The collider projections from LHC Run 2 in grey are taken from Ref.~\cite{Bauer:2017ris}, whereas the projections from searches by FCC and CLIC are from Ref.~\cite{Bauer:2018uxu}.
        \label{fig:fitHAA}
    }
\end{figure}

\bigskip 

In the aligned axion model, we expect instead that the heavy axions rapidly decay to SM particles after DW annihilation. Therefore, we do not expect constraints from $N_{\rm eff}$ or spectral distortions. Instead, one needs to make sure that the decay products of the heavy axions do not jeopardize BBN. To estimate this, we compare their decay rate into gluons and photons with the Hubble rate, i.e.\ $\Gamma_{a \rightarrow gg/\gamma \gamma} \simeq H(T)$, where the decay rate at leading order of an axion into two gluons is given by
\begin{equation}
    \Gamma_{a\rightarrow gg}\sim \frac{1}{64\pi} \left(\frac{\mathcal{C}_{gg} \alpha_{s}}{2 \pi}\right)^{2} \frac{m_{a}^{3}}{f_{a}^{2}} \simeq \SI{1.67e11}{\per\second} \left(\frac{m_{a}}{\SI{10}{\GeV}}\right)^{3}\left(\frac{\SI{e7}{\GeV}}{f_{a}}\right)^{2} \,,
\end{equation}
with $\alpha_{s} = 0.1$ and $\mathcal{C}_{gg}=1$. 
Our primary concern will be the decay into gluons for $m_{a}\geq \SI{1}{\GeV}$. This rate is fast enough to ensure decays before the onset of nucleosynthesis. In fact it guarantees that the relic axions will almost instantly decay after the DW network annihilates for all values of $m_{a}$ and $f_{a}$ considered here. 

We show the 68\,\% and 95\,\% CL contours as a function of the axion mass and decay constant in \cref{fig:fitHAA}. The technical constraints discussed for the ALP case also apply here. Again we see that the best-fit region fully agrees with the range of validity of the DW simulations we are using, and there are no conflicts with cosmological bounds. It should be noted though that the best-fits shown on the left of \cref{fig:fitALPDW,fig:fitHAA} disagree substantially. This difference is due to the heavy axion model possessing effectively only one parameter, with the surface energy $\sigma\propto f_a^2m_a$ and the annihilation temperature $T_\mathrm{ann}\propto 1/(f_a\sqrt{m_a})$ both being controlled by the same combination of parameters. This leads to the peak of the spectrum sitting at higher frequencies than the range probed by PTAs, while for the ALP model the peak can be freely adjusted and the fit prefers parameters where it falls into this range.

Furthermore, it can be interesting to ask whether the heavy axions in this model can be probed in the laboratory, in particular at the LHC. It was shown in Ref.~\cite{Bauer:2017ris} that the production of axions in the decay of electroweak bosons provide a particularly sensitive probe for heavy axions in the \SIrange{1}{100}{\GeV} mass range. 
While the projected collider reach of the LHC~(grey shaded region) is not sufficient to probe the best-fit region, it is still interesting to see that collider probes of such scenarios are in principle possible.
In particular, a future linear electron-positron collider such as CLIC with a center-of-mass energy of \SI{3}{\TeV} can explore the best-fit region for axion masses above $m_a \gtrsim \SIrange{10}{100}{\GeV}$, whereas a circular collider like FCC-ee would not be able to probe the required decay constants~\cite{Bauer:2018uxu}.

A potentially important constraint on DWs as a source for the PTA signal comes from the formation of PBH during the annihilation of the domain-wall network. Closed domain walls of typical radius $r\sim H^{-1}\sim t$ will form a BH when their linear size becomes smaller than their Schwarzschild radius $r_S\propto M \propto r^3$~\cite{Ferrer:2018uiu, Gelmini:2023ngs, Gouttenoire:2023ftk}. According to a recent study~\cite{Gouttenoire:2023ftk}, in the parameter space where the 
 PTA signal is reproduced an overabundance of PBH is produced. 
 Still, we believe that the current understanding of this process does not allow to claim the exclusion. First, the scaling regime $r \sim t$ is expected to be violated when the Schwarzschild radius becomes comparable to $r$, but dedicated simulations are lacking. Secondly, the assumption $r= t$ implies a monochromatic spectrum, which is not expected to be realized in a consistent cosmological history. Last, the evolution from the start of the annihilation process $T_\mathrm{ann}$ until the time of BH formation $T_*$ is still very uncertain, and its impact on $f_\mathrm{PBH}$ is parametrized by the ratio $(T_*/T_\mathrm{ann})^\alpha$, with $\alpha \simeq 7 - 28$. Given these large uncertainties, we refrain from showing these constraints in our figures, but we stress their potential relevance and encourage a dedicated analysis, which goes beyond our current scope.

% ----------------------------------------------------------------------
\subsection{Bosonic instabilities and late preheating}
\label{sec: Constraints AA}
% ----------------------------------------------------------------------

\begin{figure}
    \centering
    \includegraphics[width=.5\textwidth]{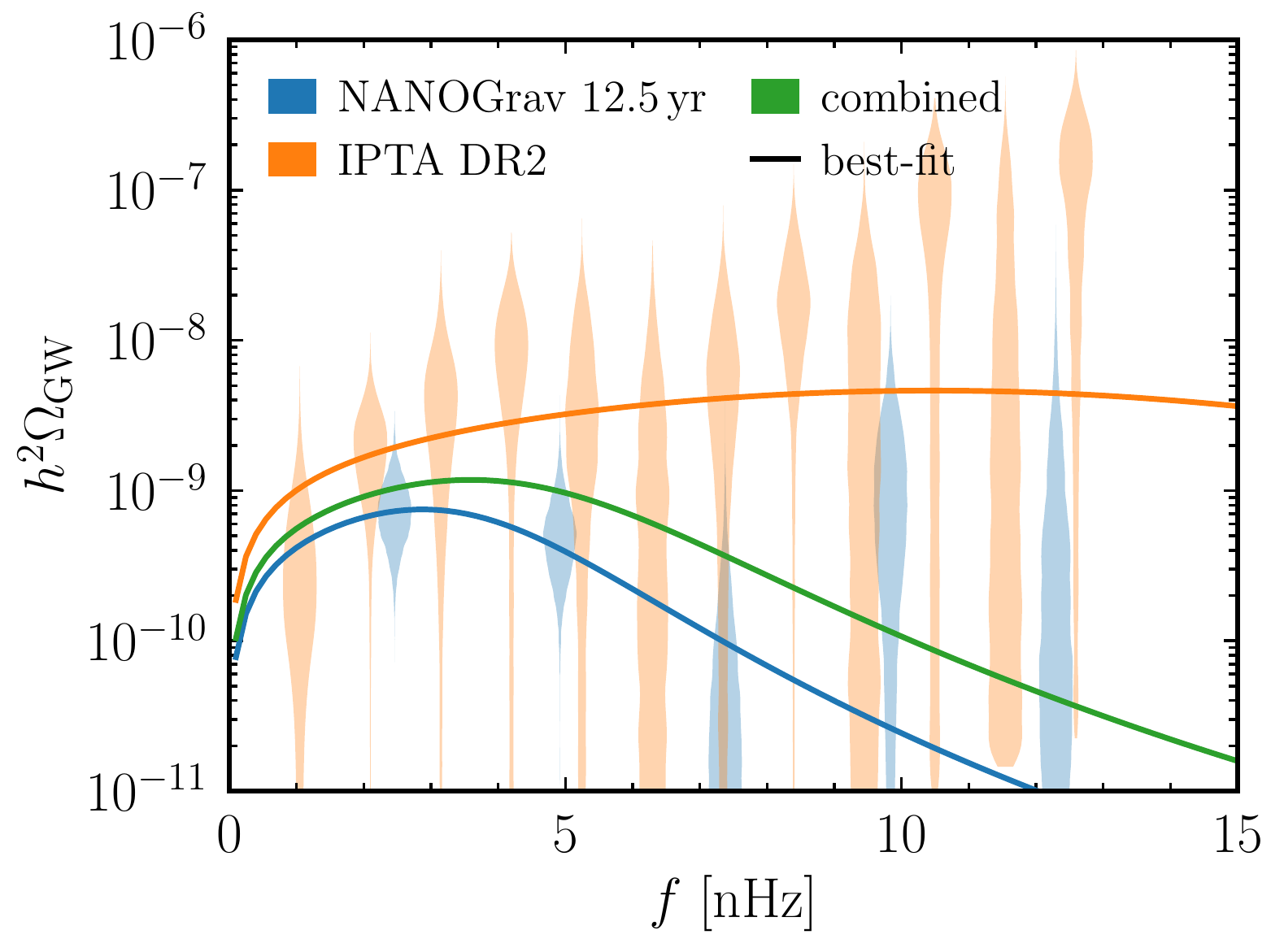}
    \hfill
    \includegraphics[width=.49\textwidth]{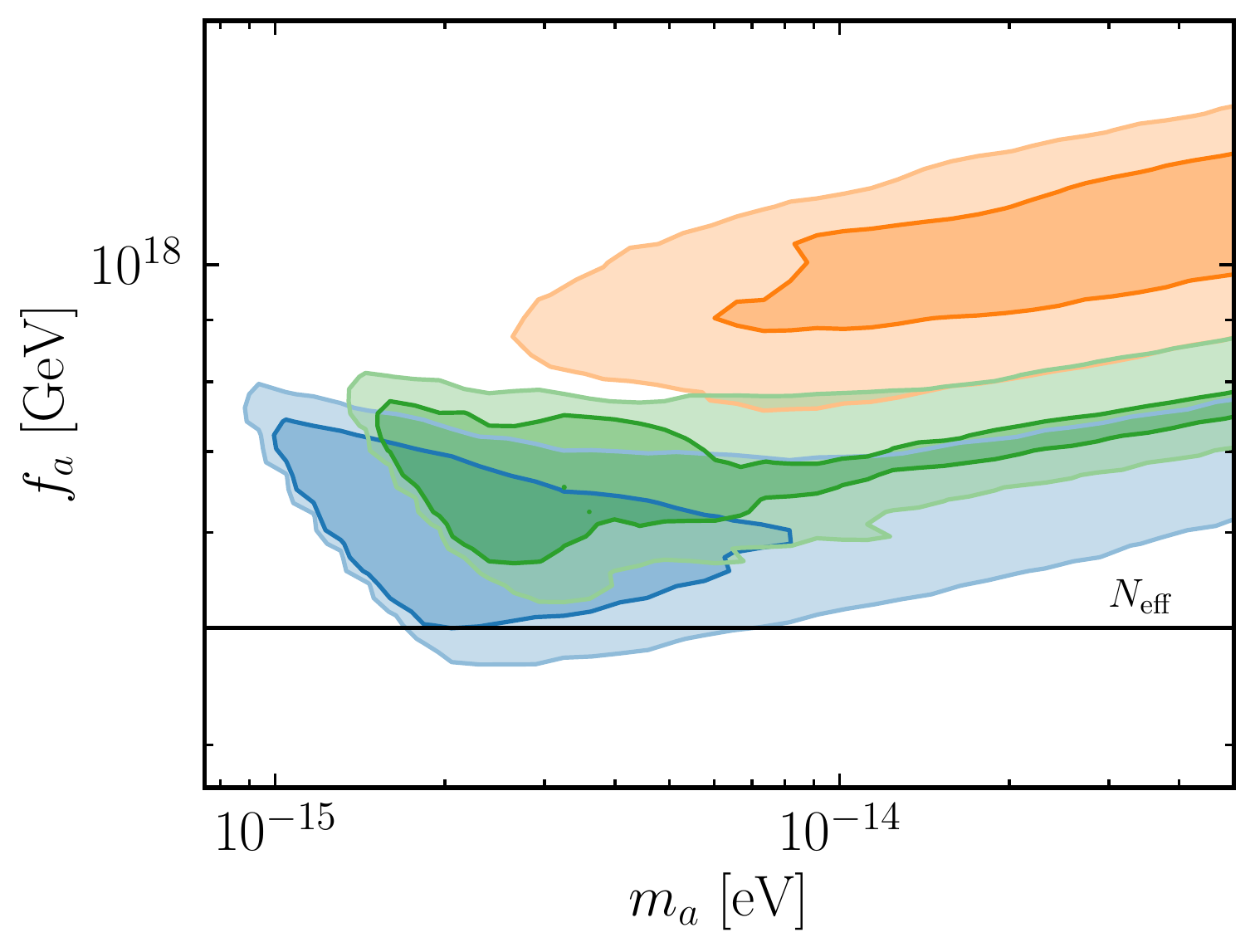}
    \caption{%
        Fit results of the audible axion model from \cref{sec:audible_axion} to NANOGrav~(blue), IPTA~(orange) and their combination~(green).
        \emph{Left:} Best-fit GW spectrum alongside the free-spectrum fit~(violins).
        \emph{Right:} 68\,\% and 95\,\%~CL fit region in terms of the axion mass~$m_a$ and decay constant~$f_a$. Decay constants above the dashed line are excluded by effective number of neutrino species~\Neff. The full triangle plot including 1D posteriors is shown in \cref{fig:aa_triangle}.
        \label{fig:fitAA}
    }
\end{figure}

Explaining the PTA signal requires the bosonic sector to comprise a non-negligible amount of the total energy. In our model of an axion coupled to a dark photon we will have two components, the axion behaving as DM and the photon contributing to \Neff, in the case where there are only gravitational interactions with the visible sector. The contribution to \Neff can be estimated as \cite{Ratzinger:2020oct}
\begin{align}
    \Delta \Neff=9.1\times\left(\frac{\theta f}{M_{\rm P}}\right)^{2}\,.
\end{align}
As one can see from \cref{fig:fitAA}, this puts the parameter space preferred by the fit in mild tension with the current bound of $\Delta \Neff\leq0.29$. Furthermore, as pointed out in Refs.~\cite{Kitajima:2017peg,Ratzinger:2020oct,Namba:2020kij}, the relic abundance of the axion is typically larger than the observed amount of dark matter. This problem has also been observed in models relying on a parametric resonance instead of tachyonic growth~\cite{Chatrchyan:2020pzh,Eroncel:2022vjg,Madge:2021abk}. A possible solution to this problem might be model extensions that allow for a time dependent axion mass as discussed in Refs.~\cite{Ratzinger:2020oct,Namba:2020kij}.

As with the other models, the second option is to deplete the energy to the SM plasma. Here this is however very challenging, and impossible to achieve via perturbative processes. 
The reason is that the axion mass sets the energy scale of all processes. The decay rate of the axion is for instance proportional to $\Gamma\propto m^3/f^2$, and therefore efficient decays can only occur long after the onset of the axion oscillations, in our case preventing the decay before BBN. The large field strengths present in this model might possibly allow for the non-perturbative production of particles through the Schwinger effect as considered in Ref.~\cite{Domcke:2019qmm}. This process, while efficient in reducing the energy in the bosonic sector, however also lowers the amount of produced GWs.

\subsection{Scalar-induced GWs}\label{sec:scalar_induced_results}

\begin{figure}
    \centering
    \includegraphics[width=.49\textwidth]{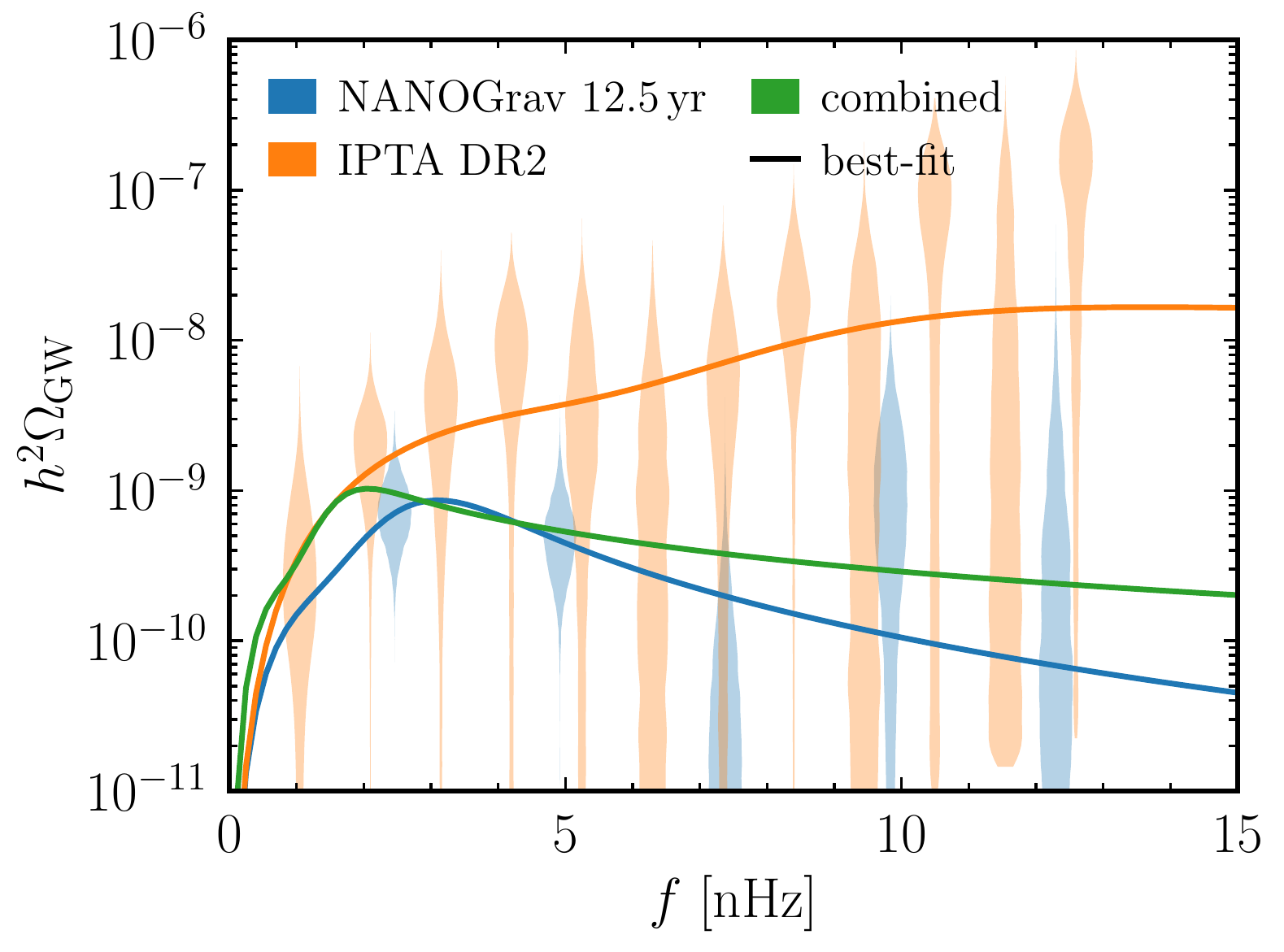}
    \hfill
    \includegraphics[width=.49\textwidth]{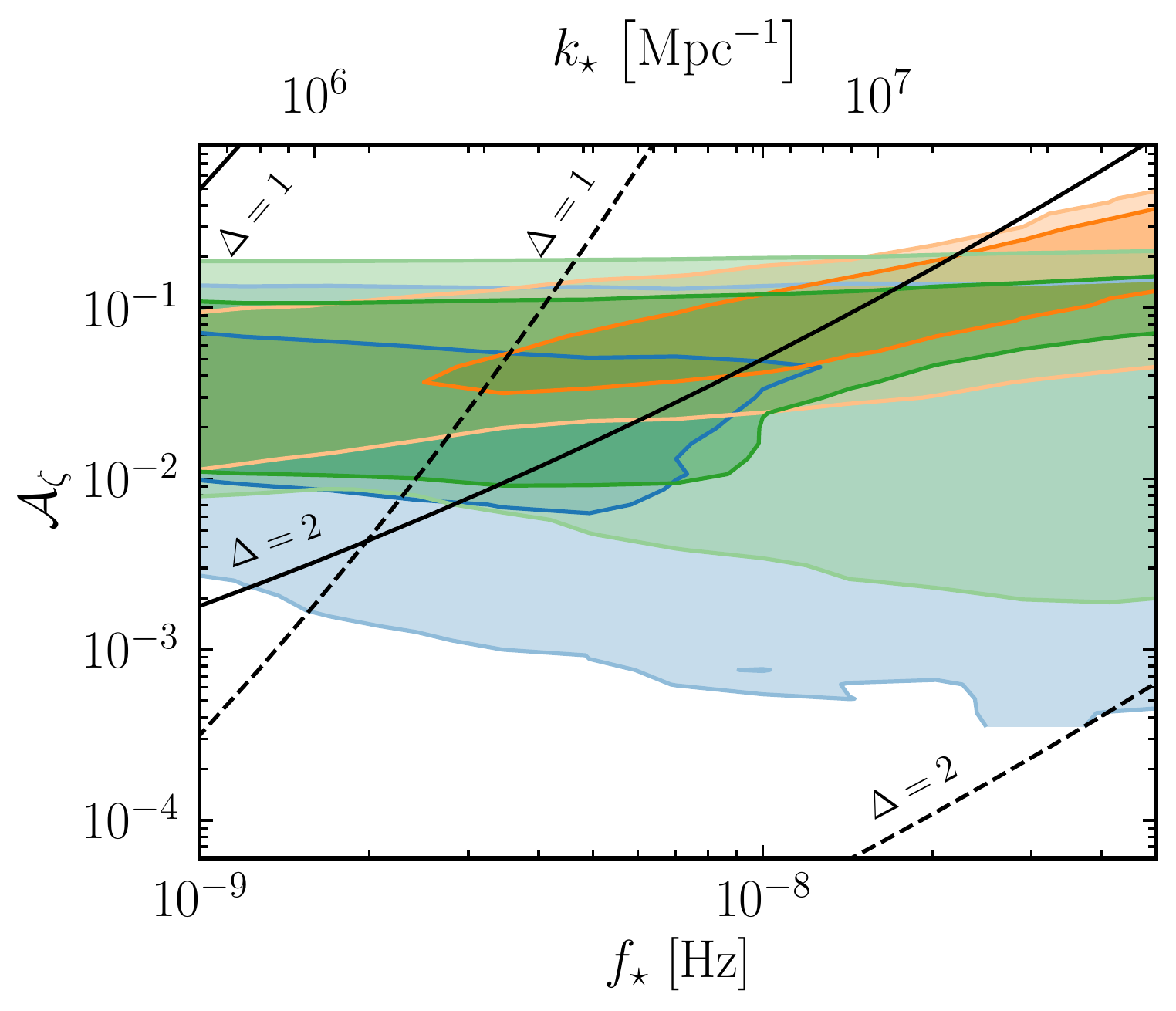}
    \caption{%
        NANOGrav~(blue), IPTA~(orange) and combined~(green) fit of the scalar-induced SGWB  (cf.\ \cref{sec:scalar_induced}). 
        \emph{Left:} Best-fit GW spectrum.
        \emph{Right:} 1$\sigma$ and 2$\sigma$ regions of the peak amplitude~$\mathcal{A}_\zeta$ and frequency~$f_\star$ of the curvature power spectrum from inflationary scalar perturbations.
        Solid lines show the upper bound on $A_\zeta$ from existing limits on $\mu$-distortions for the indicated values of $\Delta$, while dashed lines show the potential future sensitivity of PIXIE.
        The full triangle plot including 1D posteriors is shown in \cref{fig:scalar_triangle}.
        \label{fig:fitScalar}
    }
\end{figure}

Our results for scalar-induced GWs are presented in \cref{fig:fitScalar}. We see that the PTA signal can be reproduced with a rather large amplitude, $A_\zeta\sim \numrange{e-2}{e-1}$, and peak momentum around \SIrange{e6}{e7}{\per\mega\parsec}, in agreement with previous results~\cite{Dandoy:2023jot}. \Cref{fig:scalar_triangle} shows the contours and the posteriors for the other parameters of the model.

Different from the other scenarios, here the underlying source of the GWs is active during inflation. Therefore, after inflation ends and the universe reheats, only the curvature perturbations remain as traces, frozen outside the horizon. Once they reenter the horizon, besides sourcing GWs, very large curvature perturbations can lead to (over)production of PBHs. Since this has to happen at scales not too far away from the CMB, spectral distortions of the latter are also typically induced.

The production of PBHs is a delicate issue. The fraction~$f_\mathrm{PBH}$ of DM in the form of PBHs depends strongly on the choice of the window function for the variance of the density contrast, for which there is no unique prescription, and exponentially on the exact value of the critical density for collapse. This delicate sensitivity prevents a reliable calculation of $f_\mathrm{PBH}$ from the model parameters. Nevertheless, as pointed out in Ref.~\cite{Dandoy:2023jot}, the reverse approach is possible: imposing $f_\mathrm{PBH}< 1$, reliable bounds on the parameters of the model can be derived.
Assuming Gaussian distributed perturbations, the upper bound on $A_\zeta$ derived from $f_\mathrm{PBH}< 1$ is $A_\zeta < \numrange{0.01}{0.04}$~\cite{Dandoy:2023jot}, falling right inside the best-fit region shown in \cref{fig:fitScalar}. This claims for a very careful assessment of the PBH constraint.
An important role is played by non-Gaussianities~(NGs), which come from two sources: primordial NGs in the inflationary spectrum and NGs coming from non-linearities in PBH formation.
A proper inclusion of these two effects can suppress PBH formation and lift the constraint, only mildly affecting the GW spectrum~\cite{Ferrante:2023bgz}. This effect can only be computed in a model-by-model basis. We therefore refrain from presenting PBH constraints here, but we highlight their potential importance.

\medskip

The absence of spectral distortions in the CMB sets strong constraints on the amplitude of the power spectrum at low momenta. In particular, COBE/FIRAS set an upper limit on the $\mu$ parameter~\cite{Mather:1993ij, Fixsen:1996nj}, defined as
\begin{equation}\label{eq:mu parameter}
    \mu \approx \hspace{-10pt}\int\limits_{\SI{1}{\per\mega\parsec}}^\infty \hspace{-10pt}\frac{dk}{k}\, P_\zeta(k) W_\mu (k) \,,
\end{equation}
where $W_\mu$ is the window function~\cite{Chluba:2015bqa} 
\begin{align}
        W_\mu (k) &\approx 2.27\left(\exp\left[-\left(\frac{k}{\SI{1360}{\per\mega\parsec}}\right)^2\left(1+\left(\frac{k}{\SI{260}{\per\mega\parsec}}\right)^{0.3}+\frac{k}{\SI{340}{\per\mega\parsec}}\right)^{-1}\right]\right.\\
    &\hspace{1.5cm}\left.-\exp\left[-\left(\frac{k}{\SI{32}{\per\mega\parsec}}\right)^2\right]\vphantom{\left[\left(\left(\frac{1}{1}\right)^{0.3}\right)^{-1}\right]}\right).\nonumber
\end{align}

In \cref{fig:fitScalar} we show as solid lines the bounds obtained by imposing $\mu < \num{4.7e-5}$ (from a recent reanalysis of COBE/FIRAS data~\cite{Bianchini:2022dqh}) and $\mu < \num{3e-8}$ as a benchmark for future observers such as PIXIE~\cite{Kite:2020uix}.
The bounds are obtained with fixed $\Delta = 1$ or $\Delta=2$, while the tilt $n$ has no impact because the integral in \cref{eq:mu parameter} is dominated by momenta below the peak. Precisely for this reason, constraints for smaller $\Delta$ vanish, as the spectrum becomes sharper.%

Depending on the value of $\Delta$, current $\mu$-distortion constraints exclude interesting portions of the parameter space in this model, and future CMB observations have an important discovery potential, especially for spectra with broad tails in the IR, which correspond to $\Delta \gtrsim 1$ in our model.

% ======================================================================
\section{Conclusions}
\label{sec:conclusion}
% ======================================================================

In this paper, we have proposed and studied a series of benchmark models that can produce GWs in the frequency bands probed by PTA experiments. By focusing on models with a minimal set of free parameters, we were able to identify the best-fit regions for each model directly in terms of the model parameters such as the masses of the new particles and their couplings. In particular we have shown that it is possible to directly reconstruct the parameters of a model with a first order phase transition from GW data, without taking the detour of reconstructing the signal parameterization first. This has the clear advantage that one immediately knows that a given point in the fit can actually be realised by a particle physics model. 

Fitting a large variety of models to two different PTA datasets was in particular enabled by using the new tool \texttt{ceffyl}, which dramatically speeds up obtaining the viable parameter regions. We fit our models to two datasets, the 12.5~year NANOGrav search and the data from IPTA DR2 (which includes the NANOGrav nine-year data). For some models, the best-fit regions disagree at 95\% CL. This was more pronounced for models that feature a sharper peak, since IPTA seems to prefer a flatter spectrum. The discrepancy therefore also seems to grow if more frequency bins are included in the fit, see~\cref{app:IPTA_bins} for more details. It will be interesting to see if this discrepancy disappears with more data, or if there is a difference in the way the noise is subtracted.

In addition to the PTA data, we include a variety of cosmological constraints on the models. The main results are shown in~\cref{fig:fitPT,fig:fitCS,fig:fitALPDW,fig:fitHAA,fig:fitAA,fig:fitScalar}, where the constraints are superimposed on the regions preferred by PTA data. Let us briefly summarise the results for the individual models. We find that the signal can be well explained in scenarios with a supercooled phase transition that subsequently reheats the visible sector. In our concrete realisation a large initial baryon asymmetry is required due to the large dilution, but we expect that this dilution can be more moderate in variations of this scenario. GWs from topological defects also provide a good fit of the PTA data, however constraints from $\Neff$ rule out the global cosmic string scenario, while domain walls are viable both if they annihilate into a dark sector or into the visible sector. In the former case, CMB spectral distortions are predicted in range of future probes, while the latter case future colliders will be sensitive to parts of the parameter space preferred by the PTA signal. Tachyonic instabilities in an axion-dark photon model can also produce a GW background that fits the observed signal, however, the preferred parameter range is in tension with constraints from $\Neff$. Further studies are required to find out if variations of this scenario can improve this. Finally we also find that the GW spectrum produced by large scalar perturbations during inflation can provide a good fit to the data, with the caveat that the required amplitude for the perturbations cannot be obtained in the simplest single field inflationary models. 

On more general grounds, we observe the following. First, since the signal is very large, the source must at some point carry a significant fraction of the total energy density in the universe. Hiding this energy in a dark sector fails in most models due to $N_{\rm eff}$ constraints, with the exception of the dark ALP DW model, which seem to be a very efficient source of GWs, and a tiny slice of parameter space in the audible axion model. This bound might tighten in the future and will remain an important constraint (or hint) for these models. The obvious alternative is to transfer the energy back to the visible sector after the GWs are produced. If this happens through particle decays, as in the phase transition and domain wall scenarios, the models may be probed by laboratory experiments in the future. 
A third option is realised by the scalar-induced GW scenario. Here the GW source is produced and frozen in during inflation, while the energy of the original source field, the inflaton, goes into reheating of the visible sector. The large scalar fluctuations nevertheless have other observable consequences such as spectral distortions and production of primordial black holes, and the GWs would be a smoking gun for a non-minimal inflationary sector. 

One aspect that should not be neglected is naturalness, or how likely these models can be realised in nature. Often, a large separation of scales is implicitly required in these models. An example is the bias term in DW models, which is many orders of magnitude smaller than the other mass scales that appear there, and which has to be tuned such that the DWs annihilate shortly before they start to dominate the energy density of the universe. Similarly the operators that allow the decay of the lightest scalar in the phase transition model may spoil the classical scale invariance that is required in order to obtain the largest signals. There are also typically more parameters required in each model than those which determine the GW signal, but which could impact other observables.  

While here we have taken the next step in establishing simple and testable models, in most cases additional model building and more detailed phenomenological studies will be required. Furthermore, the prediction of the GW signal is still affected by large uncertainties, which are at least of order one in most models, and which often rely on extrapolation from complicated numerical simulations that are only available for a limited set of parameters. The large fluctuations in energy density associated with GW production can also have other observable consequences like the formation of dark matter substructures and mini-clusters. One may also wonder if the tension in the Hubble rate determination can be improved in one of these scenarios. 

An important question that we have not addressed here is what happens if an astrophysical background of GWs is included in the fit. It will probably open up the parameter space in most models, since not all of the GW signal has to be explained. Furthermore it will be interesting to see if a primordial component of the stochastic GW background can be distinguished from the contribution of SMBHBs. Certainly precise predictions of the spectral shape of the GW signals are essential for this task.

% ======================================================================
\section*{Note added}
% ======================================================================

After the appearance of our work, PTA collaborations have released new data in which, for the first time, they claim evidence for the Hellings–Downs correlation curve, a clear indication of a GWs signal in pulsar timing arrays~\cite{NANOGrav:2023gor, NANOGrav:2023hvm, Antoniadis:2023ott, Antoniadis:2023xlr, Reardon:2023gzh, Xu:2023wog}.
While we expect that the best fit regions would slightly shift with the new data, the overall conclusions of our work should be unaffected. Certainly, a study including the new data from all PTA experiments is beyond the scope of this work.

\acknowledgments%\addcontentsline{toc}{section}{\numberline{}Acknowledgments}

The authors would like to thank Toby Opferkuch and Kai Schmitz for very inspiring discussions. 
Work in Mainz is supported by the Cluster of Excellence “Precision Physics, Fundamental Interactions, and Structure of Matter” (PRISMA+ EXC 2118/1) funded by the German Research Foundation~(DFG) within the German Excellence Strategy (Project No. 39083149).

\appendix

% ======================================================================
\section{Priors, full fit results and best-fit parameters}
\label{app:triangle_plots}
\begin{figure}
    \begin{minipage}{.49\textwidth}
        \centering
        \includegraphics[width=\textwidth]{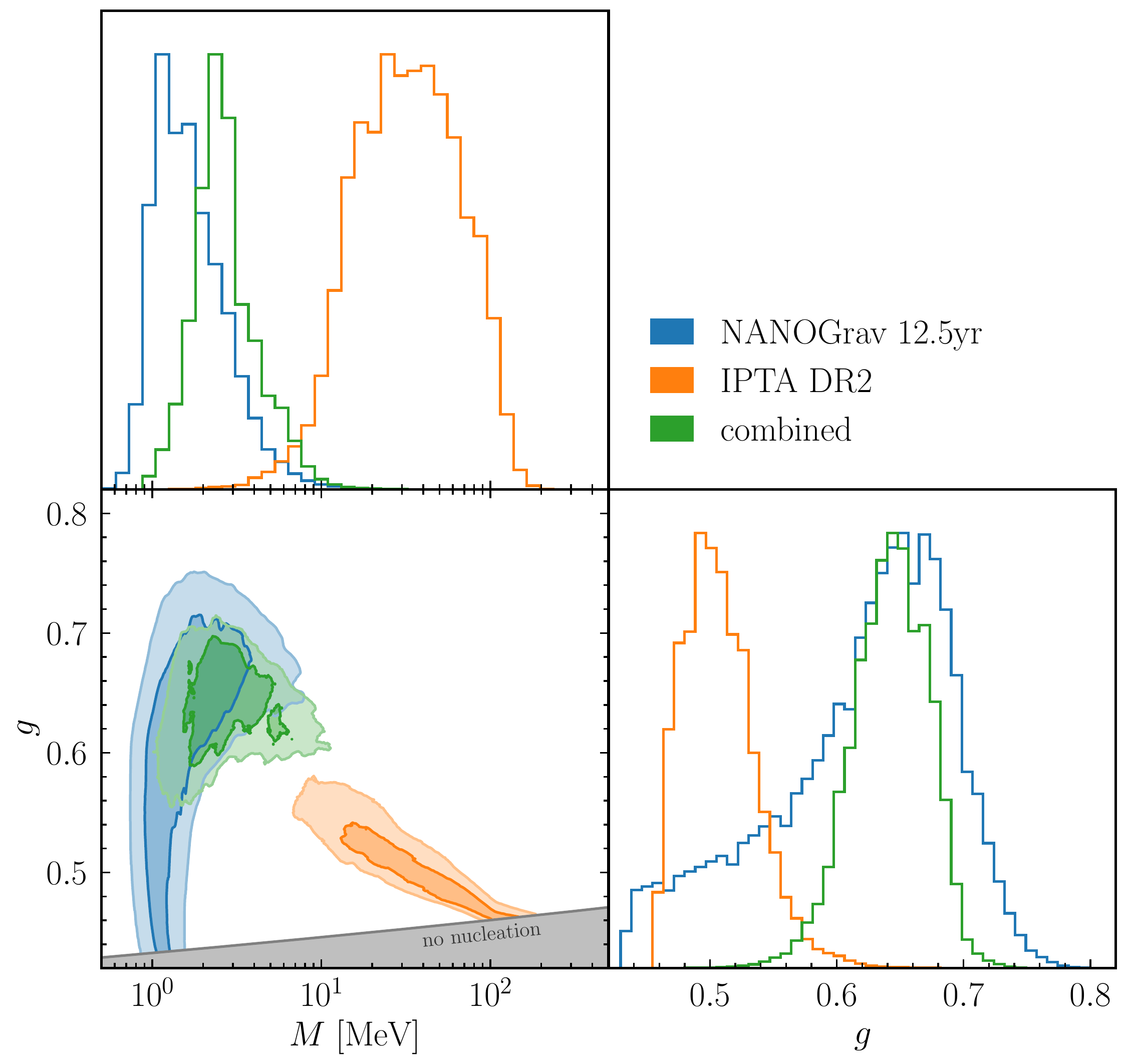}
        \caption{%
            1D and 2D posteriors for the fit to the CW model for a FOPT of \cref{sec:FOPT,sec:FOPT_result}.
        }
        \label{fig:cw_triangle}
    \end{minipage}%
    \hfill%
    \begin{minipage}{.49\textwidth}
        \centering
        \includegraphics[width=\textwidth]{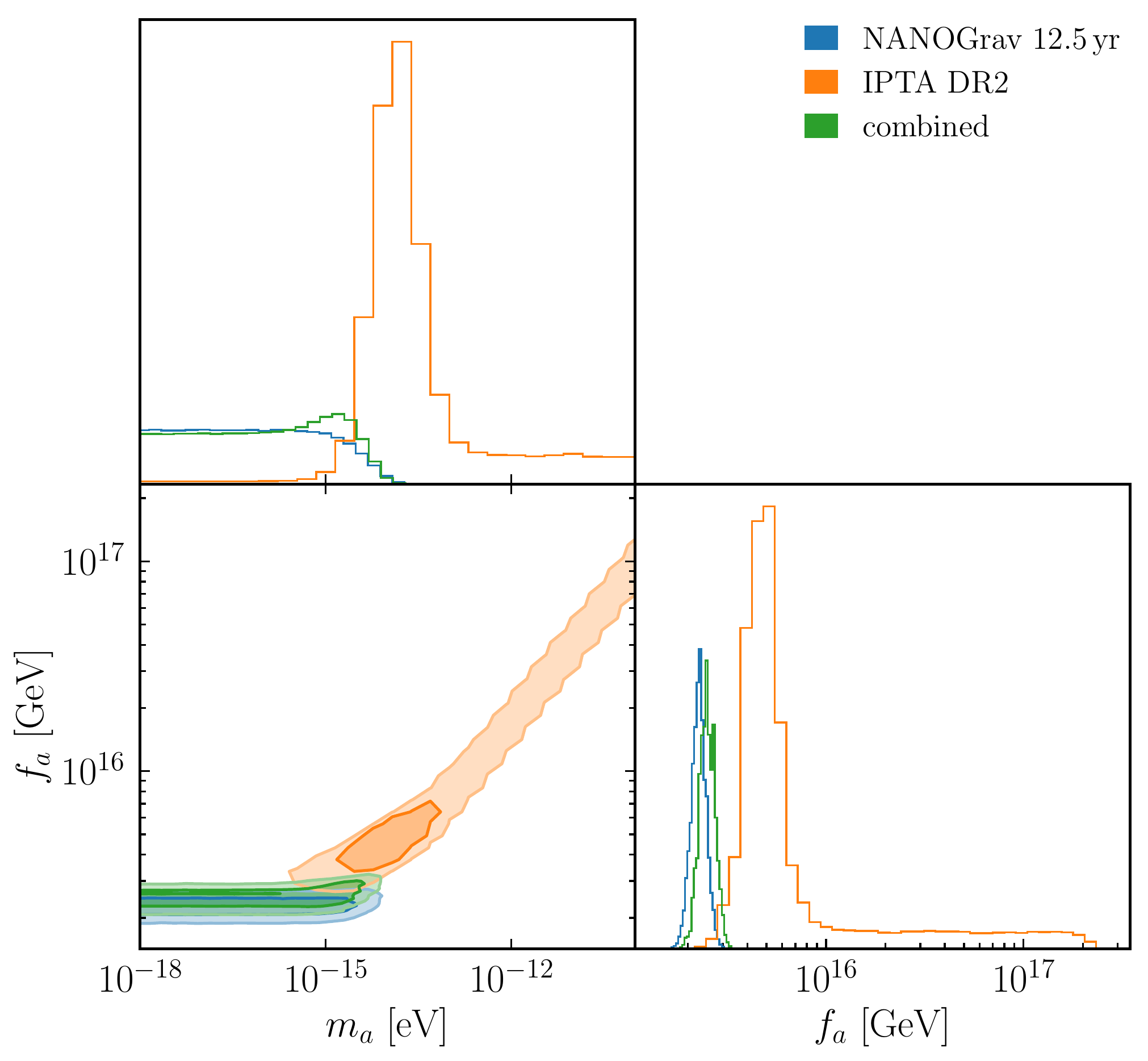}
        \caption{%
            1D and 2D posteriors for the fit to the ALP CS model of \cref{sec: Global strings,sec: Global strings results}.
            \label{fig:cs_triangle}
        }
    \end{minipage}%
\end{figure}

\begin{figure}
    \begin{minipage}{.49\textwidth}
        \centering
        \includegraphics[width=\textwidth]{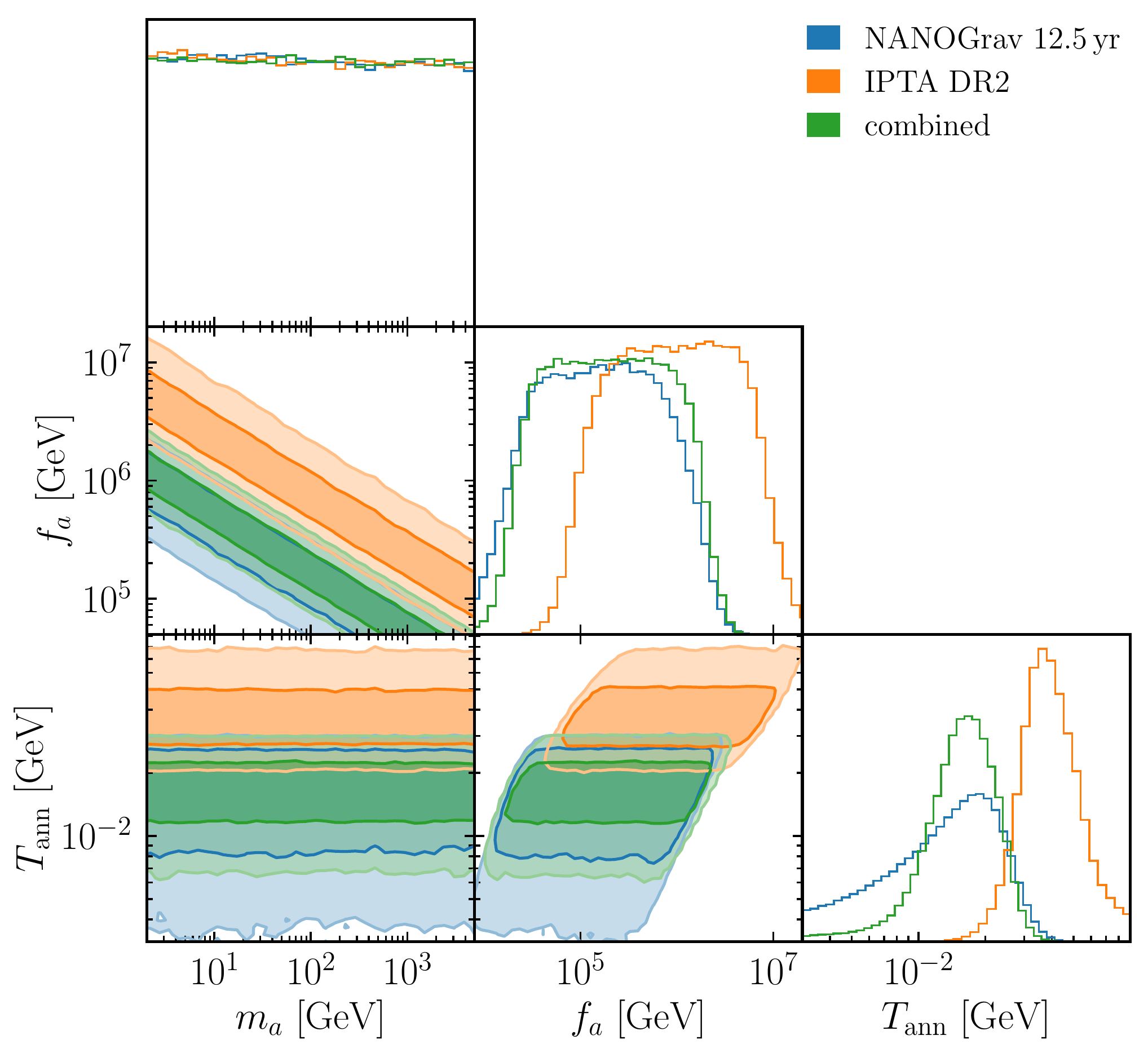}
        \caption{%
            1D and 2D posteriors for the fit to the ALP DW model of \cref{sec:ALP_DW,sec:ALP_DW_results}.
            \label{fig:alpdw_triangle}
        }
    \end{minipage}%
    \hfill%
    \begin{minipage}{.49\textwidth}
        \centering
        \includegraphics[width=\textwidth]{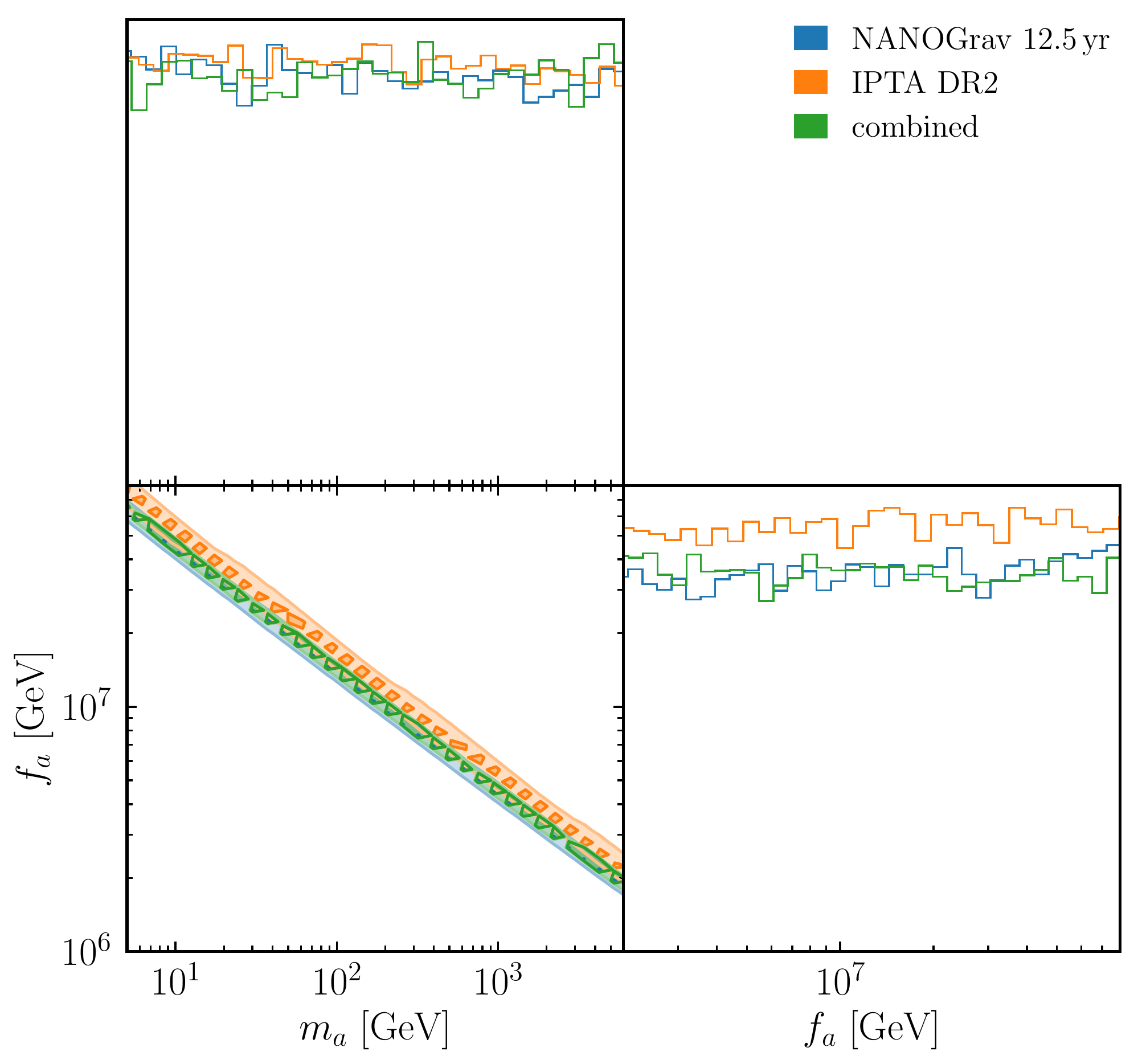}
        \caption{%
            1D and 2D posteriors for the fit to the heavy axion model generating domain walls of \cref{sec:ALP_DW,sec:ALP_DW_results}.
            \label{fig:haa_triangle}
        }
    \end{minipage}%
\end{figure}

\begin{figure}
    \begin{minipage}{.49\textwidth}
        \centering
        \includegraphics[width=\textwidth]{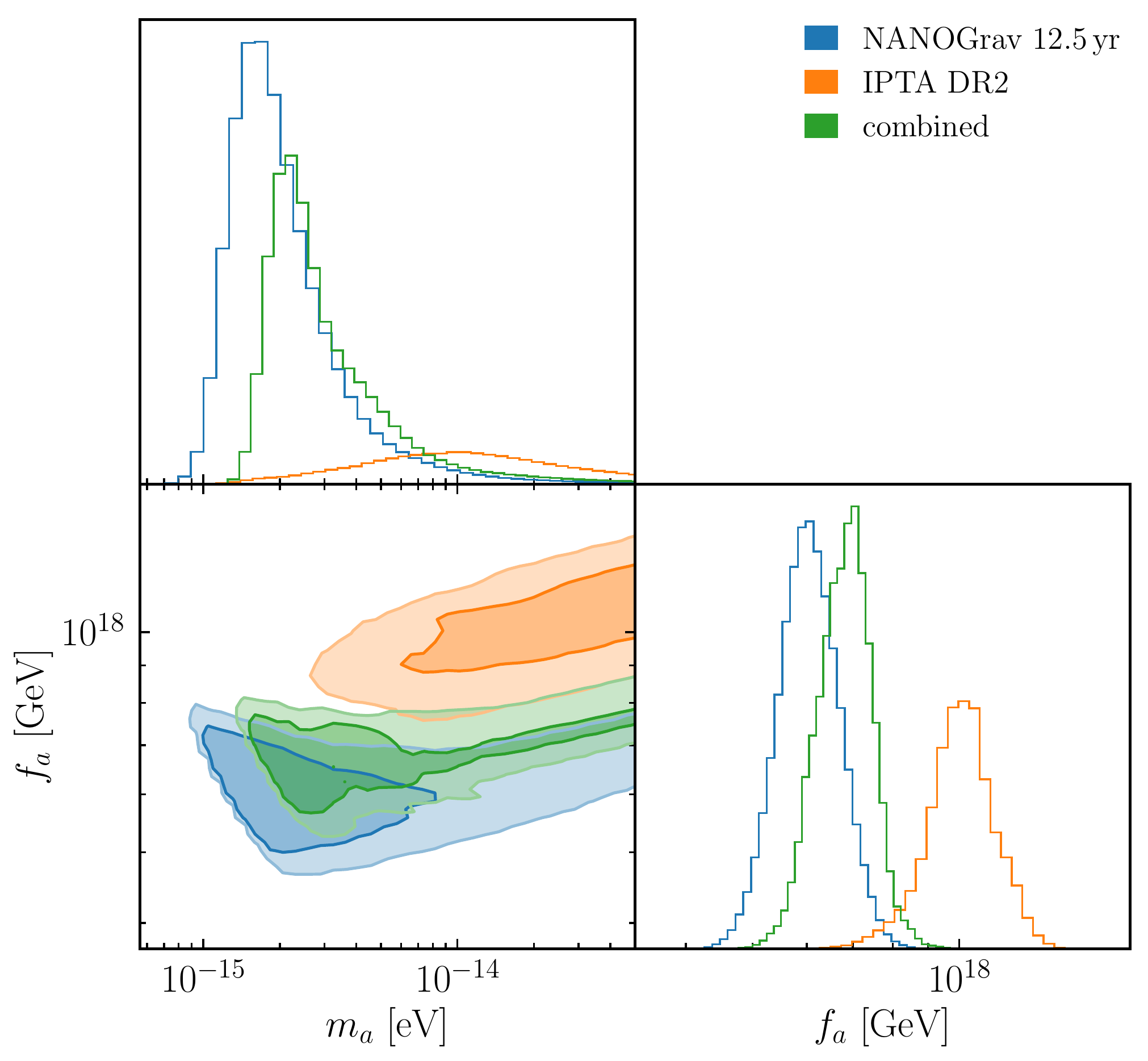}
        \caption{%
            1D and 2D posteriors for the fit to the audible axion model of \cref{sec:audible_axion,sec: Constraints AA}.
            \label{fig:aa_triangle}
        }
    \end{minipage}%
    \hfill%
    \begin{minipage}{.49\textwidth}
        \centering
        \includegraphics[width=\textwidth]{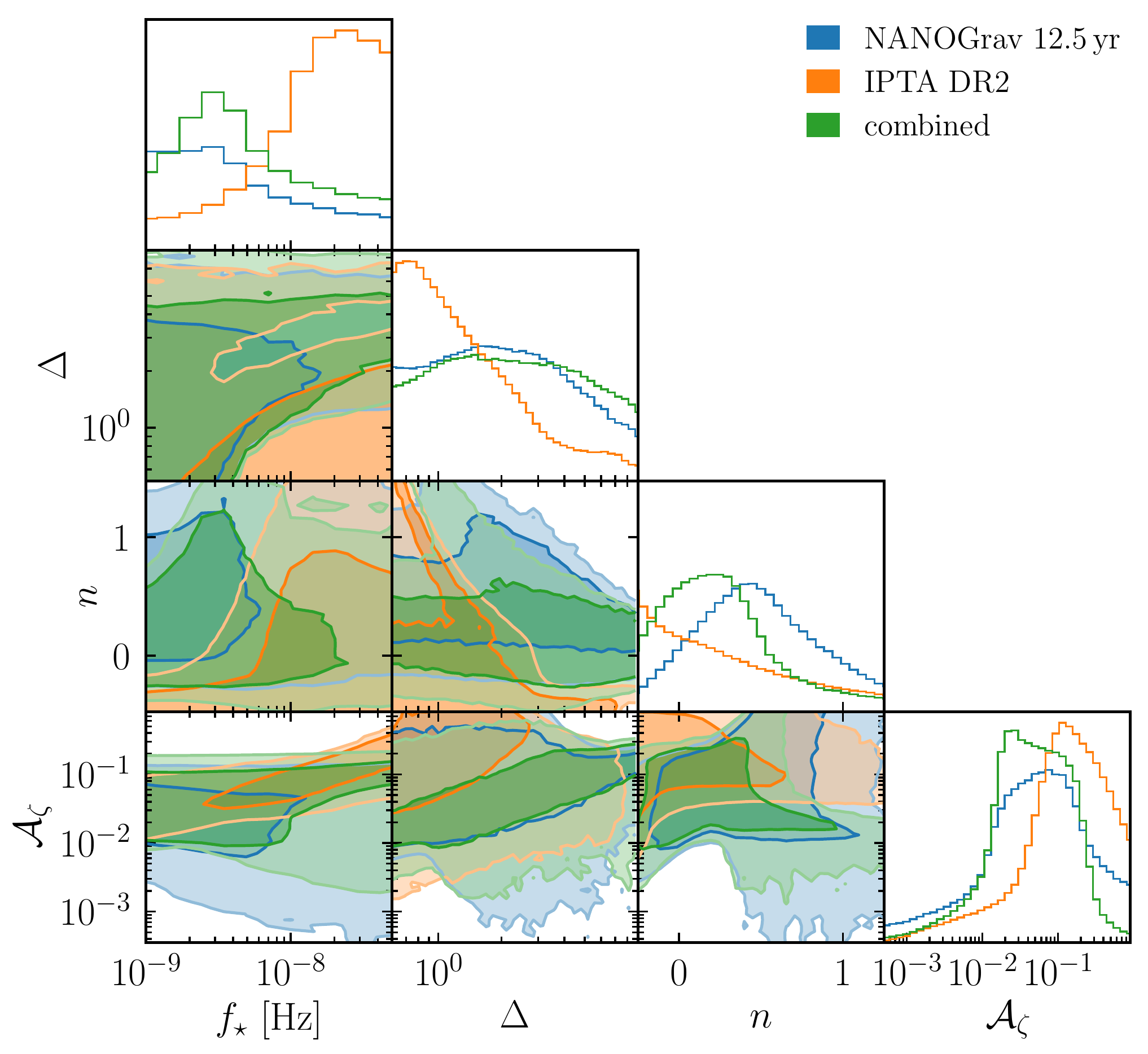}
        \caption{%
            1D and 2D posteriors for the fit to the scalar-induced GWs model of \cref{sec:scalar_induced,sec:scalar_induced_results}.
            \label{fig:scalar_triangle}
        }
    \end{minipage}
\end{figure}

% ======================================================================
%`
For completeness, here we show the full results of our fits of PTA data to the various model parameters, including the 1D posterior probabilities.
The best-fit results of the model parameters for the NANOGrav 12.5-year dataset are shown in \cref{tab:bestfits}.
The priors used for the fits in \texttt{ceffyl} are listed in \cref{tab:priors}.
\Cref{fig:cw_triangle} shows the results for a FOPT in the CW model, the results for CS from ALPs are shown in \cref{fig:cs_triangle}. The posteriors for DW networks from ALPs and heavy axions are depicted in \cref{fig:alpdw_triangle,fig:haa_triangle}, respectively, whereas \cref{fig:aa_triangle} displays the corresponding results for the audible axion model.
Last but not least, the corner plot for a scalar-induced SGWB can be found in \cref{fig:scalar_triangle}.

\begin{table}
    \centering
    \begin{tabular}{llll}
        \hline\hline
        {\bf{model}} & \multicolumn{2}{l}{\bf{parameter}} & {\bf{best-fit value}}\\\hline
        FOPT, CW & $M$ & mass scale, \cref{eq:CW_mass_scale} & $1.6_{-0.7}^{+0.6}$\,\si{\MeV}\\
            & $g$ & gauge coupling & $0.64_{-0.06}^{+0.07}$ \\
        \hline
        CS, ALP & $m_a$ & axion mass & unconstrained \\
            & $f_a$ & axion decay constant & $2.3_{-0.2}^{+0.1} \times 10^{15}$\,\si{\GeV} \\
        \hline
        DW, ALP & $m_a$ & axion mass & unconstrained \\
            & $f_a$ & axion decay constant & $3.0_{-2.6}^{+2.3} \times 10^{15}$\,\si{\GeV} \\
            & $T_{\mathrm{ann}}$ & annihilation temperature & $18.9_{-9.1}^{+7.3}$\,\si{\MeV} \\
        \hline
        DW, heavy axion & $m_a$ & axion mass & unconstrained \\
            & $f_a$ & axion decay constant & unconstrained \\
        \hline
        audible axion & $m_a$ & axion mass & $1.7_{-0.4}^{+2.1} \times 10^{-15}$\,\si{\eV} \\
            & $f_a$ & axion decay constant & $6.1_{-0.3}^{+0.5} \times 10^{17}$\,\si{\GeV} \\
        \hline
        scalar-induced & $f_\star$ & peak frequency of $P_\zeta$, \cref{eq:broken ps scalar} & unconstrained \\
            & $\Delta$ & peak width of $P_\zeta$, \cref{eq:broken ps scalar} & $1.6_{-1.0}^{+1.8}$ \\
            & $n$ & spectral tilt  of $P_\zeta$, \cref{eq:broken ps scalar}& $0.45_{-0.36}^{+0.31}$ \\
            & $\mathcal{A}_\zeta$ & amplitude  of $P_\zeta$, \cref{eq:broken ps scalar}& $7.6_{-6.1}^{+9.8} \times 10^{-2}$ \\
        \hline\hline
    \end{tabular}
    \caption{%
        Best-fit values of the model parameters obtained in \texttt{ceffyl} for the NANOGrav 12.5-year dataset. The corresponding models are presented in \cref{sec:sources,sec:results}. 
    }
    \label{tab:bestfits}
\end{table}

\begin{table}
    \centering
    \begin{tabular}{lllll}
        \hline\hline
        {\bf{model}} & \multicolumn{2}{l}{\bf{parameter}} & {\bf{prior type}} &{\bf{prior range}}\\\hline
        FOPT, CW & $M$ & mass scale, \cref{eq:CW_mass_scale} & log-uniform & \SIrange{e-1}{e3}{\MeV} \\
            & $g$ & gauge coupling & uniform & \numrange{0.4}{0.8} \\
        \hline
        CS, ALP & $m_a$ & axion mass & log-uniform & \SIrange{e-21}{e-9}{\eV} \\
            & $f_a$ & axion decay constant & log-uniform & \SIrange{e12}{e18}{\GeV} \\
        \hline
        DW, ALP & $m_a$ & axion mass & log-uniform & \SIrange{1}{e4}{\GeV} \\
            & $f_a$ & axion decay constant & log-uniform & \SIrange{e3}{e8}{\GeV} \\
            & $T_{\mathrm{ann}}$ & annihilation temperature & uniform & \SIrange{3}{100}{\MeV} \\
        \hline
        DW, heavy axion & $m_a$ & axion mass & log-uniform & \SIrange{1}{e4}{\GeV} \\
            & $f_a$ & axion decay constant & log-uniform & \SIrange{e5}{e8}{\GeV} \\
        \hline
        audible axion & $m_a$ & axion mass & log-uniform & \SIrange{6e-17}{6e-14}{\eV} \\
            & $f_a$ & axion decay constant & log-uniform & \SIrange{1.3e17}{4e18}{\GeV} \\
        \hline
        scalar-induced & $f_\star$ & peak frequency of $P_\zeta$, \cref{eq:broken ps scalar}& log-uniform & \SIrange{e-12}{e-6}{\Hz} \\
            & $\Delta$ & peak width of $P_\zeta$, \cref{eq:broken ps scalar} & log-uniform & \numrange{0.5}{10} \\
            & $n$ & spectral tilt  of $P_\zeta$, \cref{eq:broken ps scalar}& uniform & \numrange{-0.5}{1.5} \\
            & $\mathcal{A}_\zeta$ & amplitude  of $P_\zeta$, \cref{eq:broken ps scalar}& log-uniform & \numrange{3e-4}{1} \\
        \hline\hline
    \end{tabular}
    \caption{%
        Parameter priors used in \texttt{ceffyl} for the models presented in \cref{sec:sources,sec:results}.
    }
    \label{tab:priors}
\end{table}

% ======================================================================
\section{Choosing the IPTA frequency range}
\label{app:IPTA_bins}
% ======================================================================

In the main body of this work, we present results from IPTA data based on twelve frequency bins. 
We here briefly investigate the effects of changing the number of bins in the analysis.

We use the second IPTA data release~\cite{Perera:2019sca}, consisting of 65 pulsars with an observation time of up to 30~years.
The dataset comprises data from the first data release of EPTA~\cite{Desvignes:2016yex}, the extended first data release of PPTA~\cite{Manchester:2012za,Reardon:2015kba}, and the NANOGrav nine-year dataset~\cite{NANOGrav:2015qfw}.
Similar to the NANOGrav collaboration~\cite{NANOGrav:2020bcs}, IPTA has performed a free spectrum fit with 30 frequency bins~\cite{Antoniadis:2022pcn}.
We repeat this fit using the \texttt{enterprise} code~\cite{enterprise} with 
\texttt{enterprise\_extensions}~\cite{enterprise_extensions} and process the result for further fitting in \texttt{ceffyl}~\cite{Lamb:2023jls}.

We here follow the NANOGrav search for a SGWB from a FOPT~\cite{NANOGrav:2021flc} and focus on the first five frequency bins of the 12.5~year dataset, as these are most sensitive to a cosmological background and give the dominant  contribution the SNR of the observed common red process signal~\cite{NANOGrav:2020bcs}.
This hence includes frequencies below \SI{12.5}{\nano\Hz}, corresponding to the twelfth frequency bin of the IPTA data. 
Including all IPTA bins with frequencies below the sixth NANOGrav bin at about \SI{15}{\nano\Hz}, on the other hand, increases the number of bins to up to fourteen.\footnote{IPTA uses thirteen bins in their SGWB search~\cite{Antoniadis:2022pcn}.}

\begin{figure}
    \includegraphics[width=.33\textwidth]{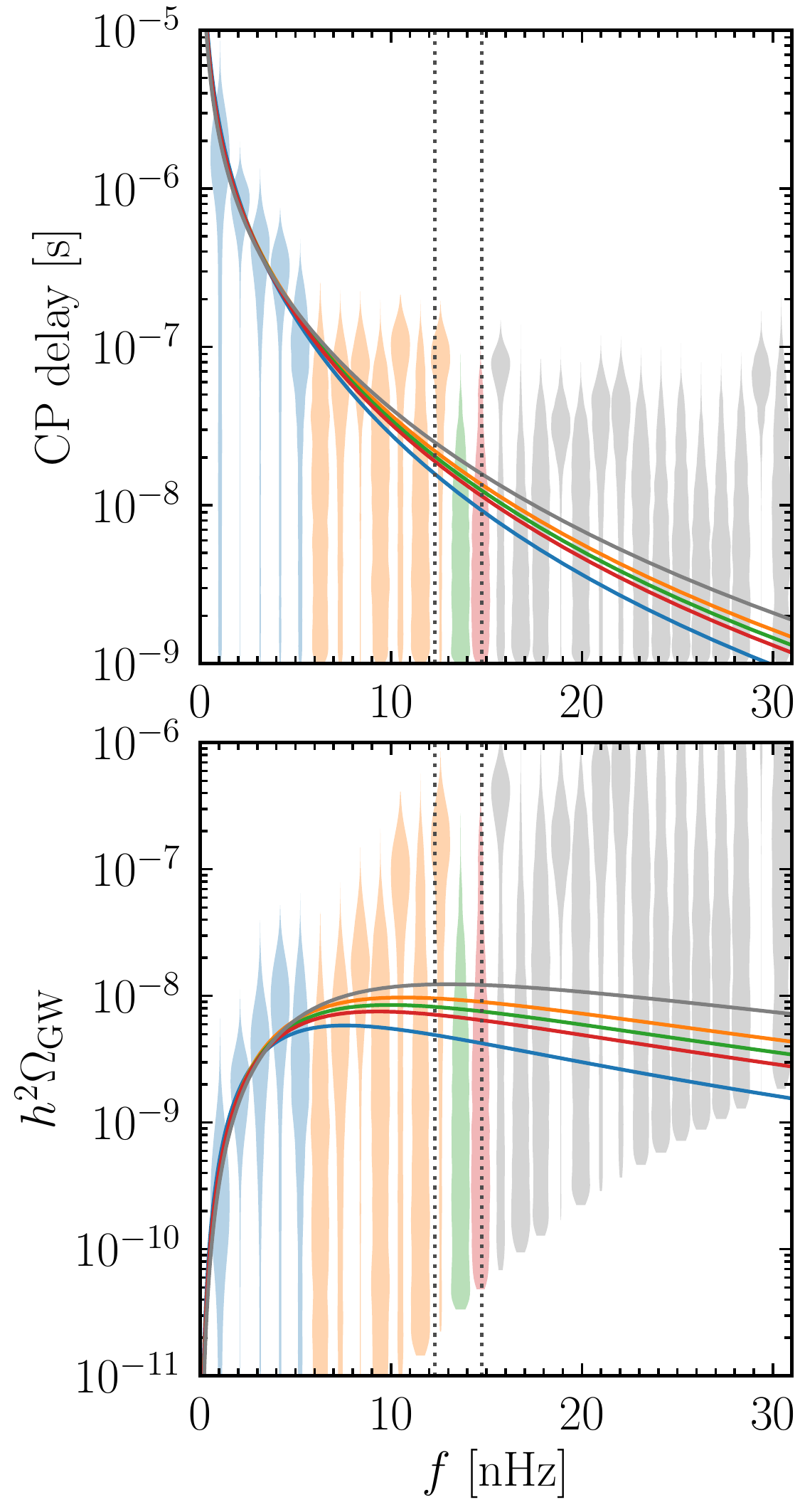}
    \hfill
    \includegraphics[width=.66\textwidth]{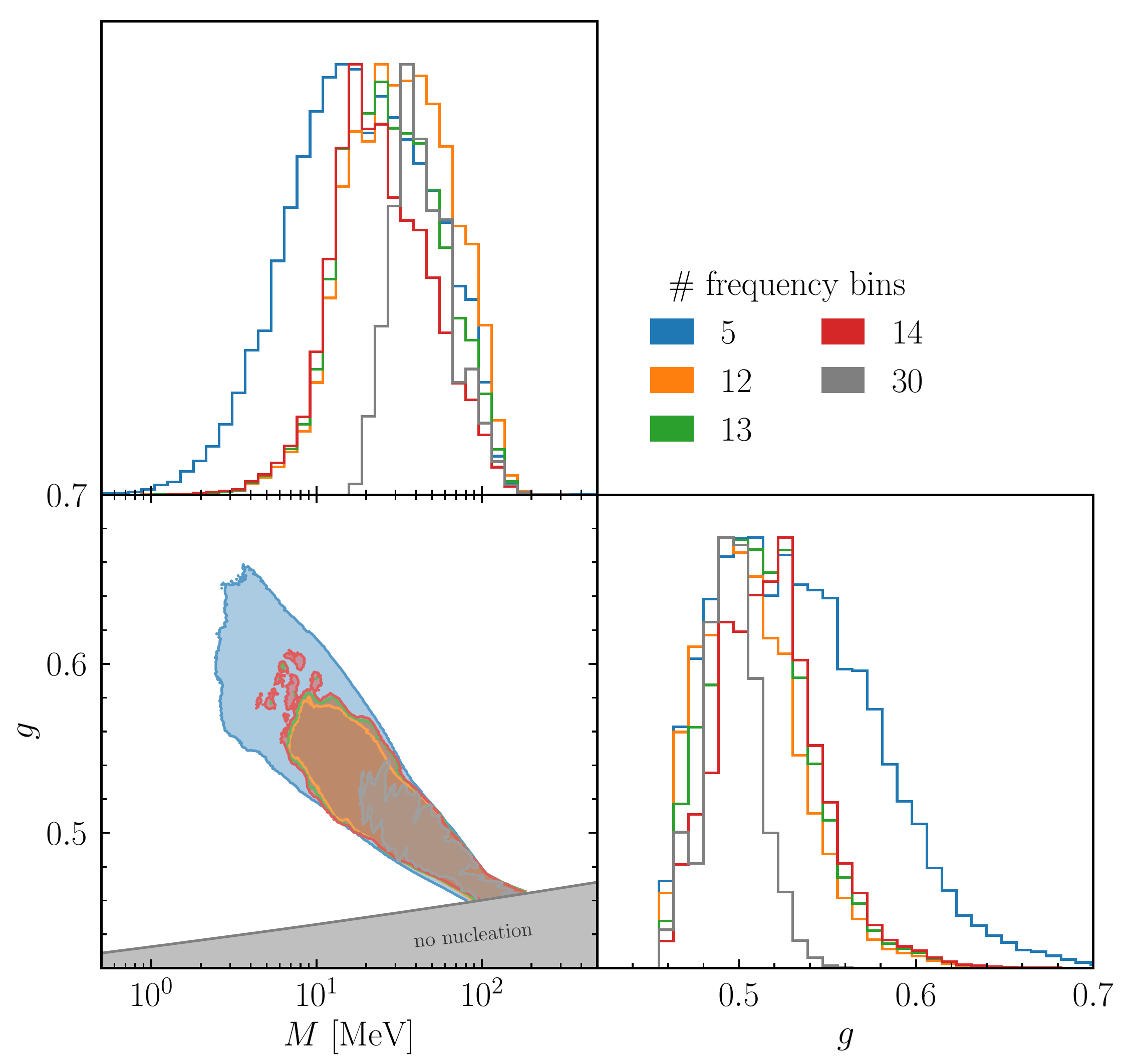}
    \caption{%
        \textit{Left:} Fit of the GW spectum of the CW model to IPTA data, using different numbers of frequency bins. The two vertical dotted lines indicate the frequency of the fifth and sixth frequency bin of NANOGrav.\\
        \textit{Right:} Corresponding posteriors of the CW model parameters and 95\,\% contours.
    }
    \label{fig:IPTAfreqs}
\end{figure}

In \cref{fig:IPTAfreqs} we hence show results for fits of the SGWB from a FOPT in the CW model taking into account five, twelve, thirteen, fourteen, as well as all thirty frequency bins of IPTA. 
The left panel depicts the resulting mean spectrum~(solid lines) in fractional energy density~$\Omega_\GW = \rho_\GW/\rho_\mathrm{crit}$ (bottom) and timing residual amplitude~(top).
The violins represent the result of the thirty-bin free-spectrum fit, and the vertical dotted lines indicate the position of the fifth and sixth NANOGrav frequency bin.
The corner plot in the right panel depicts the posterior probabilities in terms of the mass scale~$M$ and gauge coupling~$g$. 
The colored regions in the lower left corner correspond to the 95\,\%~CL region.

An analysis including only the first five bins (i.e.\ the same number of bins as NANOGrav) yields on average a GW spectrum peaking at a few \si{\nano\Hz}, corresponding to transitions at temperatures and mass scales of a few \si{\MeV}. 
Including more frequencies moves the peak to higher frequencies and scales around \SIrange{10}{100}{\MeV}, leading to a slight tension with our NANOGrav 12.5~year fit (cf.\ also \cref{fig:fitPT}).
As the thirteenth and fourteenth bin, however, favor lower amplitudes, adding these bins shifts the spectrum back to somewhat lower frequencies, alleviating the tension a bit.
Including further bins then again moves the spectrum back to higher scales. 

% ======================================================================
\section{Details on the Coleman-Weinberg model}
\label{app:CW_details}
% ======================================================================

Neglecting the tree-level contribution to the potential, the leading-order effective potential consists of the Coleman-Weinberg zero-temperature and the thermal one-loop potential.
The former is given by~\cite{Coleman:1973jx}
\begin{equation}
    V_{CW} = \sum_{i} N_i \frac{m_i^4(\phi)}{64\pi^2}\left[\log\left(\frac{m_i^2(\phi)}{\mu_R^2}\right)-c_i\right] \,,
    \label{eq:VCW}
\end{equation}
where the sum runs over species with field-dependend masses $m_i^2(\phi)$, $N_i$ counts the degrees of freedom for each species, $c_i=5/2$ ($3/2$) for gauge bosons (scalars and fermions) respectively, and $\mu_R$ is the renormalization scale.
In our case, the only species contributing are the gauge bosons with $N_A = 3$ and $m_A^2 = g^2 \phi^2$.
Using this, we obtain a zero-temperature vacuum expectation value $\phi_0=(e^{1/6} \mu_R)/g$.
For the thermal corrections, we use the high-temperature expansion, which for bosonic DOFs becomes
\begin{align}
    V_T = T^4 \sum_i N_i \left[\frac{1}{24} \frac{m_i^2(\phi)}{T^2} - \frac{1}{12 \pi} \left(\frac{m_i^2(\phi)}{T^2}\right)^\frac{3}{2} - \frac{m_i^4(\phi)}{64 \pi^2 T^4} \log\frac{m_i^2(\phi)/T^2}{16 \pi^2 e^{\frac{3}{2}-2\gamma_E}} \right] \,
    \label{eq:VT}
\end{align}
where $\gamma_E$ is the Euler-Mascheroni constant. 
Summing \cref{eq:VCW,eq:VT}, the logarithms of the field cancel, and one obtains the polynomial potential in \cref{eq:V1loop_approx}.
Note that we here neglect the contribution from Daisy resummations.
These become important for $g<0.53$~\cite{Levi:2022bzt}, and should hence be negligible for most of the parameter space we consider.

It is possible to further redefine the potential \cref{eq:V1loop_approx} in terms of a single parameter $\kappa\equiv\frac{\lambda(T) m^2(T)}{\delta^2(T)}$, leading to a simplified form
\begin{equation}
    \label{eq:Vtilde}
    \Tilde{V}_1(\phi,T)= \frac{1}{2}\phi^2-\frac{1}{3}\phi^3+\frac{\kappa(T)}{4}\phi^4
\end{equation}
for which the tunneling action can be calculated as a function of a $\kappa$ only.
See Ref.~\cite{Levi:2022bzt} for further details.
For the CW model in the high-temperature expansion, $\kappa = \log(T/M)/6$.

Now we want to calculate the microphysics inputs to describe PTs, i.e.\ the PT strength $\alpha$, the duration of the PT $\beta$ and the temperature at which the PT occurs $T_*$. 
For this we use the approximations in Ref.~\cite{Levi:2022bzt} and calculate the three-dimensional Euclidean tunneling action~$S_3(T)$ of the bounce solution~\cite{Coleman:1977py,Callan:1977pt,Linde:1980tt,Linde:1981zj} as a function of $\kappa$ from \cref{eq:Vtilde}.
The corresponding expression for $S_3/T$ in terms of $\kappa$ can be found in Ref.~\cite{Levi:2022bzt}.
We here employ the tunneling potential method~\cite{Espinosa:2018hue}.
The transition strength and inverse timescale can be calculated from the potential and tunneling actions as
\begin{align}
    \alpha &\approx \frac{\Delta V(T_*)}{\rho_R(T_*)} \,,&
    \beta &= \left[ T \frac{d}{dT} \frac{S_3}{T}\right]_{T=T_*} \,,
\end{align}
where we need to calculate the parameters of the PT at the temperature $T_*$. A good approximation in the case of a supercooled sector is using the percolation temperature $T_p$.
In the expressions above $\rhoR\equiv\pi^2(\geff^\mathrm{SM}+\geff^\mathrm{BSM}) T^4/30$ is the energy density of the universe in radiation at the time of the transition, and $\Delta V(T)$ is the positive potential difference between the false and true vacuum. 
While \cref{eq:V1loop_approx} gives a good approximation around the barrier and hence also for the tunneling action, we obtain $\kappa<0$ in a large fraction of the parameter space.
It can thus not be used to determine the finite-temperature minimum.
As we are considering supercooled transitions, we however expect that the corrections around the true vacuum at the percolation time are small.
Therefore, we approximate the transition strength using the  difference of the potential at zero temperature, $\alpha = \Delta V_0/\rhoR$. For $\Delta V_0$ we obtain $\Delta V_0 =43.5\,M^4$~\cite{Levi:2022bzt} using the CW potential.

% ======================================================================
\section{FOPT GW spectrum and PT parameter fit}
\label{app:FOPT_SGWB}
% ======================================================================

We here present results for supercooled phase transitions in terms of the thermodynamics parameters describing a FOPT, i.e.\ the percolation temperature $T_p$, the transitions strength~$\alpha$, and its inverse timescale~$\beta$.
As before, the wall velocity is fixed to $v_w=1$.

The present-day GW spectrum of a supercooled PT is the sum of the contributions from vacuum bubble and fluid shell collisions, $\Omega_\GW h^2 = \Omega_\GW^\mathrm{vac} h^2 + \Omega_\GW^\mathrm{fl} h^2$.
The latter two take the form~\cite{Lewicki:2022pdb}
\begin{align}\begin{aligned}
	\label{eq:FOPT_spectrum}
        \Omega_\GW(f) \,h^2 &= \num{1.67e-5} \left[\frac{\geff(\Trh)}{100}\right] \!\left[\frac{100}{\gs(\Trh)}\right]^\frac{4}{3} \left[\frac{H_*}{\beta}\right]^2 \left[\frac{\kappa \,\alpha}{1+\alpha}\right]^2 
	\frac{A\, (a+b)^c}{\left[ b \left(\frac{f}{f_p}\right)^{-\frac{a}{c}} + a \left(\frac{f}{f_p}\right)^{\frac{b}{c}}  \right]^c} \,,
	\\
	\text{with}\quad f_p &= \SI{16.5}{\nano\Hz} \left[\frac{\geff(\Trh)}{100}\right]^\frac{1}{2} \left[\frac{100}{\gs(\Trh)}\right]^\frac{1}{3} \left[\frac{\Trh}{\SI{100}{\MeV}}\right] \left[\frac{\beta}{H_*} \right] \,r_p\,,
\end{aligned}\end{align}
where \geff and \gs are the effective number of relativistic and entropic degrees of freedom, and the reheating temperature $\Trh = T_p (1+\alpha)^{1/4}$ is used for the redshift factors~\cite{Athron:2022mmm}.
The coefficients~$A$ and~$r_p$ as well as the spectral slope parameters~$a$, $b$ and~$c$ depend on the contribution under consideration and, in case of the vacuum bubble collisions, on whether the broken symmetry is gauged or global.
We refer the reader to table~I of Ref.~\cite{Lewicki:2022pdb} for the corresponding numerical values.
The respective efficiency factors are $\kappa_\mathrm{vac} = 1/(1+\Reff/\Req)$ for the vacuum bubble collisions and $\kappa_\mathrm{fl} =  \left(1-\kappa_\mathrm{vac}\right) \kappa_\mathrm{sw}(\alpha)$ for the fluid shells, with the sound wave efficiency~$\kappa_\mathrm{sw}(\alpha)$ taken from Ref.~\cite{Espinosa:2010hh}. 
We use  $\Reff = 5/\beta$ for the bubble radius upon collision~\cite{Lewicki:2022pdb}.
The radius~$\Req$ at which the leading and next-to-leading order pressures~\cite{Bodeker:2009qy,Bodeker:2017cim} exerted on the bubble wall are equal depends on the underlying particle physics model.

\begin{figure}
    \includegraphics[width=.33\textwidth]{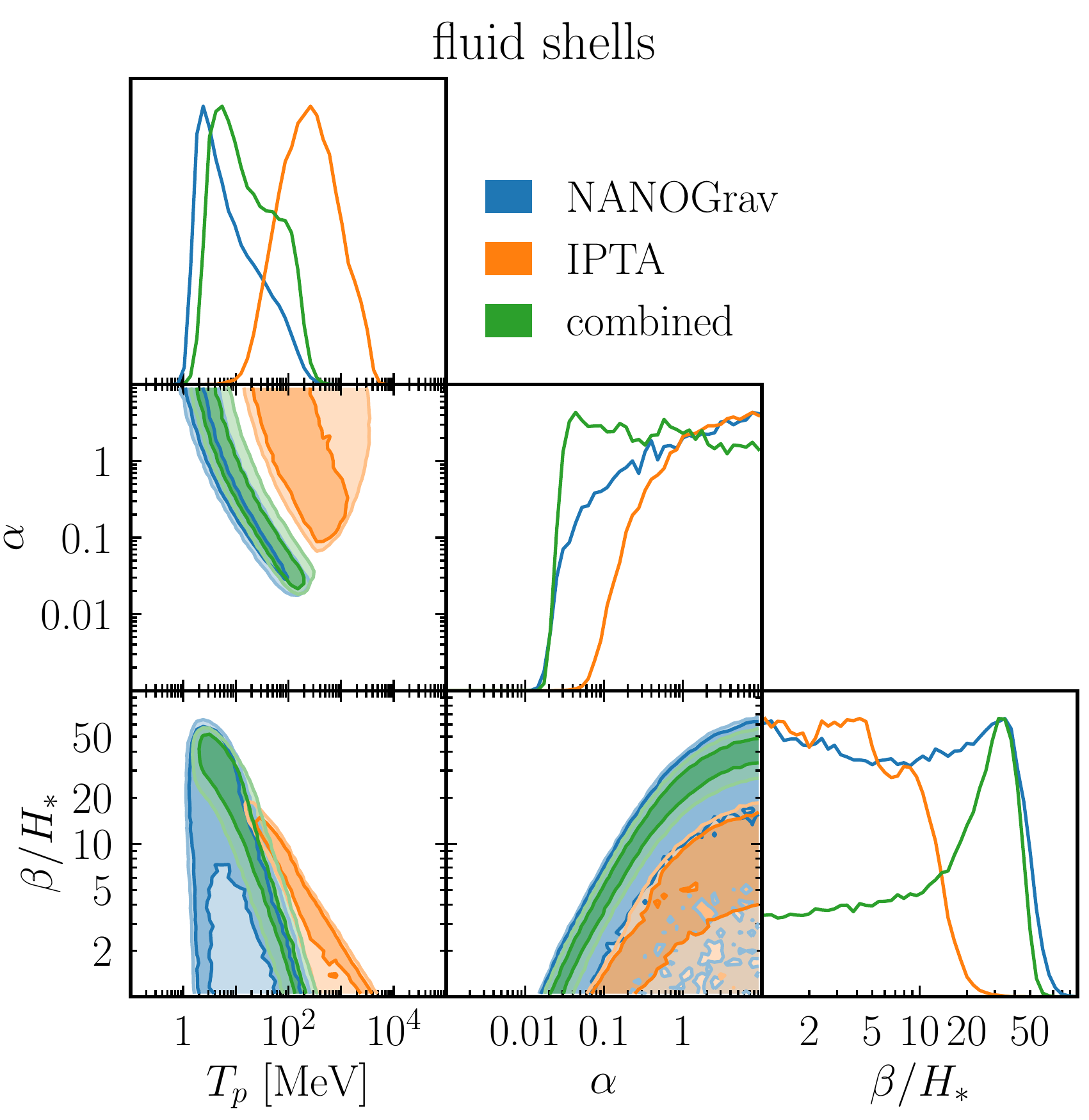}\hfill
    \includegraphics[width=.33\textwidth]{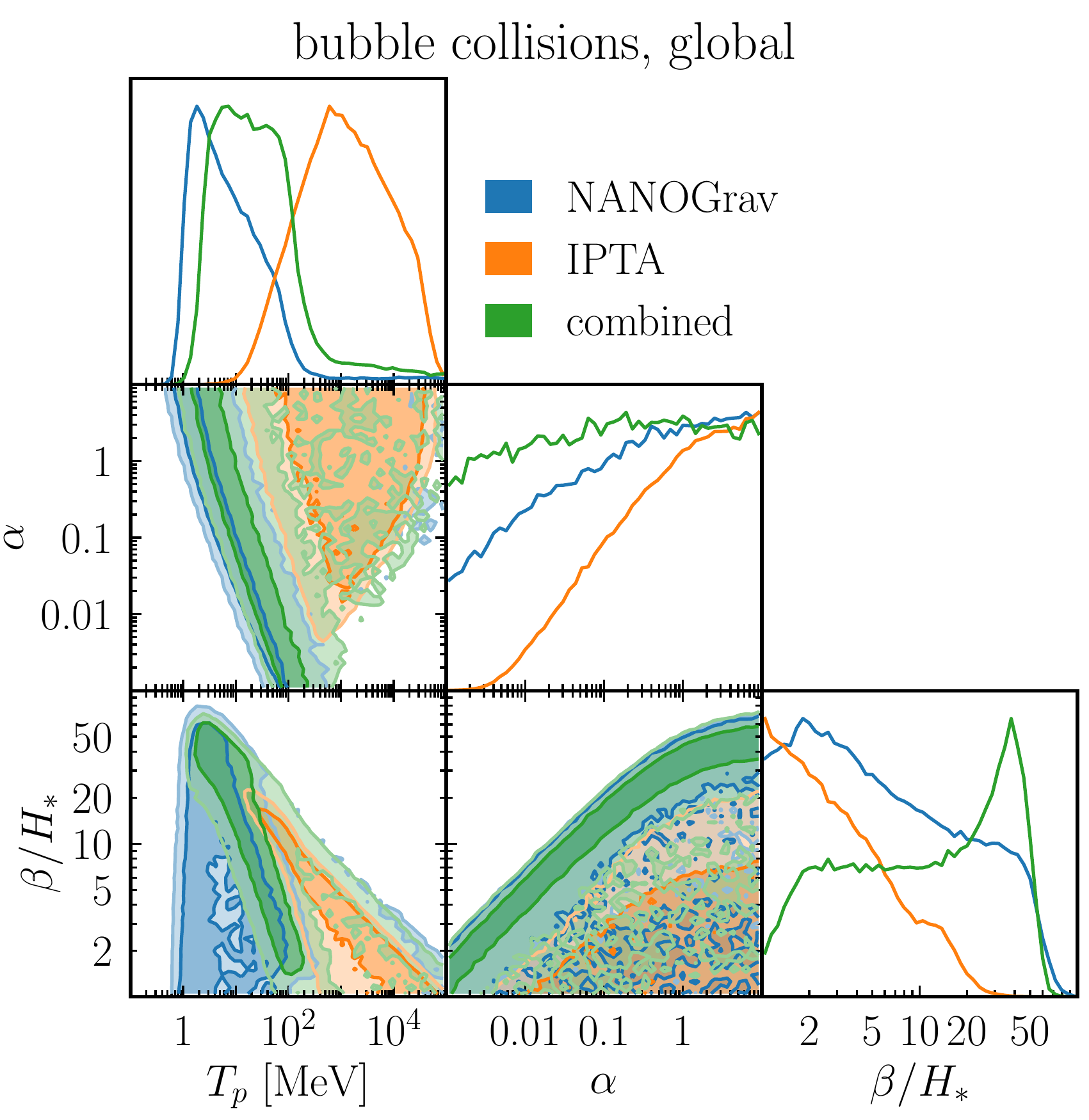}\hfill
    \includegraphics[width=.33\textwidth]{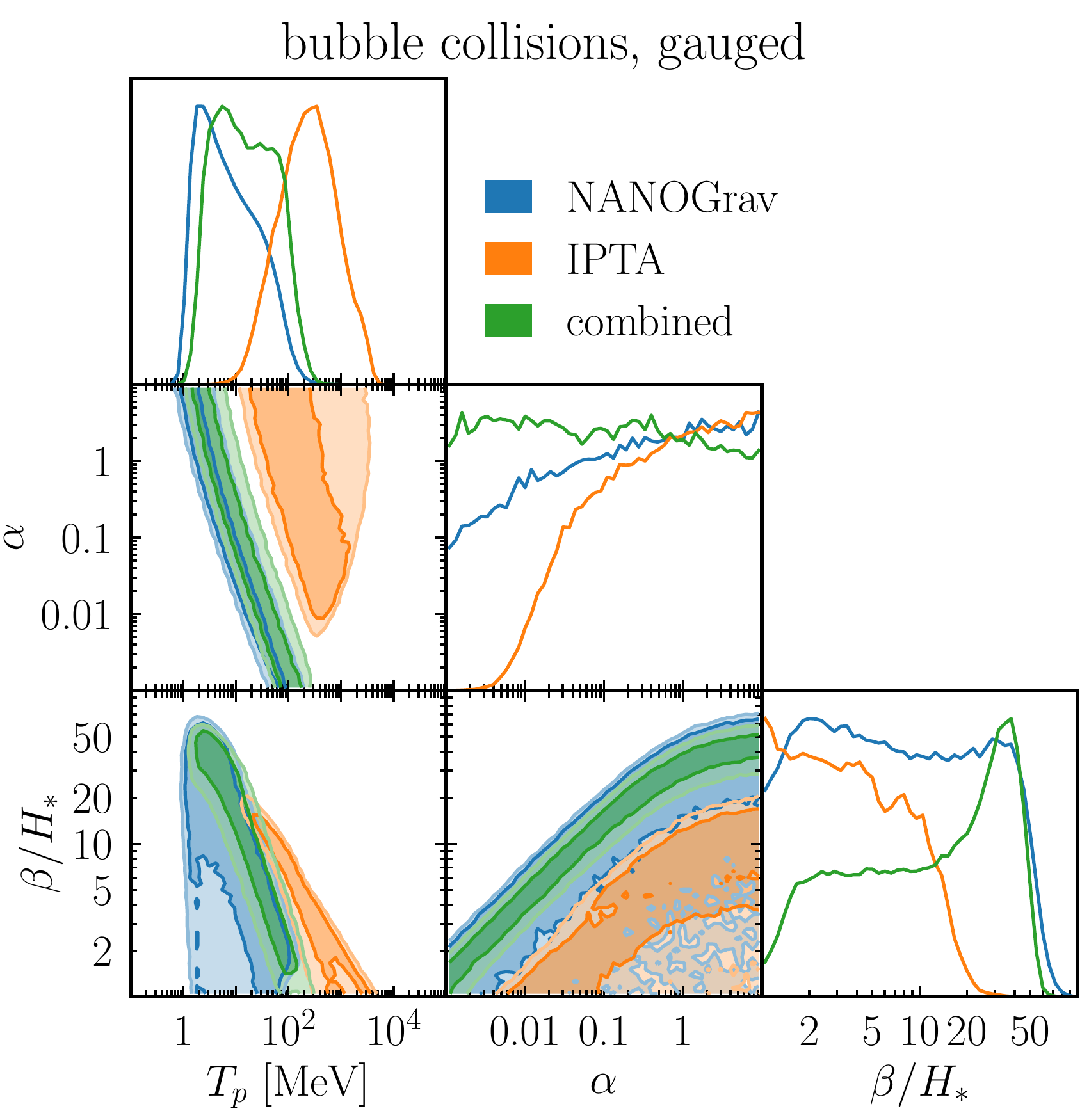}
    \caption{%
        1D and 2D posteriors for the fit to the SGWB contributions from a supercooled PT of \cref{app:FOPT_SGWB} in terms of the PT parameters for collisions of fluid shells~(left) as well as vacuum bubbles from breaking a global~(center) or gauged~(right) symmetry.
    }
    \label{fig:pt_param_fit}
\end{figure}

\Cref{fig:pt_param_fit} depicts a fit of the GW spectrum contributions in \cref{eq:FOPT_spectrum} in terms of the PT parameters with \texttt{ceffyl}, using the NANOGrav 12.5~year dataset~(blue) and IPTA DR2~(orange).\footnote{%
	A similar fit to the NANOGrav data has also been performed by the collaboration itself~\cite{NANOGrav:2021flc}, considering the sound wave contibution to the GW spectrum, as well as a slightly different shape for the bubble collisions.
}
We also show the naive combination of the two datasets~(green).
The fit considers only one contribution at a time, setting the bubble collision efficiency factors to $\kappa_\mathrm{vac}=1$.
We here employ uniform priors on $\log_{10}(T_p/\si{\GeV}) \in (-4,2)$, $\log_{10}(\alpha) \in (-3,1)$ and $\log_{10}(\beta/H_*) \in (0,3)$. 
Note that, although large values of $\alpha$ can be obtained for instance in the CW model, we here only consider $\alpha<10$ as the kinetic energy fraction $K\equiv\alpha/(1+\alpha)$ becomes constant~($K\to 1$) for large $\alpha\gg 1$.

NANOGrav favors MeV-scale transition temperatures, fitting the common red process with the high-frequency part of the spectrum with the peak around or below the first frequency bin, whereas IPTA prefers percolation temperatures around the GeV scale, using the rising part of the spectrum to fit the data; cf.~discussion in \cref{app:IPTA_bins}.
The NANOGrav fit further allows for lower transition strengths and larger inverse timescales.

% ======================================================================
\section{Examples of scalar-induced GWs}
\label{app:scalar_induced}
% ======================================================================

\begin{figure}
    \centering
    \includegraphics[width=.3\textwidth]{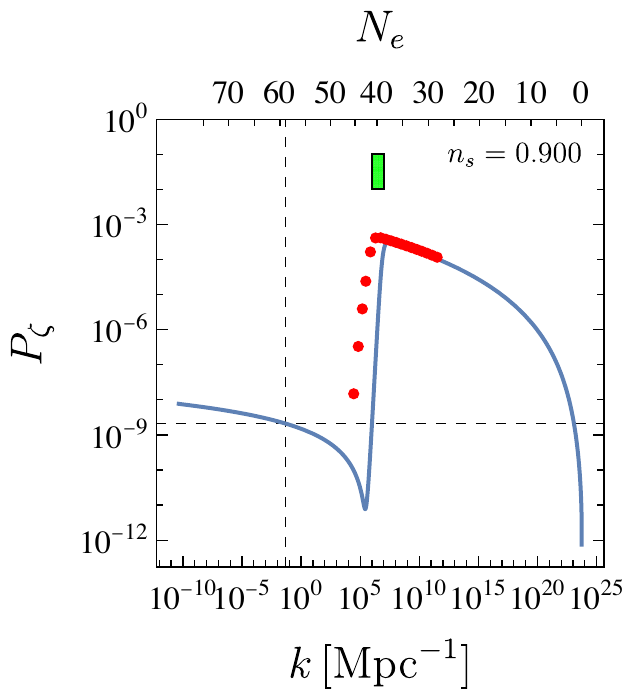}
    \includegraphics[width=.3\textwidth]{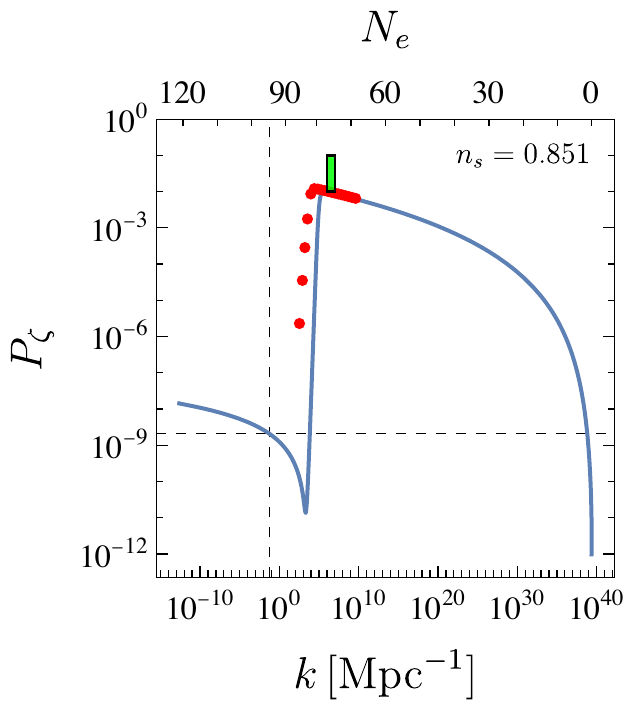}
    \includegraphics[width=.3\textwidth]{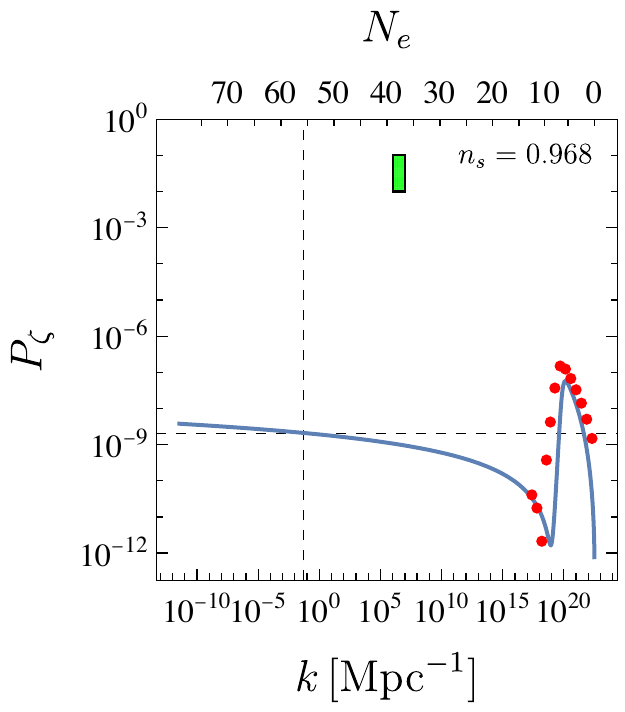}
    \caption{Scalar power spectrum for the three benchmark points in \cref{tab:scalar induced params}. The blue continuous lines are obtained with the approximate slow-roll result, \cref{eq:scalar PS SR}. The red dots around the peak region are obtained by solving the Mukanov-Sasaki equation for the perturbations. The green vertical bands identify the target region that could fit PTA data~\cite{Dandoy:2023jot}. The black dashed lines identify the CMB pivot scale. We indicate in the plots the value of $n_s$ at the CMB scale. \textit{Left:} a spectrum with a peak in the PTA band and $N_e \simeq 60$ efolds of inflation always have too small amplitude and low $n_s$. \textit{Center:} A point that could generate the PTA signal requires a much more than 60 efolds of inflation. \textit{Right:} Matching CMB observations in both $P_\zeta$ and $n_s$ leads to very small peaks at much smaller scales.}
    \label{fig:scalar induced benchmarks}
\end{figure}

We choose three representative benchmark points, with parameters listed in \cref{tab:scalar induced params}, that exemplify the three possible behaviours. The power spectra are shown in \cref{fig:scalar induced benchmarks}.
In the first, the peak is at the correct position and the pivot scale exits the horizon about 60 efolds before the end of inflation. The spectral index, $n_s\simeq 0.9$, is as high as it can be in this case, and the amplitude of the signal is too low to be visible. In the second example, the spectrum can lead to a measurable signal, but the number of efolds of inflation turns out to be unacceptably large and the spectral index is even lower than in the former case. The last benchmark, instead, reproduces CMB observations (spectral index and amplitude), but the peak has a very small amplitude and is located at much smaller scales. 
\begin{table}
    \centering
    \begin{tabular}{cccccc}
    \hline\hline
         & $\log_{10}\lambda_0$ & $\xi_0$ & $\phi_0/M_P$ & $x_c$ & $\log_{10}\beta$ \\
    \hline
       priors & $[-7.25,-5.5]$ & $[3,9]$ & $[0.1,0.65]$ & $[0.7,0.875]$ & $[-6.5,-4]$ \\
       1 & $-6.75$ & $4.5$ & $0.45$ & $0.775$ & $-5.5$ \\
       2 & $-7.0$ & $2.5$ & $0.6$ & $0.775$ & $-6.5$ \\
       3 & $-7.0$ & $5.5$ & $0.3$ & $0.8$ & $-6.5$ \\
    \hline\hline
    \end{tabular}
    \caption{Range for the flat priors used in the scan of inflection point inflation scenarios and parameter choice for the benchmark points. We scan a total of roughly \num{2.3e5} points. The exact value of $\beta$ becomes irrelevant below $\sim 10^{-6.5}$, thus smaller values are not considered.}
    \label{tab:scalar induced params}
\end{table}

\clearpage

% ======================================================================

\bibliographystyle{JHEP}
\bibliography{sample}

% %%%%%%%%%%%%%%%%%%%%%%%%%%%%%%%%%%%%%%%%%%%%%%%%%%%%%%%%%%%%%%%%%%%%%%
\end{document}